\documentclass[review,12pt]{elsarticle}

\usepackage[top=1in, bottom=1in, left=1in, right=1in]{geometry}

\usepackage{lineno,hyperref}
\modulolinenumbers[5]

\usepackage{natbib}

\usepackage[T1]{fontenc}

\biboptions{authoryear}

\begin{document}
\begin{frontmatter}
	
\title{ {\scriptsize   A. Kuzmin, X-ray absorption spectroscopy in high-entropy material research, \\
in {\it High-Entropy Alloys: Design, Manufacturing, and Emerging Applications}, \\
G. Yasin, M. Abubaker Khan, M. Afifi, T. Anh Nguyen, Y. Zhang (Eds.),  pp. 121-155 (Elsevier, 2024). \\
 Doi: 10.1016/B978-0-443-22142-2.00006-5.} \\  
{\normalsize     }
 \textbf{6 -- X-ray absorption spectroscopy in high-entropy material research}}
 
\author{Alexei Kuzmin} 

\address{Institute of Solid State Physics, University of Latvia, Riga, Latvia}

\begin{abstract}
	
This chapter introduces the use of X-ray absorption spectroscopy (XAS) in studying the local electronic and atomic structure of high-entropy materials. The element selectivity of XAS makes it particularly suitable to address the challenges posed by the study of multicomponent compounds. By analysing different parts of the X-ray absorption spectra for each element, one can obtain information on its chemical state from the X-ray absorption near-edge structure (XANES) and its local environment, distortions, and lattice dynamics from the extended X-ray absorption fine structure (EXAFS). The theoretical background underlying X-ray absorption spectra and existing data analysis procedures are briefly described, with particular emphasis on advanced atomistic modelling techniques that enable more reliable extraction of structural information. Finally, an overview of the applications of the XAS technique in studying high-entropy materials is presented.  \\
	
\end{abstract}


\end{frontmatter}

\newpage


\subsection{Introduction}

High-entropy materials (HEMs) are a class of materials that contain five or more principal elements in equal or near-equal atomic fraction (\cite{Brechtl2021,George2019}). The first HEMs were high-entropy alloys (HEAs) or multicomponent alloys, discovered in the early 2000s (\cite{Yeh2004,Cantor2004}). The entropy engineering concepts were later extended to many other materials (\cite{Lei2019}), including oxides (\cite{Rost2015,Sarkar2018,Albedwawi2021,Xiang2021}),  
chalcogenides (\cite{Fan2016,Wei2020,Cavin2021,Chen2022,Buckingham2022,Nemani2023}),
hydroxides (\cite{Teplonogova2022,Kim2022,Ritter2022}), 
metal-organic frameworks (\cite{Ma2021_2101342,Hu2021,Yuan2022}), and polymers (\cite{Qian2021,Huang2021}).
One of the key features of HEMs is their high configurational entropy, which arises from the mixing of multiple elements (\cite{Brechtl2021,George2019}) and favors the formation
of a compositionally and site disordered atomic structure (\cite{Miracle2017,Zhang2022}). At the same time, the local atomic structure in HEMs can exhibit a certain degree of clustering or chemical short-range ordering (SRO), which depends strongly on their composition. The local coordination environment around each element is determined by its size, charge, and electronegativity, as well as the presence of other elements. The bond lengths between atoms depend on the types of atoms involved, as well as their relative positions in the structure. Therefore understanding of the local atomic structure of
HEMs and how it is related to their chemical composition and properties is essential for designing and optimizing these materials for various applications.

X-ray absorption spectroscopy (XAS) is a well-established experimental method that provides information on the local electronic and atomic structure of a particular element in a material.  The method is complementary to X-ray diffraction and is especially suitable for investigating local lattice distortions or the local environment around impurities with concentrations down to the ppm level, in crystalline, nanocrystalline and disordered materials. XAS is well-adapted to the study of multicomponent compounds because of its element selectivity, which is achieved by tuning the photon energy in the X-ray absorption edge region of the desired element. Samples in different aggregate states, such as solids, liquids, and gases, can be probed equally well under 
a wide range of external conditions, including temperature, pressure, external magnetic or electric fields,  as well as in oxidising or reducing environments. Suitable solid samples can be prepared in the form of single crystals, powders, thin films or nanoparticles. Samples of metals or alloys can be
used either in the form of foil or as a single piece.
Time-dependent X-ray absorption experiments are possible using fast scanning techniques, special experimental setups (such as dispersive ones), or pulsed X-ray sources (such as X-ray free electron lasers). 

Recent advances in XAS are related to the progress and availability of synchrotron radiation sources, which ensure high-quality experimental data and open up new possibilities for time-dependent and extreme condition studies. In addition, laboratory XAS spectrometers using different X-ray tubes as a source of X-rays are also available today and are becoming increasingly popular. While they cannot compete with synchrotron radiation sources in terms of intensity and various other parameters, their daily availability and relative affordability make X-ray tubes attractive for such applications as prescreening of samples, industrial use, and user
training.

\subsection{Basics of X-ray Absorption Spectroscopy}

X-ray photons with energies sufficient to excite an electron from the core level (1s, 2s or 2p) in an atom are usually employed in XAS when one is interested in the structural information. However, the upper levels (3s, 3p, 3d, …) can be also used in the case of transition metals and more heavy elements to probe the local electronic structure. 

The X-ray absorption spectra are designated following to the absorption edges as K, L, M, etc., which are classified according to the electron principal quantum number $n$ values of 1, 2, 3, respectively. In the X-ray absorption spectra, the electric-dipole-induced transitions play the most important role, so that the electric dipole selection rules apply for the orbital angular momentum $\Delta l = \pm$1. 
For example, the K-edge X-ray absorption spectrum corresponds to transitions from 1s$_{1/2}$ level to empty $n$p states, L$_1$-edge from 2s$_{1/2}$ level to empty $n$p states, L$_2$-edge from 2p$_{1/2}$ level to empty $n$d$_{3/2}$ states, and L$_3$-edge from 2p$_{3/2}$ level to empty $n$d$_{3/2}$ and $n$d$_{5/2}$ states. The transitions from the 2p levels to $n$s states are also allowed but have weak intensity and, therefore, are usually neglected.  

The electron excited by an X-ray photon is often called the photoelectron. The minimum (threshold) energy $E_0$ required to eject the photoelectron is equal to the electron binding energy. Therefore, the wavenumber $k$ of the photoelectron is related to the incident X-ray photon energy $E$ and to the threshold energy $E_0$  as $k=\sqrt{(2m_e/\hbar^2)(E-E_0)}$, where $m_e$ is the electron mass, and $\hbar$ is Plank's constant.  

When a core-shell electron is excited by an X-ray photon, the absorbing atom goes to an excited state with a positively charged core hole (inner shell vacancy) located at the electron initial core level. The excitation lifetime $\Delta t$ can be estimated from Heisenberg’s uncertainty principle $\Delta E \Delta t \sim \hbar/2$, where $\Delta E$ is the core shell natural width, and is equal to about 10$^{-15}$-10$^{-16}$ seconds.  Note that this time is much smaller than the characteristic time of thermal vibrations equal to 10$^{-13}$-10$^{-14}$ seconds, so that the absorbing and surrounding atoms remain frozen on their positions during the photoabsorption process. At the same time, the presence of the core hole leads to a relaxation of other (passive) electrons in the absorbing atom.  Finally, the return of an excited atom to the ground state occurs through various relaxation channels.
Light elements with small atomic numbers $Z$ relax mainly nonradiatively via the production of Auger electrons, whereas relaxation of heavy elements is mostly radiative via the emission of X-ray fluorescence (XRF) photons. 

When monochromatic X-rays with an energy $E$ travel through a homogeneous sample of thickness $x$, they are absorbed in a material so that their incident $I_0(E)$ and the transmitted $I(E)$ intensities are given by the Beer-Lambert law:  
\begin{equation}
	I(E) = I_0(E) e^{- \mu(E) x}
\label{eq1}
\end{equation}
where $\mu(E)$ is the linear X-ray absorption coefficient. 

\begin{figure}[t]
	\centering
	\includegraphics[width=1\textwidth]{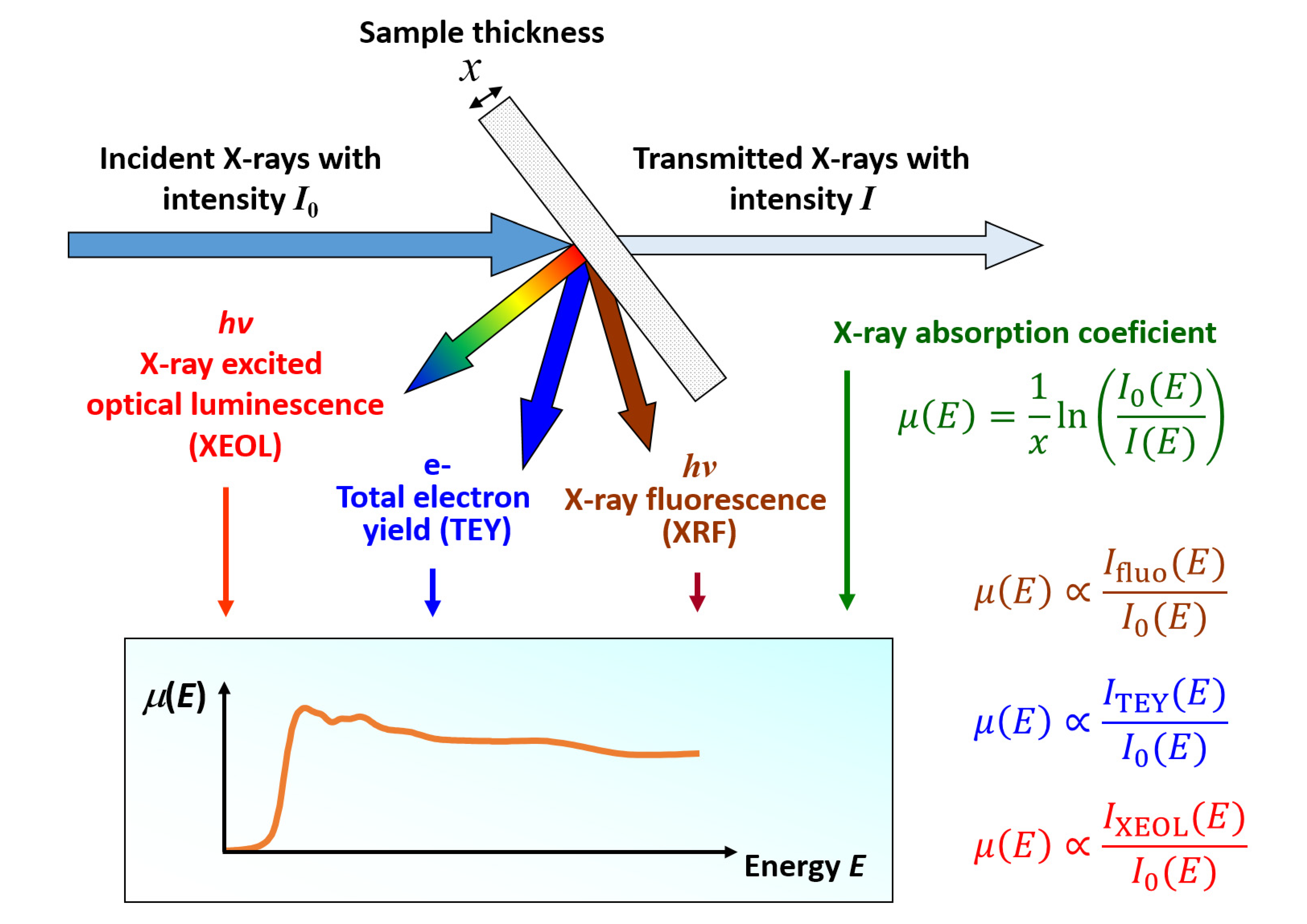}
	\caption{Four different detection modes of X-ray absorption spectrum. X-ray absorption experiment in transmission, XRF, TEY, and XEOL modes. \textit{TEY}, Total electron yield; \textit{XEOL}, X-ray-excited optical luminescence; \textit{XRF}, X-ray fluorescence. }	
	\label{fig1}	
\end{figure}

The experimental measurements of the X-ray absorption coefficient $\mu(E)$\ can be performed in transmission mode or  using indirect detection modes, such as XRF mode, Auger electron yield and total electron yield (TEY) modes, and X-ray excited optical luminescence (XEOL) mode (Fig.\ \ref{fig1}). The fluorescence detection is advantageous for dilute samples and heavy elements. The Auger and TEY modes are suitable for light elements and surface limited (depths up to about 100~\AA) studies. XEOL mode provides site-selective information on the local structure of luminescent centres.

The X-ray absorption coefficient $\mu(E)$ in the one-electron approximation is proportional to the transition rate between the initial core-state $i$ and the final excited-state $f$ 
of an electron, which is given by Fermi’s Golden rule: 
\begin{equation}
	\mu(E) \propto \sum_f \left | \langle f| \hat{H} |i \rangle \right |^2 \delta (E_f - E_i - E)
	\label{eq2}  
\end{equation}
where $E=\hbar \omega$ is the X-ray photon energy, and the transition operator 
$\hat{H} = \hat{\epsilon} \cdot \vec{r}$ in the dipole approximation. Note that the final state of the electron is the relaxed excited state  in the presence of the core-hole screened by other electrons. 

It is common to divide the X-ray absorption spectrum into two parts: X-ray absorption near edge structure (XANES), located near the absorption edge, and extended X-ray absorption fine structure (EXAFS), which extends far beyond the edge. Though the two parts have the same physical origin, their distinction is convenient for the interpretation.
The XANES region (typically within 30-50 eV of the absorption edge) contains information on the local electronic structure and is sensitive to the oxidation state of the absorbing atom, the type and distribution of empty electron states, the local symmetry and multi-electron effects. 
The EXAFS part contains information on the local atomic structure, including pair and many-atom distribution functions. Its analysis can provide structural information such as the coordination numbers, interatomic distances and their variations due to static and thermal disorder. 

\begin{figure}[t]
	\centering
	\includegraphics[width=1\textwidth]{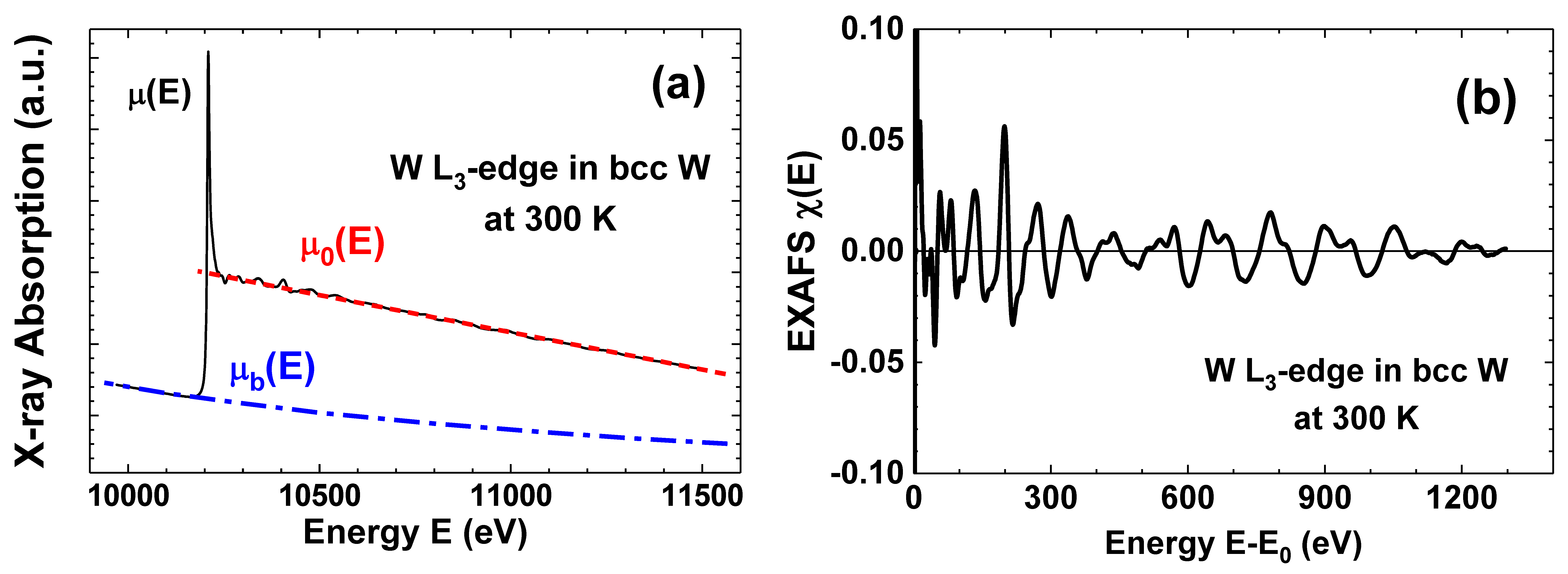}
	\caption{Extraction of the EXAFS from the X-ray absorption coefficient. The W L$_3$-edge X-ray absorption spectrum $\mu(E)$ (a) and extracted W L$_3$-edge EXAFS $\chi(E)$ (b) of BCC tungsten at 300~K. Background $\mu_b(E)$ and atomic $\mu_0(E)$ contributions are shown in (a). \textit{BCC}, body-centered cubic; \textit{EXAFS}, extended X-ray absorption fine structure.}	
	\label{fig2}	
\end{figure}

\begin{figure}[t]
	\centering
	\includegraphics[width=1\textwidth]{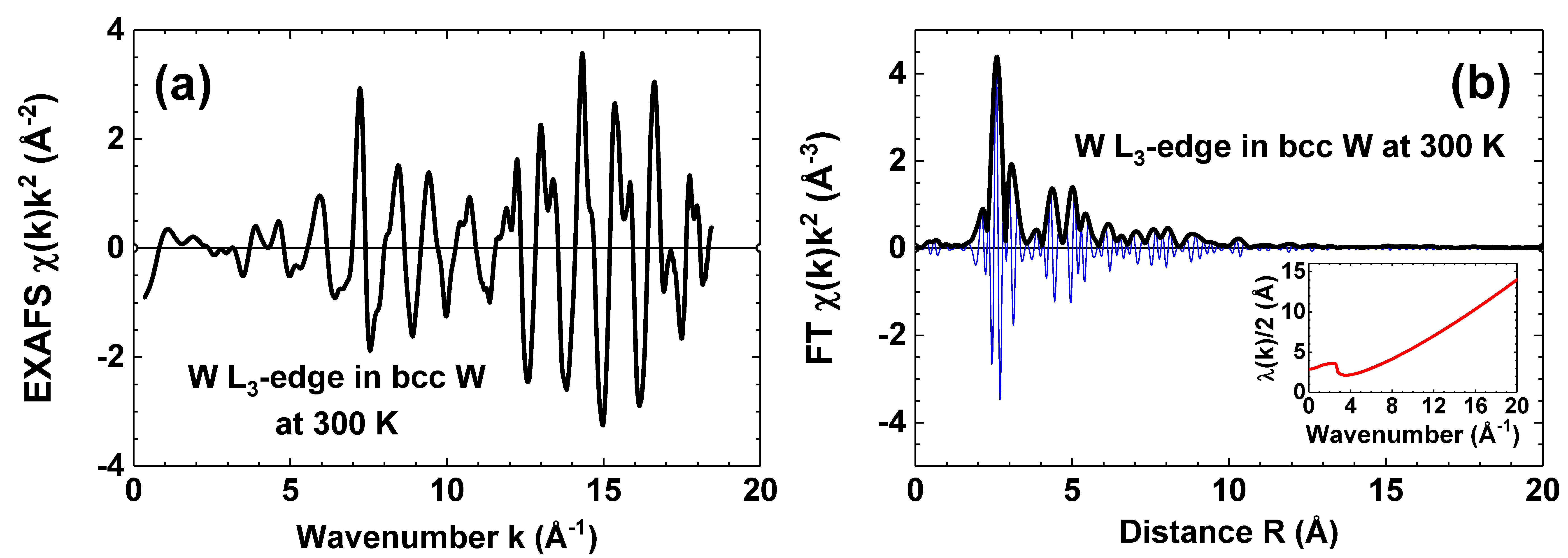}
	\caption{The W L$_3$-edge EXAFS spectrum and its FT for BCC tungsten at 300~K. The W L$_3$-edge EXAFS spectrum $\chi(k)k^2$ (a) and its Fourier transform (b) for BCC tungsten at 300~K. Both the modulus (\textit{thick line}) and imaginary (\textit{thin line}) parts are shown (b). The inset in (b) shows the calculated mean free path $\lambda(k)$ of a photoelectron, including the core–hole effect,  plotted versus wavenumber $k$ for the W L$_3$-edge.
	\textit{BCC}, Body-centered cubic; \textit{EXAFS},
	extended X-ray absorption fine structure; \textit{FT}, Fourier transform.}
	\label{fig3}	
\end{figure}

The  oscillating part of the absorption coefficient, that is,  EXAFS, $\chi^l(E)$ (Fig.\ \ref{fig2}) located above the absorption edge of orbital type $l$ is defined as 
\begin{equation}
	\chi^l(k)= \frac{\mu(E)-\mu_0(E)-\mu_b(E)}{\mu_0(E)}
	\label{eq3}
\end{equation}
where $\mu_b(E)$ is the background absorption, and $\mu_0(E)$ is the atomic-like absorption due to an isolated absorbing atom (\cite{Lee1981}). It is common to display the EXAFS spectrum as a function of the wavenumber $k$, often multiplied by $k^n$ (where $n = 1, 2, 3$) (Fig.\ \ref{fig3}(a)). 
The contributions of different coordination shells to the EXAFS spectrum 
can be visualized  in the $R$-space by calculating the direct Fourier transform (FT) (Fig.\ \ref{fig3}(b)). 
The peaks in FT can be observed up to certain values of $R$, which correlate with half of the mean-free path (MFP) of the photoelectron.  However, the presence of static and/or thermal disorder can significantly reduce contributions from the distant coordination shells in the FT.

It is important to note that while the shape of the Fourier transform may appear similar to that of the radial distribution function (RDF), there is a fundamental difference between these two functions. The positions of peaks in the FT are always shifted to shorter distances compared to their crystallographic values due to the presence of the phase shift in the EXAFS equation, and the shape of peaks in the FT is distorted due to the presence of the scattering amplitude. Furthermore, the FT also includes the contributions from high-order distribution functions, known as  multiple-scattering (MS) effects (\cite{Natoli1990,Rehr2000}).

The theoretical description of EXAFS in the framework of the MS theory (\cite{Rehr2000,Rehr2009}) is  given using a series: 
\begin{eqnarray}
	\chi^l(k) &=& \sum_{n=2}^{\infty} \chi^l_n(k), \nonumber \\
	\chi^l_n(k)  &=& \sum_{j}A_n^l(k,R_j) \sin[2kR_j+\phi_n^l(k,R_j)]  
	\label{eq4}
\end{eqnarray}
which includes contributions $\chi^l_n(k)$  from the $(n-1)$th-order
scattering processes of the excited photoelectron by the
neighbouring atoms, before it returns to the absorbing atom (\cite{Rehr2000,RuizLopez1988}). 
The fast convergence of the MS series occurs at least at high-$k$ values due to the finite lifetime of the excitation, the scattering path lengths, interference cancellation effects, and path disorder. In practice, the MS contributions up to the eight-order can be calculated, for example, using {\it ab initio} FEFF code (\cite{FEFF8,FEFF9}).

An alternative description of the EXAFS $\chi^l(k)$  in terms of the $n$th-order distribution functions $g_n(R)$ is also known:
\begin{eqnarray}
	\chi^l(k) &=& \int 4\pi R^2 \rho_0 g_2(R)
	[\chi_2^{oio}(k) + 
	\chi_4^{oioio}(k) 
	+  \ldots] dR \nonumber \\
	& + &  \int\!\!\!\int\!\!\!\int 8 \pi^2 R_1^2 R_2^2 \sin(\theta)
	\rho_0^2 g_3(R_1, R_2, \theta) \nonumber \\
	&\times& [2\chi_3^{oijo}(k) +
	2\chi_4^{oiojo}(k) + 
	\chi_4^{oijio}(k) +  \chi_4^{ojijo}(k) +
	\ldots ] dR_1 dR_2 d\theta  \nonumber \\
	 &+&  
	 \int\!\!\!\int\!\!\!\int\!\!\!\int\!\!\!\int 8 \pi^2 R_1^2 R_2^2 R_3^2 \sin[\theta]
	 \rho_0^3  g_4(R_1, R_2, \theta, R_3, \Omega)  \nonumber \\
	 & \times& [ 2\chi_4^{oijko}(k) + 2\chi_4^{oikjo}(k) +
	 2\chi_4^{ojiko}(k)  + \ldots ]
	 dR_1 dR_2 d \theta dR_3 d \Omega  \nonumber \\
	& +&  \ldots
	\label{eq5}
\end{eqnarray}
where $\rho_0$\ is the average density of a system, and $\chi_m(k)$
are the MS EXAFS contributions of the $(m-1)$\ order
generated within a group of atoms (o, i, j, \ldots) described by
$g_n$ (\cite{Filipponi1995a,Filipponi1995b}).
This approach was implemented in the GNXAS code (\cite{DiCicco1995,GNXAS}), which 
is able to account for the two-body ($g_2$), three-body ($g_3$), and four-body ($g_4$) distribution functions.

\begin{figure}[t]
	\centering
	\includegraphics[width=0.8\textwidth]{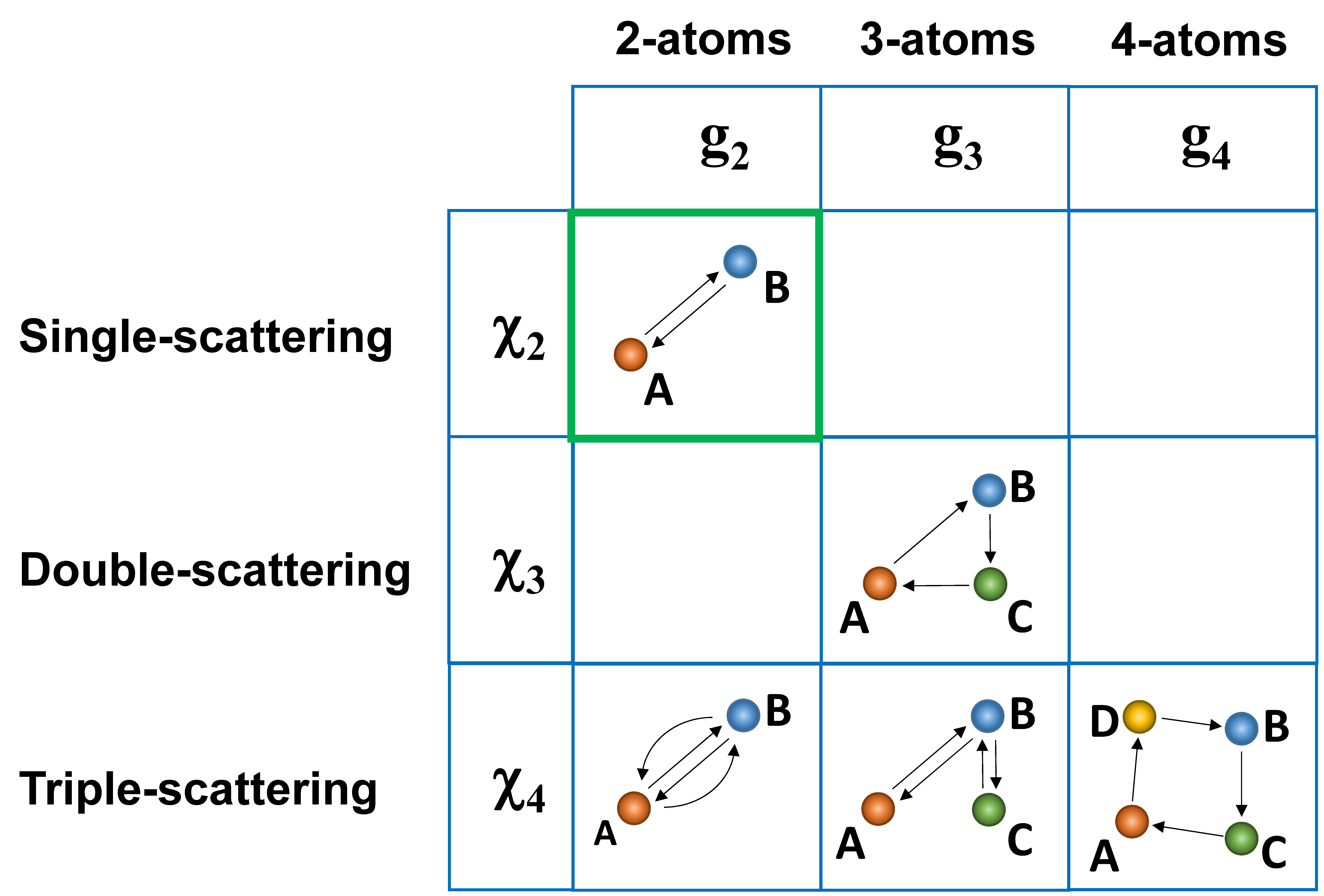}
	\caption{The relation between the scattering paths due to the multiple-scattering contributions $\chi_n$  and  many-body distribution functions $g_n$. The scattering paths involving two atoms (A and B), tree atoms (A--C), and three atoms (A--D) are shown.  $\chi_2$ corresponds to the SS contribution for a pair of atoms A and B. $\chi_3$ represents the double-scattering contribution generated within a group of three atoms A--C.   $\chi_4$ includes  three  triple-scattering contributions generated within three groups of two (A and B), three (A--C), and four (A--D) atoms. \textit{SS}, Single-scattering.}	\label{fig4}	
\end{figure}

Note that there is a relation between the MS contributions $\chi_m$ and many-body distribution functions $g_n$ (Fig.\ \ref{fig4}). Each function $g_n$ accounts for all scattering processes involving the respective group of $n$-atoms,
whereas each MS contribution $\chi_m$ accounts for the MS processes of the $(m-1)$th-order involving all possible atomic configurations. 

\subsection{Multicomponent extended X-ray absorption fine structure analysis}

The analytical expression for EXAFS can be greatly simplified 
when one needs to extract information only from the first coordination shell of the absorbing atom.

The contribution of the first coordination shell to the total EXAFS spectrum can be isolated  by the Fourier filtering procedure and analysed within the single-scattering approximation, since the lengths of all MS paths are longer than the first coordination shell radius.  Thus only the first term of the series given by Eq.~(\ref{eq4}) remains. In the case of a Gaussian distribution (or in the harmonic approximation), the EXAFS expression takes a simple form 
\begin{eqnarray}
	\chi^l_2(k) & = & S_0^2 \sum_{i} N_i \frac{|f^l_{\rm eff}(k,R_i)|}{kR_i^2} 
	\exp\left[-\frac{2 R_i}{\lambda(k)}\right] \nonumber \\
	& \times &  \sin[2kR_i+\phi^l(k,R_i)]  \exp(-2\sigma_i^2 k^2)
	\label{eq6}
\end{eqnarray}
where $S_0^2$ is the amplitude reduction factor (typically between 0.8 and 1.0); $N_i$ is the coordination number; 
$R_i$ is the interatomic distance; $\lambda(k)$ is the photoelectron MFP; 
$f^l_{\rm eff}(k,R)$ and $\phi^l(k,R)$ are the photoelectron effective scattering amplitude and phase shift
functions; $\sigma^2(T) = \sigma_{st}^2 + \sigma_{th}^2(T)$  is the Debye-Waller (DW) factor or mean-squared relative displacement (MSRD), accounting for static $\sigma_{st}^2$ and thermal $\sigma^2(T)$ disorder effects (\cite{Sayers1971,Lee1975}). The sum in Eq.~(\ref{eq5}) is taken over groups of atoms located at different distances from the absorber.

For moderate disorder, when the distribution  of interatomic distances becomes asymmetric, the EXAFS equation can be expressed using the cumulant decomposition (\cite{Bunker1983,Dalba1993}).  
The cumulant model is often useful for the analysis of anharmonic and thermal expansion effects  (\cite{Tranquada1983,Fornasini2017}), nanoparticles
(\cite{Clausen2000,Sun2017}), and  disordered materials (\cite{Dalba1995age,Okamoto2002}).

Temperature-dependent MSRD $\sigma_{ij}^2(T)$ for the $i$-$j$ pair of atoms with the mean-square displacement (MSD) amplitudes 	MSD$_i(T)$ and MSD$_j(T)$ are related as follows:
\begin{equation}
	 \sigma_{ij}^2(T)={\rm MSD}_i(T)+{\rm MSD}_j(T)-2 \phi \sqrt{{\rm MSD}_i(T)} \sqrt{{\rm MSD}_j(T)},
	\label{eq7}
\end{equation}
where $\phi$ is a dimensionless correlation parameter (\cite{Booth1995}). Note that for atoms located at large distances (in distant coordination shells), the correlation effects become negligible
(\cite{Jeong2003,Sapelkin2002}) so that the value of $\sigma_{ij}^2(T)$ 
can be used to estimate the sum of two  mean-square displacement amplitudes. The MSD values are traditionally obtained from  diffraction measurements or lattice dynamics calculations, but molecular dynamics (MD) simulations and reverse Monte Carlo analysis of EXAFS data can be also used for this purpose (\cite{Jonane2018}).

\begin{figure}[t]
	\centering
	\includegraphics[width=1\textwidth]{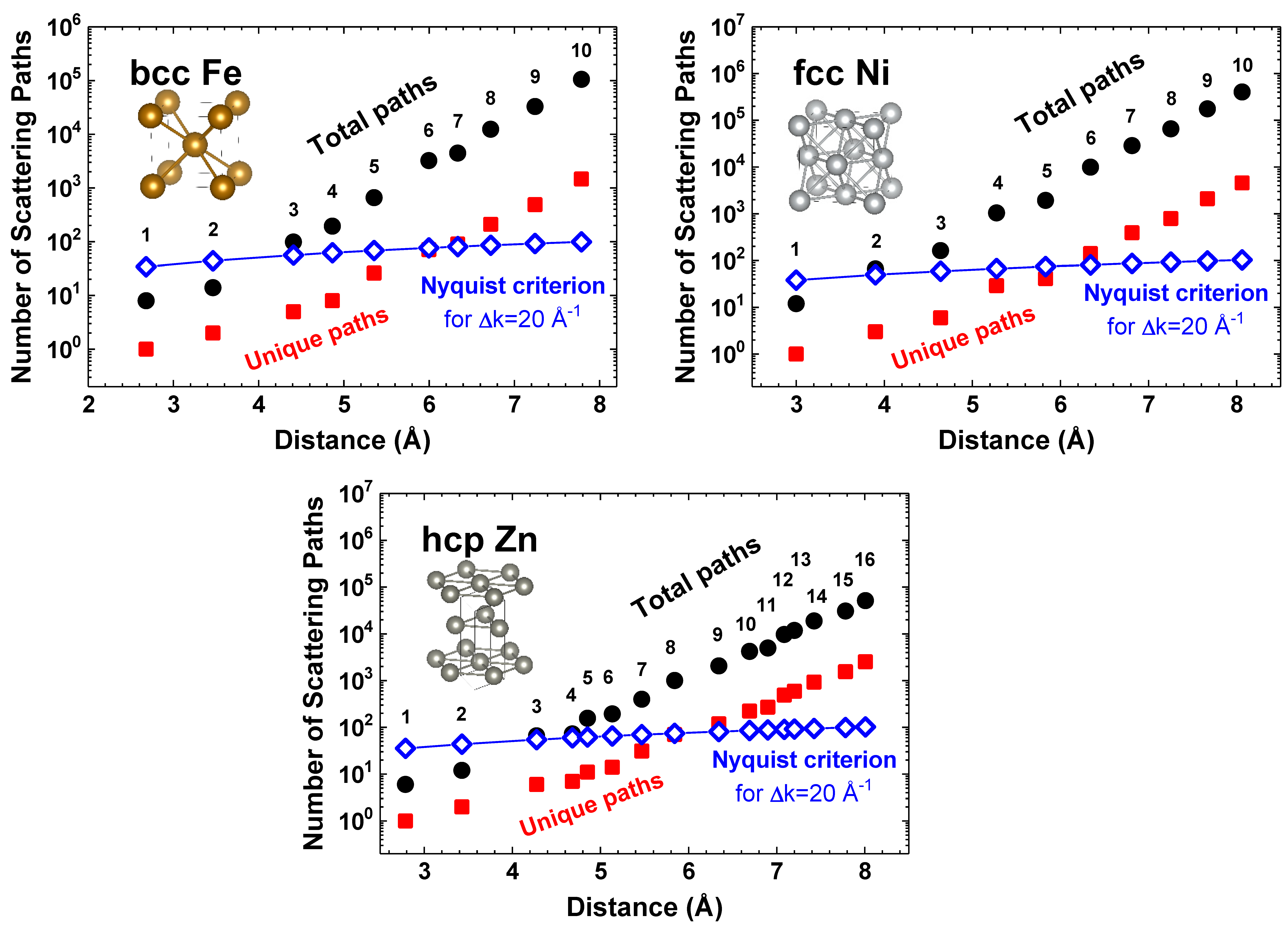}
	\caption{The dependence of the number of scattering paths on cluster size for the FCC Fe, BCC Ni, and HCP Zn structures. Solid circles -- total number of paths, solid squares -- unique number of paths, open diamonds -- number of fitting parameters according to the Nyquist criterion.  Note the logarithmic scale on the vertical axes. \textit{BCC}, Body-centered cubic; \textit{FCC}, face-centered cubic; \textit{HCP}, hexagonal close-packed.}	
	\label{fig5}	
\end{figure}

The analysis of the outer coordination shells using Eq.\ \ref{eq6} is often inaccurate due to the presence of the MS contributions. In fact,  when MS effects are taken into account, the number of scattering paths increases rapidly for distant shells resulting in a huge number of fitting parameters required. To illustrate this problem, three cases of the face-centered cubic (FCC) Fe, body-centered cubic (BCC) Ni, and hexagonal close-packed (HCP) Zn structures were consider  (Fig.\ \ref{fig5}), and the numbers of unique (accounting for a structure symmetry) and total scattering paths were compared with those evaluated according to the Nyquist criterion [$N_{\rm par} = 2 \Delta k \Delta R / \pi$ (\cite{Bordiga2013})] for a long EXAFS spectrum with $\Delta k = 20$~\AA$^{-1}$, as a function of radial distance (cluster radius around the absorbing atom). Note that the Nyquist criterion estimates the maximum number of independent variables for EXAFS model. As one can see, the Nyquist criterion is not satisfied already at $R \sim 4$~\AA\ when symmetry is broken as is in the case of HEAs due to the compositional disorder.

Note that only the effect of radial disorder is accounted in Eq.\ \ref{eq6}, while both scattering amplitude and phase shift functions of a  photoelectron demonstrate  non-linear angular dependence and are sensitive even to small variations of angles
along the scattering path, especially in the case of linear atomic
chains (\cite{Kuzmin1993,Teo1986}).  This problem has been addressed
in the past for small disorder using the low-order Taylor expansion
for the amplitude and phase of the EXAFS signal (\cite{Filipponi1995a,Filipponi1995b}).

\subsection{Advanced extended X-ray absorption fine structure analysis}

\begin{figure}[t]
	\centering
	\includegraphics[width=0.8\textwidth]{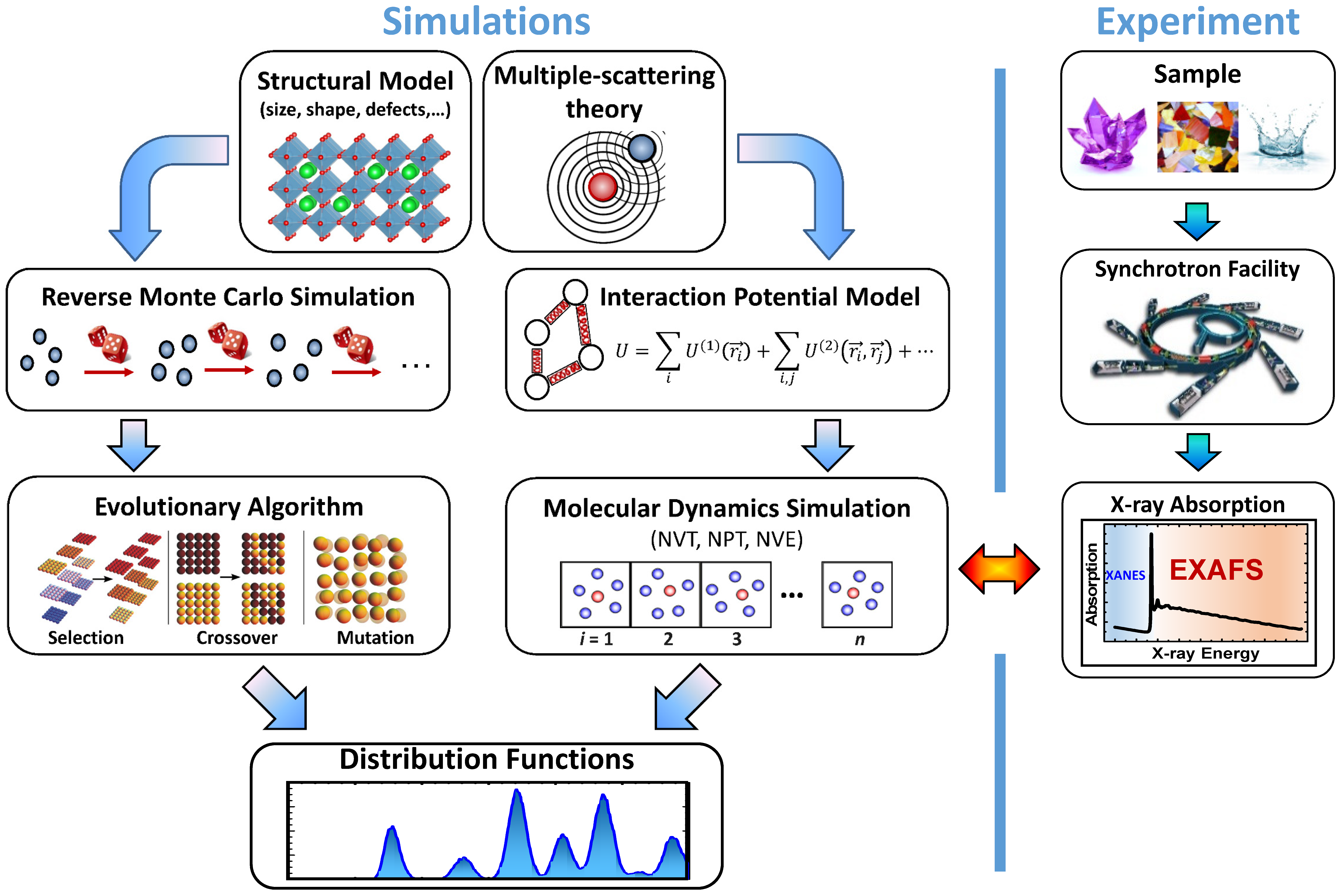}
	\caption{Scheme of EXAFS analysis using RMC and MD methods. 
	The simulations are performed based on the initial structural model and the ab initio multiple-scattering formalism. In the case of the RMC approach, the agreement between the experimental and calculated data  is used to optimize the structural model, while  the configuration-averaged EXAFS spectrum obtained after MD simulation  is employed to validate the interatomic potential model.  In both cases, the atomic configurations are finally used to determine different distribution functions that describe the structure of a material, including disorder effects. \textit{EXAFS}, Extended X-ray absorption fine structure; \textit{MD}, molecular dynamics; \textit{RMC},
	reverse Monte Carlo.}	
	\label{fig6}	
\end{figure}

To overcome the problems related to disorder effects and properly account for the MS contributions,
the methods based on atomistic simulations (Fig.\ \ref{fig6}), such as MD
(\cite{Angelo1994,Merkling2001,Cabaret2001,Angelo2002,Okamoto2004,Farges2004,Ferlat2005,Kuzmin2009,Price2012,Yancey2013}) and RMC
(\cite{Winterer2000,McGreevy2001,DiCicco2005,Gereben2007,Krayzman2009,Krayzman2010,Levin2014,Timoshenko2014rmc,Timoshenko2014zno,Timoshenko2014cuwo4}),  have been developed in the past. 
In both methods, the configuration-averaged (CA) EXAFS signal is calculated using coordinates of atoms obtained from one or more atomic configurations (``snapshots'') during the simulation (\cite{KUZMIN2020rev}). It is important that these configurations naturally include  both static and dynamic disorder.
The CA-EXAFS spectra for different absorption edges can be calculated from the same set of atomic coordinates and used in the analysis. 
Note that in addition to atomic coordinates, two nonstructural parameters ($\Delta E_0$ and $S_0^2$) should be also provided for comparison with experimental EXAFS spectra (\cite{Kuzmin2014exafs}). 
These parameters can be determined a priori by analyzing reference materials or by achieving the best match between experimental and calculated EXAFS spectra.

Apart from the similarities, there is also a principle difference between using MD and RMC methods to interpret EXAFS, which will be considered below.

\subsubsection{Molecular Dynamics}

In MD simulation, the 3D atomistic model of a material evolves for a fixed period of time following the  trajectories which are determined by numerically solving classical Newton's equations of motion (\cite{GULP2003,Gowthaman2023}). 
As a result, MD technique tends to be effective at high temperatures, enabling the modeling of the anharmonic motion of atoms. At the same time, the calculated amplitudes of thermal vibrations at low temperatures are  underestimated due to a neglect of the zero-point (quantum) atomic motion (\cite{Markland2018,Yang2012}).  To address this problem, alternative methods such as, for example, path integral MC or path integral MD, should be employed instead (\cite{Berne1986,Marx1996,Tuckerman1996}).

MD simulations are performed using different ensembles (\cite{Abraham1986}) such as the microcanonical ensemble (NVE), the canonical ensemble (NVT), or the isothermal-isobaric (NPT) ensemble, in which the number of atoms (N), the system's volume (V), the temperature (T), the pressure (P), or the energy (E) remain conserved during the simulation. 
The MD method provides a dynamic representation of the system's evolution, effectively simulating thermal disorder. 
Depending on the description of atomic interactions, the method can be classified as either classical or ab initio MD. 

In classical MD, the forces between the atoms and their potential energies are  calculated using empirically defined interatomic potentials (frequently called force fields).  This significantly reduces the total computational time and required resources but allows one to model large systems, including millions of atoms on
a long time-scale up to milliseconds (\cite{Delaye2001,Pierce2012}). 
At the same time, the reliability of such MD simulations depends directly on the accuracy of the empirical potential. Designing the interatomic potential is a complex task, involving choosing its functional form and obtaining a set of parameters on which it depends. Conventionally, two strategies are used to optimize the potential parameters: They are fitted to reproduce material properties such as structure, elastic constants, and phonon frequencies, or  the potential-energy surface determined directly from  ab initio density functional theory (DFT) simulations (\cite{GULP2003}). 

The EXAFS spectrum, which includes information on both structural and dynamic disorder,  represents an additional ``property'' that is well-suited for testing the quality of MD simulation or interatomic potentials. For this, a set of atomic configurations obtained during MD simulation is used to calculate the CA-EXAFS spectrum for an element of interest based on the ab initio multiple-scattering theory using, for example,  FEFF (\cite{FEFF8,FEFF9}) or GNXAS (\cite{GNXAS}) code.   
It can then be compared with the experimental data, and the degree of agreement between the two EXAFS spectra can be used as a criterion to validate the interatomic potential model. Examples of such applications  to SrTiO$_3$, ZnO, UO$_2$, BCC W, and ScF$_3$ can be found in (\cite{Bocharov2017,Bocharov2020,Jonane2018,Kuzmin2009,Kuzmin2016}). 

\begin{figure}[t]
	\centering
	\includegraphics[width=0.6\textwidth]{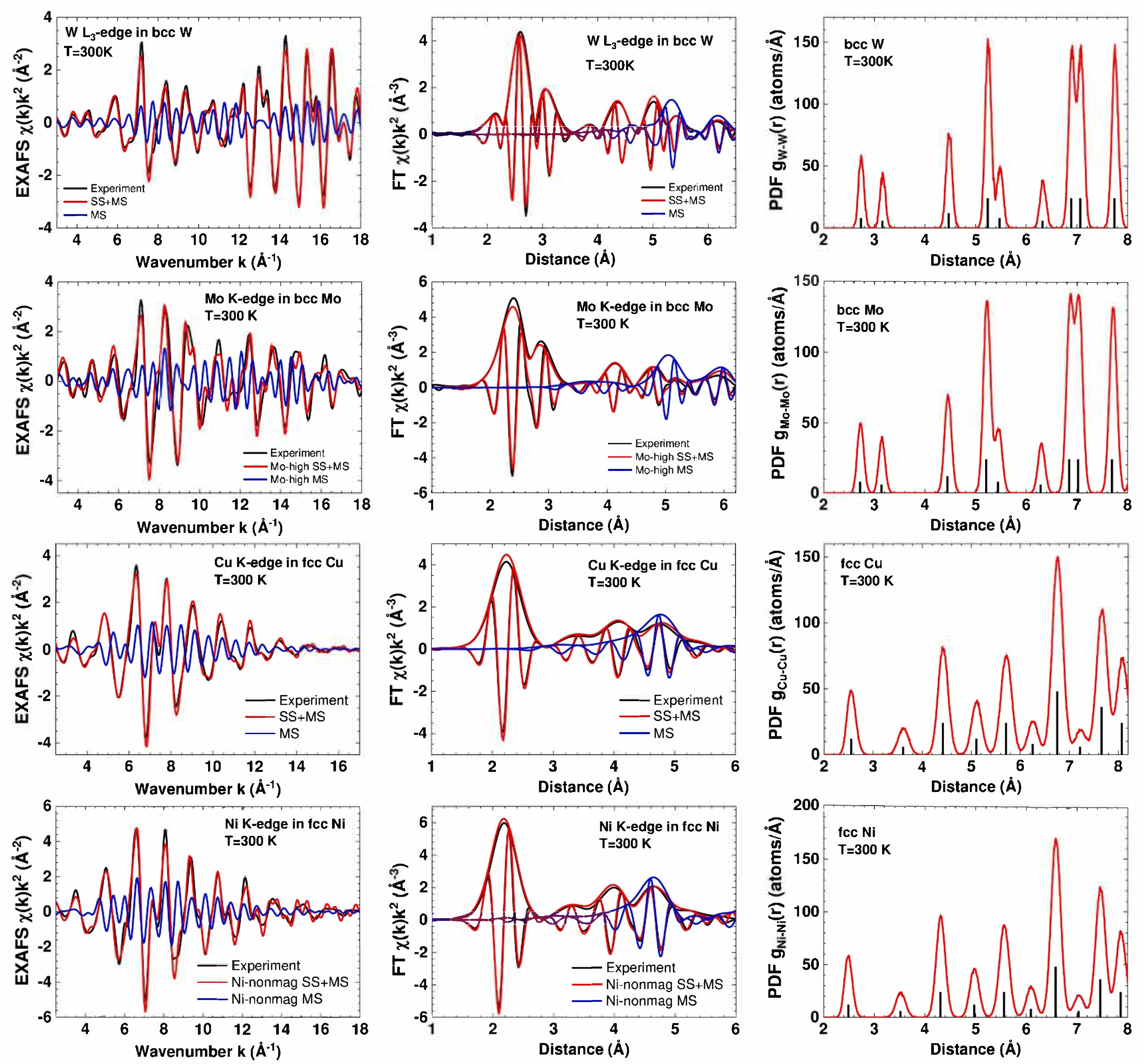}
	\caption{Comparison of the results of the MD--EXAFS simulations using the moment tensor potentials for BCC W and Mo and FCC Cu and Ni metals: the experimental and calculated EXAFS spectra (left panels) and their Fourier transforms (middle panels) at 300~K for BCC W and Mo and FCC Cu and Ni metals. The total calculated spectra, including SS and MS contributions, are shown by red lines, whereas the MS contributions are separately shown by blue lines. The PDFs $g(r)$\ (\textit{solid curves}) are given in the right panels together with the positions of crystallographic shells (\textit{vertical bars}).   \textit{BCC}, Body-centered cubic; \textit{EXAFS}, extended X-ray
     absorption fine structure; \textit{FCC}, face-centered cubic; \textit{MD}, molecular dynamics; \textit{MS}, multiplescattering; \textit{PDFs}, pair distribution functions; \textit{SS}, single-scattering.
     From (\protect\cite{Shapeev2022}). Copyright (2022), with permission from Elsevier.}
     \label{fig7}	
\end{figure}

Machine learning (ML) (\cite{Behler2016}) has recently emerged as an alternative method for representing potential-energy surfaces, wherein large datasets derived from electronic structure calculations are fitted. ML potentials combine the strengths of empirical potentials and first-principles models, making them a valuable tool for atomistic simulations. However, the need for verification of ML potentials remains. Recently,  the accuracy of moment tensor potentials has been successfully validated by comparing the EXAFS spectra of four metals (BCC W and Mo, FCC Cu and Ni) obtained experimentally and computed from the results of MD simulations (Fig.\ \ref{fig7}) (\cite{Shapeev2022}). The method allows one to account for  both SS and MS contributions to the total EXAFS spectrum while also considering disorder effects.  This is important for achieving good agreement in the region of distant coordination shells (above  3.5--4~\AA), especially in materials with many linear atomic chains in their structure, such as BCC and FCC metals.  Note that the MS contributions extend over the entire $k$-range, whereas the MS peaks appear at long distances in the $R$-space due to the large lengths of the MS paths.

\subsubsection{Reverse Monte Carlo Simulations}	   

The RMC simulation method employs a variation of the Metropolis algorithm (\cite{Metropolis1953}) and aims to identify an atomic configuration which is consistent with specific material properties, such as, for example, an EXAFS spectrum, neutron, or X-ray scattering data (\cite{Woicik2023}). It is important that the selected properties are directly connected to the material structure.  In this scenario, the structural model is adjusted  at each simulation step to match experimental data, and no information on interatomic potentials is required, which is an advantage over MD simulation. 

The RMC method was originally proposed in the late 1980s as an approach to extract structural information such as atomic pair distribution functions from X-ray or neutron scattering data (\cite{Keen1990,McGreevy1988}).
Shortly after, its application expanded to include the analysis of EXAFS spectra  (\cite{Gurman1990rmc}). A recent review discussing the strengths and weaknesses of the method for the modeling of EXAFS data can be found in (\cite{DiCicco2022}).

The use of the RMC method in EXAFS analysis additionally benefits from the sensitivity of EXAFS spectra to high-order atomic distribution functions, which contribute through MS effects (\cite{Rehr2000}).  This additional information, exclusively present in the EXAFS spectra, has a positive effect on the stability of the solution and, ultimately, on the reliability of the resulting structural model.  Moreover, the element selectivity of XAS allows one to obtain experimental EXAFS spectra of different chemical elements present in a multicomponent compound independently. These spectra can be simultaneously employed in RMC simulation to determine a single structural model that is in agreement with all available experimental data. 

In an RMC simulation, the model is represented by atoms placed according to the material density within a simulation box (often referred to as a cell or supercell) that matches the desired size and shape. 
Periodic boundary conditions (PBC) are commonly employed to mitigate surface-related effects. However, it is important to note that PBC imposes a limitation on the maximum cluster radius for EXAFS calculations, restricting it to half the minimum size of the box.  This restriction helps to avoid the occurrence of artificial correlation effects.

During RMC simulation, the positions of all atoms in the box are typically randomly modified in each iteration (simulation step), and the CA--EXAFS signal is calculated. The decision to accept or reject the new atomic configuration is made based on the Metropolis algorithm, taking into account the difference (residual) between the experimental and simulated data in either $k$ or $R$ space, or both simultaneously  through the wavelet transformation (\cite{Timoshenko2009wt}). Additionally, chemical or geometrical constraints can be incorporated into the residual value to prevent scenarios where atoms excessively approach or distance themselves from each other, certain bond angles acquire non-physical values, or unexpected deviations occur in the coordination number of specific atoms  (\cite{Tucker2007rmcprofile}).    

The efficiency of RMC simulation can be significantly improved  using an evolutionary algorithm alongside a simulated annealing scheme (\cite{Timoshenko2014rmc,Timoshenko2012rmc}). 
The RMC technique  is based  on stochastic processes and typically converges towards the most disordered solution that aligns with the experimental data  (\cite{Tucker2007rmcprofile}). Executing the simulation multiple times with various initial conditions yields distinct sets of final atomic coordinates, which are then  used to obtain information about configuration-averaged atomic and bond-angle distribution functions. 

To perform RMC analysis of EXAFS spectra, several software packages exist, including RMC-GNXAS (\cite{DiCicco2022,DiCicco2005}), RMC++/RMC\_POT (\cite{Gereben2012rmc_pot,Gereben2007}), EvAX (\cite{Timoshenko2014rmc}), RMCProfile (\cite{Tucker2007rmcprofile,Zhang2020}), RMCXAS (\cite{Winterer2022,Winterer2000}), m-RMC (\cite{Fujikawa2014}), EPSR-RMC (\cite{Bowron2008epsr}), and SpecSwap-RMC (\cite{Leetmaa2010specswap}).

\subsection{Applications to High-Entropy Materials}

The use of XAS to investigate the local environment in HEMs  dates back to 2009 (\cite{Chen2009}), but until recently it has been rather sporadic (\cite{Rost2015,Tamm2015,Maulik2017,Zhang2017,Zhang2018,Braun2018,Harrington2019}).  
Most studies published in the past have been devoted to HEAs (\cite{Chen2009,Tamm2015,Maulik2017,Zhang2017,Zhang2018,Fantin2020,Tan2021,He2021,MORRIS2021,Wu2021,Fantin2022,Smekhova2022a,Smekhova2022b,Zhang2022a,Pugliese2023,Tan2023,Smekhova2023,Gornakova2023}) 
and high entropy oxides (HEOs)  (\cite{Rost2015,Braun2018,Ghigna2020,Tavani2020,Tavani2021,DUPUY2021,Sushil2021,Han2022,Walczak2022,Luo2022,Wang2022,Molenda2023,Jacobson2023,Bakradze2023}), but a few works
have been dedicated to high-entropy metal carbides (\cite{Harrington2019}), diborides (\cite{GABOARDI2022}), and metallic glasses (\cite{Zhang2022b}).
Below several examples of XAS studies of HEA and HEO materials will be discussed.   

It should be mentioned that the high concentration of principal elements in HEMs simplifies the measurement  of their experimental X-ray absorption spectra. However, the compositional disorder and often proximity of the  elements constituting HEM  in the periodic table of elements make the analysis of these spectra challenging, requiring the use of simple structural models or specific constraints between model parameters. Therefore, existing and emerging advanced methods of data analysis will contribute greatly to the development of the field in the future.

\subsubsection{High-Entropy Alloys}

In the earlier study (\cite{Chen2009}), the authors successfully prepared a binary to octonary Cu-–Ni-–Al–-Co-–Cr–-Fe–-Ti-–Mo alloy series using mechanical alloying (MA) and investigated their phase evolution and amorphisation behavior during milling. The formation of FCC and BCC structures was observed in the binary Cu$_{0.5}$Ni alloy and the ternary Cu$_{0.5}$NiAl alloy, respectively, both of which remained crystalline even after 60~h of milling. However, the FCC phase was first formed in the quaternary Cu$_{0.5}$NiAlCo alloy and alloys with more components, but it transformed into the amorphous phase after prolonged milling. Qualitative analysis of Fourier transformations of the Cu, Ni, Co, Cr, Fe, and Ti  K-edge EXAFS spectra was employed to examine the local atomic configuration of each element in the alloys (\cite{Chen2009}). The presence or reduction of peaks in the FTs was used to confirm the formation of the FCC or amorphous phase (\cite{Chen2009}).

The SRO of equimolar ternary NiCrCo and quaternary NiCrCoFe alloys in the FCC phase was theoretically studied using a lattice MC method combined with DFT calculations by  \cite{Tamm2015}. The structural model consisted of a supercell with 108 atoms, and the MC simulation was followed by an MD run to allow for atomic relaxations. The MC simulations suggested a significant degree of SRO in both alloys compared to fully random distributions (\cite{Tamm2015}). Specifically, in the ternary NiCrCo alloy, a 40\% decrease in the number of Cr--Cr pairs was observed, accompanied by an increase in the number of Ni--Cr and Cr--Co atom pairs (\cite{Tamm2015}).  Similarly, the quaternary NiCrCoFe alloy exhibited a decrease in the number of Cr--Cr, Fe--Fe, Ni--Ni, and Co--Co pairs, along with an increase in Ni--Cr, Ni--Fe, Cr--Co, and Co--Fe atom pairs (\cite{Tamm2015}). An attempt was also made to validate the theoretical models using the experimental Fe K-edge EXAFS spectrum of the NiCrCoFe alloy. However, the difference between the calculated Fe K-edge EXAFS spectra for quasi-random and SRO structures was negligible, and both spectra aligned quite well with the experimental data (\cite{Tamm2015}).

The local atomic structure in a medium entropy alloy (MEA) NiCoCr was studied using three structural methods such as X-ray or neutron total scattering and XAS at the Cr, Co, and Ni K-edges in (\cite{Zhang2017}). Total scattering methods are complementary to XAS and provide the pair distribution functions (PDFs) extending up to 20--30~\AA.   However,  the analysis of the total PDFs was not able to discriminate partials contributions and  to evidence any detectable lattice distortion  (\cite{Zhang2017}). At the same time, the SRO was revealed from the analysis of EXAFS spectra using the multicomponent analysis (\cite{Zhang2017}). It was found that the bond lengths and DW factors of Ni--Cr and Co--Cr pairs are different from those of Ni--Ni(Co) and Co--Co(Ni) pairs. Also, some evidence of the SRO enhancement by ion irradiation was proposed based on the EXAFS results (\cite{Zhang2017}).  The local structure and lattice distortion in NiCoFeMnCr HEA were also studied by the same group (\cite{Zhang2018}).  Independent analysis of the K-edge EXAFS spectra for five metallic elements indicates close (within the error reported) nearest interatomic distances of 2.53--2.54$\pm$0.01~\AA\ and  the DW factors of  $\sigma^2$=0.007$\pm$0.001~\AA$^2$ (\cite{Zhang2018}).

The local lattice distortions and their correlations with chemical compositions were investigated in CrCoNi MEA, CrFeCoNi and CrMnFeCoNi HEAs  using synchrotron radiation based X-ray diffraction and EXAFS techniques by \cite{Tan2021}.
The distortion of the first coordination shell around each alloying element was estimated from the analysis of the EXAFS contribution from the first peak in the FTs (\cite{Tan2021}). The study revealed that 
the  distortions are not severe and are highly element dependent, following the order of Ni > Co > Fe > Cr > Mn in all three alloys (\cite{Tan2021}). Additionally, the averaged strain around each alloying element was also calculated; it was positive around Ni, while, negative around Cr (\cite{Tan2021}). However, it is important to approach these findings with caution due to the simplicity of the model used.

To overcome the limitations of single-element catalysts, FeCoNiXRu (X = Cu, Cr, or Mn) HEA nanoparticles with different active sites for water splitting in alkaline conditions were reported by \cite{Hao2022}. The authors found that the Co and Ru sites in the HEA can simultaneously stabilize OH$^*$ and H$^*$ intermediates, thus enhancing the efficiency of water dissociation (\cite{Hao2022}).
XAS was used to investigate the chemical states and bond structures of Co and Ru in the HEA before and after the stability test. The analysis of the Co and Ru K-edges XANES shows that Co and Ru are in metallic states and surrounded by different metallic species, suggesting the formation of a single-phase HEA (\cite{Hao2022}). The EXAFS results also confirm that Co and Ru keep their metallic states and bond lengths after the long-term stability test, demonstrating the excellent durability of the HEA catalyst (\cite{Hao2022}).

The local environment of six alloying elements in the  Al$_8$Cr$_{17}$Co$_{17}$Cu$_8$Fe$_{17}$Ni$_{33}$ (at.\%) compositionally complex alloy (CCA) with the FCC-type lattice was studied by \cite{Fantin2020}. The analysis of the contribution of the first coordination shell to the metal K-edge EXAFS spectra, based on the multiparameter fit approach, allowed the determination of coordination numbers, interatomic distances, and DW factors. A higher affinity of Al for the heavier 3d metals, especially Cu, and the absence of Al-–Al pairs were discovered  (\cite{Fantin2020}). Thus the EXAFS results indicated the presence of SRO in the alloy (\cite{Fantin2020}).
A comparative analysis of the K-edge XANES in the alloy and pure metals suggested charge redistribution between Ni/Cu and Al due to orbital hybridization, which is responsible for the shrinkage of the Al metallic radius  (\cite{Fantin2020}).
In the subsequent study  (\cite{Fantin2022}), the effect of heat treatment up to 1150$^\circ$C on the  Al$_8$Cr$_{17}$Co$_{17}$Cu$_8$Fe$_{17}$Ni$_{33}$ CCA was addressed, since the alloy exhibits  
a high temperature single-phase $\gamma$ state above 900$^\circ$C and a two-phase state with $\gamma$' precipitates below that temperature. 
A multi-edge analysis of the first coordination shell for each transition metal edge was performed, utilizing simultaneously two data sets for the transition metal and aluminium  (\cite{Fantin2022}).  Such an approach allows one to constrain the total coordination number to 12 nearest neighbours, as expected in the FCC lattice, and account for mixed coordination  (\cite{Fantin2022}). 
The obtained results suggest that the heat treatment at 910$^\circ$C, which is just above the $\gamma$' formation temperature, does not affect the microstructure, hardness, local atomic or electronic structure of the alloy (\cite{Fantin2022}). 

\begin{figure}[t]
	\centering
	\includegraphics[width=0.7\textwidth]{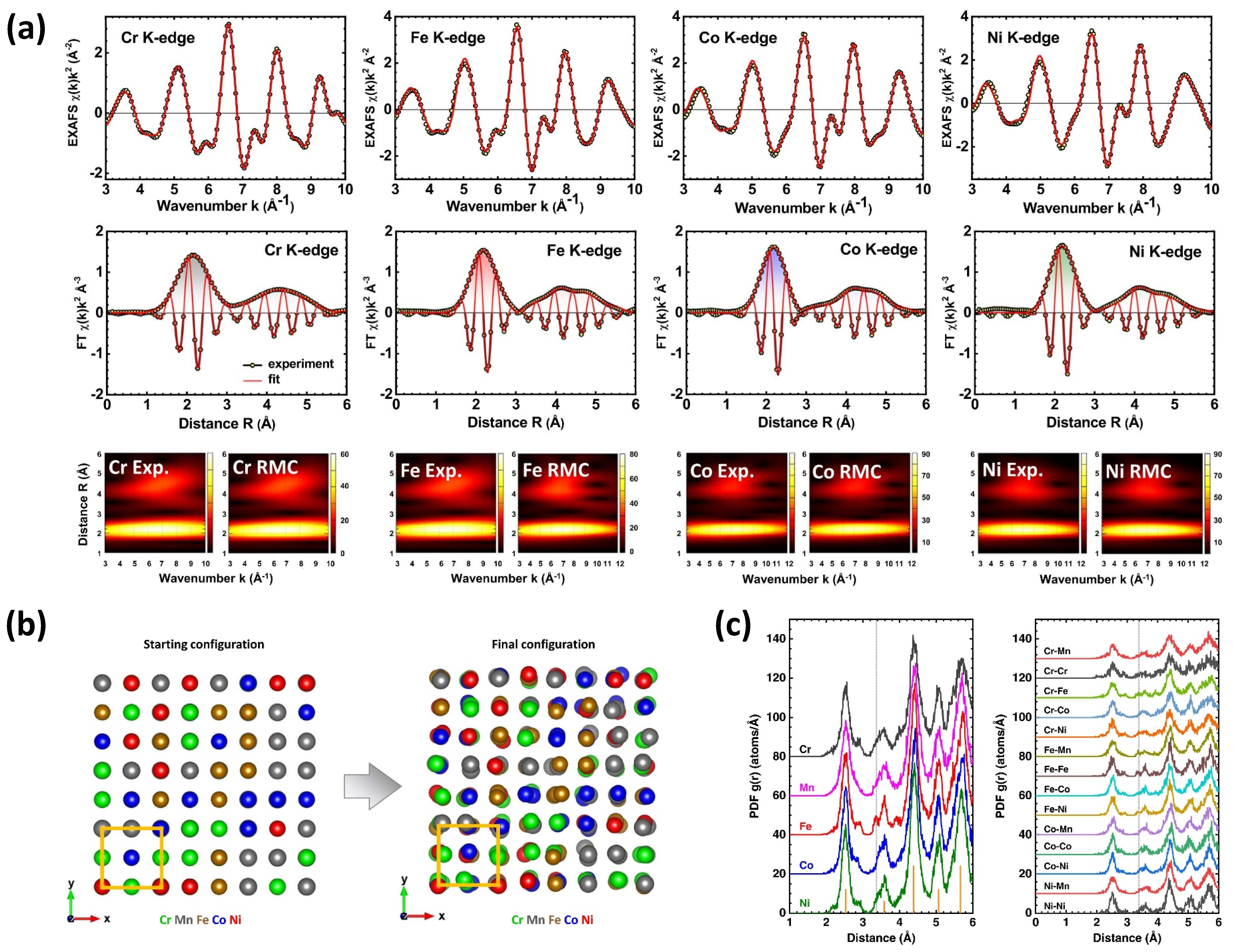}
	\caption{Analysis of the Cr, Fe, Co, and Ni K-edge EXAFS spectra of the FCC CrMnFeCoNi HEA at 300~K using the RMC method: 		
(a) Comparison of the experimental and RMC-calculated Cr, Fe, Co, and Ni K-edge EXAFS spectra $\chi(k)k^2$\ and their Fourier and Morlet wavelet transforms.
(b) Starting and final atom configurations used in the RMC simulations for simultaneous fit to EXAFS spectra 
at four absorption edges. 
(c) Total and partial pair distribution functions (PDFs) $g(r)$ for the FCC CrMnFeCoNi HEA calculated from the final atom configuration.
\textit{EXAFS},
Extended X-ray absorption fine structure; \textit{FCC}, face-centered cubic; \textit{HEA}, high-entropy alloys;
\textit{PDFs}, pair distribution functions; \textit{RMC}, reverseMonte Carlo.
	From (\protect\cite{Smekhova2022a}). Copyright (2022), with permission from Elsevier.}	
	\label{fig8}	
\end{figure}

The combination of XAS with the RMC simulations was utilized to investigate the local crystallographic ordering and specific structural relaxations of each constituent in the equiatomic single-phase FCC CrMnFeCoNi (Cantor) HEA at room temperature (Fig.\ \ref{fig8}) by \cite{Smekhova2022a}. The study demonstrated the reliability of the RMC method  (\cite{Timoshenko2014rmc}) in analysing complex multicomponent systems like HEAs. This method enabled multi-edge analysis and simultaneous fitting of the same structural model to wavelet transforms  (\cite{Timoshenko2009wt}) of all available experimental EXAFS spectra (\cite{Smekhova2022a}). Consequently, a set of partial and total distribution functions was generated for a detailed analysis from the final atomic configuration obtained in the RMC simulations. The findings revealed that all five elements of the alloy are distributed at the nodes of the FCC lattice with statistically similar averaged interatomic distances (2.54--2.55~\AA), without exhibiting any signs of atomic-scale ordering  (\cite{Smekhova2022a}). Notably, Cr atoms displayed larger structural displacements compared to the other elements, potentially influencing the magnetic properties of the alloy (\cite{Smekhova2022a}).

The same group (\cite{Smekhova2022b})  employed a similar approach to explore the Al-driven peculiarities of the local structure in single-phase FCC  Al$_{0.3}$-CrFeCoNi and bcc Al$_{3}$-CrFeCoNi HEAs.   
The analysis of the Cr, Fe Co, and Ni K-edges EXAFS spectra by the RMC method revealed unimodal and bimodal distributions of all five elements, along with their correlation with the magnetic properties of the alloys  (\cite{Smekhova2022b}).  Furthermore, soft X-ray XANES at the L$_{2,3}$ -edges of the metals uncovered a degree of surface atoms oxidation, suggesting different kinetics of oxide formation for each type of constituents  (\cite{Smekhova2022b}). 

In a recent study (\cite{Smekhova2023}), element-specific XAS techniques were employed to probe the structural, electronic, and magnetic properties of  individual constituents in a nanocrystalline Cr$_{20}$Mn$_{26}$Fe$_{18}$Co$_{19}$Ni$_{17}$ thin film with a single-phase FCC structure. 
The EXAFS analysis, using the RMC method, revealed a homogeneous short-range FCC atomic environment around each absorber with close interatomic distances (2.54-–2.55~\AA) to their nearest-neighbors and enlarged structural relaxations of Cr atoms (\cite{Smekhova2023}). The XANES analysis indicated that Cr and Mn atoms were oxidized at the surface, whereas Fe, Co, and Ni atoms remained predominantly metallic (\cite{Smekhova2023}). Furthermore, the analysis of X-ray magnetic circular dichroism spectra revealed  that Fe, Co, and Ni atoms exhibit significant spin and orbital magnetic moments, whereas Cr and Mn atoms possess small magnetic signals with opposite signs (\cite{Smekhova2023}).

The element-specific local lattice distortions in the BCC TiZrHfNbTa refractory HEA were quantified before and after a tensile test using a combination of synchrotron radiation X-ray diffraction and EXAFS methods  by \cite{Tan2023}.  
The averaged atom pair distances obtained from diffraction patterns  using  Rietveld refinement and the distances for atom pairs determined from the Zr and Nb  K-edge  EXAFS spectra were used to estimate 
element specific local distortions  (\cite{Tan2023}). It was found that the local environment around Zr experienced stronger distortion  than that around Nb in the as-prepared HEA, and this distortion increases after the tensile test (\cite{Tan2023}).  

A temperature-dependent  Zr K-edge EXAFS study was conducted on  polycrystalline superconductors by \cite{Pugliese2023}, including CoZr$_2$ ($T_c = 6.2$~K), medium entropy Co$_{0.33}$Rh$_{0.34}$Ir$_{0.33}$Zr$_2$ ($T_c = 8.7$~K), and high entropy Co$_{0.2}$Fe$_{0.2}$Ni$_{0.2}$Rh$_{0.2}$Ir$_{0.2}$Zr$_2$  ($T_c = 5.3$~K). 
It was found that the temperature effect on the interatomic distances was marginal, however, an increase in static disorder around Zr was observed with increasing mixing entropy (\cite{Pugliese2023}). 
The temperature dependences of the EXAFS Debye-Waller factors for near neighbor distances, such as   Zr--Co(Fe,Ni), Zr--Ir(Rh), Zr--Zr$_1$, Zr--Zr$_2$, Zr--Zr$_3$, were determined, demonstrating close behaviour except for the longest Zr--Zr$_2$  and Zr--Zr$_3$ bonds, which exhibited increased stiffness in the HEA (\cite{Pugliese2023}).

\begin{figure}[t]
	\centering
	\includegraphics[width=0.7\textwidth]{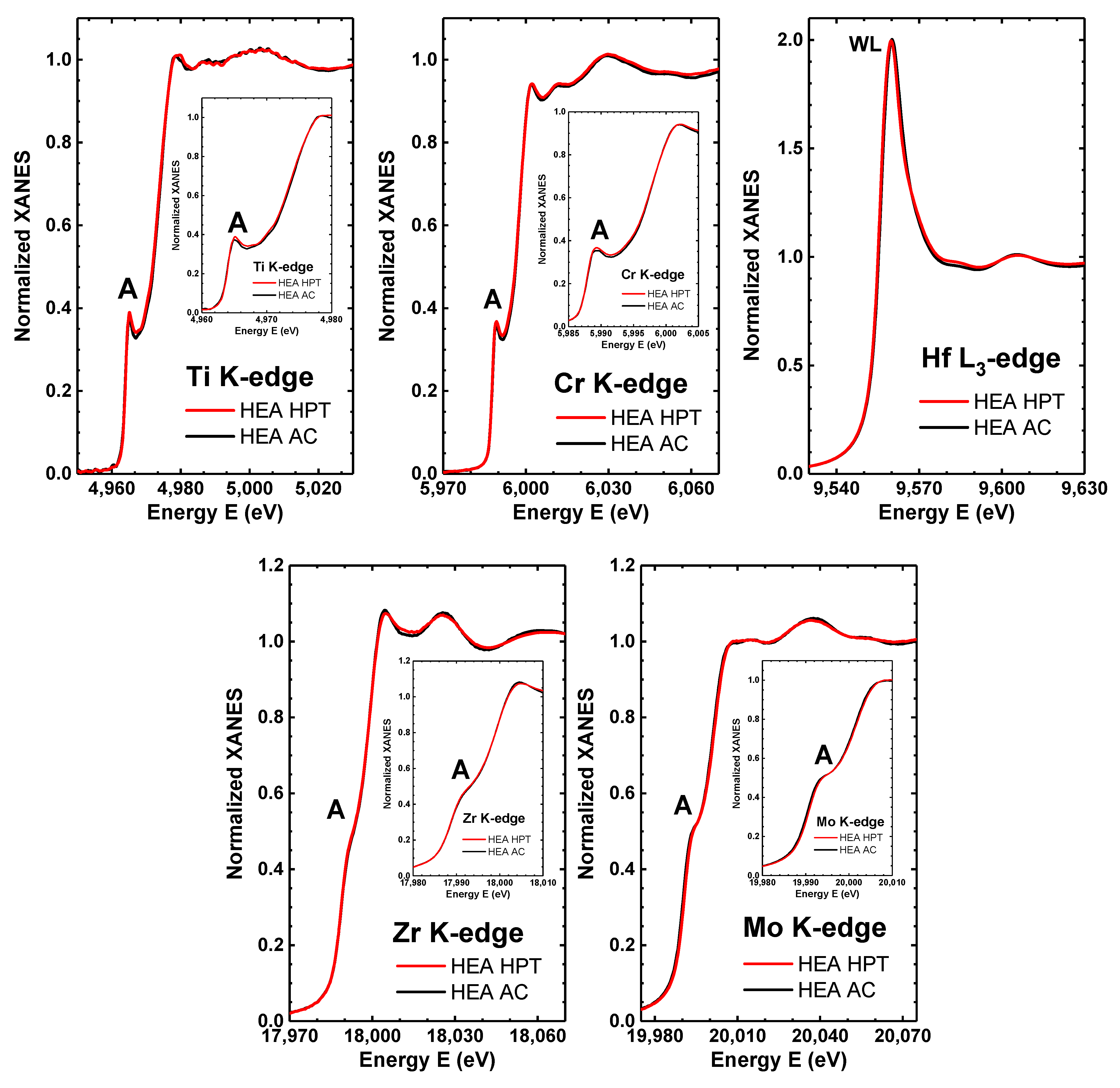}
	\caption{Normalized XANES spectra of TiZrHfMoCr HEA sample (as-cast (AC) and after high-pressure torsion (HPT) treatment) measured at the Ti, Cr, Zr, and Mo K-edges and Hf L$_3$-edge. The insets show an enlarged view around the absorption edge. The pre-edge peak A at K-edges and the white line (WL) at the L$_3$-edge are indicated.
	\textit{AC}, As-cast; \textit{HEA}, high-entropy alloys; \textit{HPT}, high-pressure
	torsion; \textit{WL}, white line; \textit{XANES}, X-ray absorption near-edge structure. From (\protect\cite{Gornakova2023}).	 Copyright (2023), with permission from MDPI. }	
	\label{fig9}	
\end{figure}

The influence of heat treatment and high-pressure torsion (HPT) on the structure of the TiZrHfMoCr high-entropy alloy was investigated by \cite{Gornakova2023}. XAS was employed at the Ti, Cr, Zr, and Mo K-edges, as well as at the Hf  L$_3$-edge, to probe the local atomic structure of two samples: one in the as-cast (AC) state and another after HPT treatment. 
The normalized XANES spectra of the HEAs (Fig.\ \ref{fig9})  indicate the similarity of the local environment around metal atoms before and after HPT treatment. A pre-edge peak (or shoulder) A arising from the 1s\textrightarrow$n$d  ($n$ = 3 for Ti and Cr, n = 4 for Zr and Mo) transition (\cite{Muller1978, Muller1982}) is present in all K-edge XANES spectra. It  becomes less prominent at higher excitation energies due to an increase in the natural line width of the 1s core level for heavier elements (\cite{KESKI1974}). The strong resonance just above the Hf L$_3$-edge, known as the white line (WL), is produced by the dipole-allowed transition 2p$_{3/2}$(Hf) \textrightarrow 5d (\cite{Qi1987}).
The EXAFS results also indicated that the local environment around metal atoms was preserved after HPT, except for some increased static disorder, presumably in the grain boundary region   (\cite{Gornakova2023}). 
Additionally, distinct local environments around Mo/Cr, Zr/Hf, and Ti atoms were revealed through the analysis of their EXAFS spectra, which were consistent with the crystallographic phases identified by XRD  (\cite{Gornakova2023}).

\subsubsection{High-Entropy Oxides}	

In a pioneering study of HEO (\cite{Rost2015}), five binary oxides were equimolarly mixed, forming an entropy-stabilized (Mg,Ni,Co,Cu,Zn)O system with the rock-salt structure, confirmed by diffraction.  
It is important to note  that pure MgO, NiO, and CoO oxides have the crystal structures identical to HEO, but CuO and ZnO possess different tenorite and wurtzite structures, respectively. Therefore a structural transition takes place upon mixing (\cite{Rost2015}). EXAFS results obtained at the metal K-edges  show that the local structures of each cation species in HEO are similar and consistent with a random cation distribution, supporting the entropy stabilization hypothesis (\cite{Rost2015}).
The metal-to-oxygen nearest-neighbour distances for each metal species were found to be close within experimental error and comparable to the values reported for rocksalt structures (\cite{Rost2015}).
The absence of SRO or clustering in HEO was demonstrated based on the shape of the EXAFS oscillations 
(\cite{Rost2015}).

In a subsequent study (\cite{Braun2018}), the effect of an additional sixth element (Sc, Sb, Sn, Cr, Ge) on the  properties of the (Mg,Ni,Co,Cu,Zn)O system was investigated.  
It was observed that entropy-stabilized oxides have very low thermal conductivity and high elastic modulus attributed to a disorder in the interatomic forces caused by local ionic charge disorder (\cite{Braun2018}). The EXAFS results for the first and second coordination shells showed that the addition of a sixth cation to a five-cation HEO leads to a significant distortion of the oxygen sublattice and a change in the local coordination environment around the cobalt absorber (\cite{Braun2018}). These structural changes indicate that the ionic bonds in HEOs are highly disordered and affect the phonon scattering rate, which in turn determines the thermal conductivity (\cite{Braun2018}).

The lithiation/delithiation mechanism in the (Mg,Ni,Co,Cu,Zn)O HEO anode  for application in Li-ion batteries  was elaborated upon by \cite{Ghigna2020}. 
It was shown that the insertion of lithium is a complex and  irreversible process, resulting in the reduction of transition metals (Co, Ni, Cu, Zn) to their metallic state and the formation of an alloy with Li ions (\cite{Ghigna2020}). 
Operando XAS studies at the metal K-edges supported this finding by demonstrating the changes in the local structure and oxidation state of the metal cations during the lithiation/delithiation process (\cite{Ghigna2020}). 
The multivariate curve resolution (MCR) and principal component analysis (PCA) methods were employed  
to quantitatively determine the number and identity of key metal-related species that appear during the HEO anode evolution and to monitor their concentration changes throughout the reaction (\cite{Tavani2020}). 
XANES and EXAFS results also revealed that the HEO rock-salt structure collapses after a certain degree of lithiation, leading to the segregation of metals and oxides (\cite{Ghigna2020}). 

In the recent paper, the XAS method was employed to further advance our understanding of the behaviour of the (Mg,Ni,Co,Cu,Zn)O HEO during electrochemical cycling with lithium (\cite{Wang2023}). The remarkable electrochemical performance was demonstrated to result from the synergetic effect of the cations in the HEO during the conversion reaction with lithium. The analysis of the Co, Ni, Cu, and Zn K-edge XANES and EXAFS  spectra revealed that the metal ions in the HEO are reduced to metallic state during the initial discharge, forming FCC Co, Ni, and Cu phases, as well as an FCC LiZn alloy (\cite{Wang2023}). After recharging to 3.0~V, the metal ions exhibit different behaviors: Ni and Cu remain in the metallic state, whereas a significant  fraction of Co participates in the redox reaction, with some Co remaining in a metallic state.  Zn is almost fully deoxidized to the 2+ state (\cite{Wang2023}). Thus, the elements with higher electronegativity (Cu and Ni) are responsible for creating an electrochemically inert conductive  network, whereas the Zn cations stabilize an oxide nanophase, allowing for the accommodation of lithium ions (\cite{Wang2023}).

A novel catalyst (Pd$_1$@HEFO), combining Pd  single-atom with a high-entropy fluorite oxide (CeZrHfTiLa)O$_x$ (HEFO) used as a support, was synthesized  using a mechanochemical-assisted method by \cite{Xu2020}. It demonstrated superior CO oxidation activity as well as thermal and hydrothermal stability compared to a conventional Pd@CeO$_2$ catalyst. The formation of single Pd atoms was confirmed by an atomic-resolution transmission electron microscopy image and the energy-dispersive X-ray spectroscopy  mapping image of Pd (\cite{Xu2020}).  XANES and EXAFS measurements performed at the Pd K-edge were used to confirm the formation of single Pd atoms in Pd$_1$@HEFO and to investigate their electronic structure and coordination environment (\cite{Xu2020}). The EXAFS spectra showed that the Pd atoms are incorporated into the HEFO sublattice by forming stable Pd--O--M bonds (M = Ce, Zr, La), with no Pd--Pd or Pd--O--Pd bonding (\cite{Xu2020}). The XANES results revealed that the valence state of Pd ranges between 0 and +2. These features contribute to the enhanced catalytic performance of Pd$_1$@HEFO (\cite{Xu2020}).

A cobalt-free spinel (CrMnFeNiCu)$_3$O$_4$ HEO is a  candidate for the use as an anode material in lithium-ion batteries. Its charge–discharge mechanism was studied using operando quick-scanning XAS by \cite{Luo2022}. Valence and coordination states of transition metal elements were determined from the edge positions and shapes of their K-edge XANES spectra.  
The results reveal that all the constituent elements (Cr, Mn, Fe, Ni, and Cu) in the HEO participate in the redox reactions during Li$^+$ uptake/release, but with different transition steps, redox sequence, reversibility, and overpotential (\cite{Luo2022}).

The concept of high entropy was employed by \cite{Han2022} to synthesize nanoporous NiFeCoMnAl oxide with an  amorphous structure on carbon paper through electrochemical deposition followed by the dealloying technique aimed at reducing the amount of aluminium and creating a nanoporous structure. This HEO was proposed as an efficient electrocatalyst for the oxygen evolution reaction (OER) (\cite{Han2022}). The authors attributed the high OER activity to the synergistic effect of multiple elements, especially Mn doping, which promoted the formation of $\beta$-NiOOH intermediates with higher intrinsic activity compared to  $\gamma$-NiOOH intermediates. The Ni K-edge XANES and EXAFS were employed to understand the influence of Mn doping on the valence state of nickel ions in NiFeCoMnAl before and after OER (\cite{Han2022}). 
The emergence of a higher oxidation state of Ni and shorter Ni--O distances after the OER were observed (\cite{Han2022}).

\begin{figure}[t]
	\centering
	\includegraphics[width=0.9\textwidth]{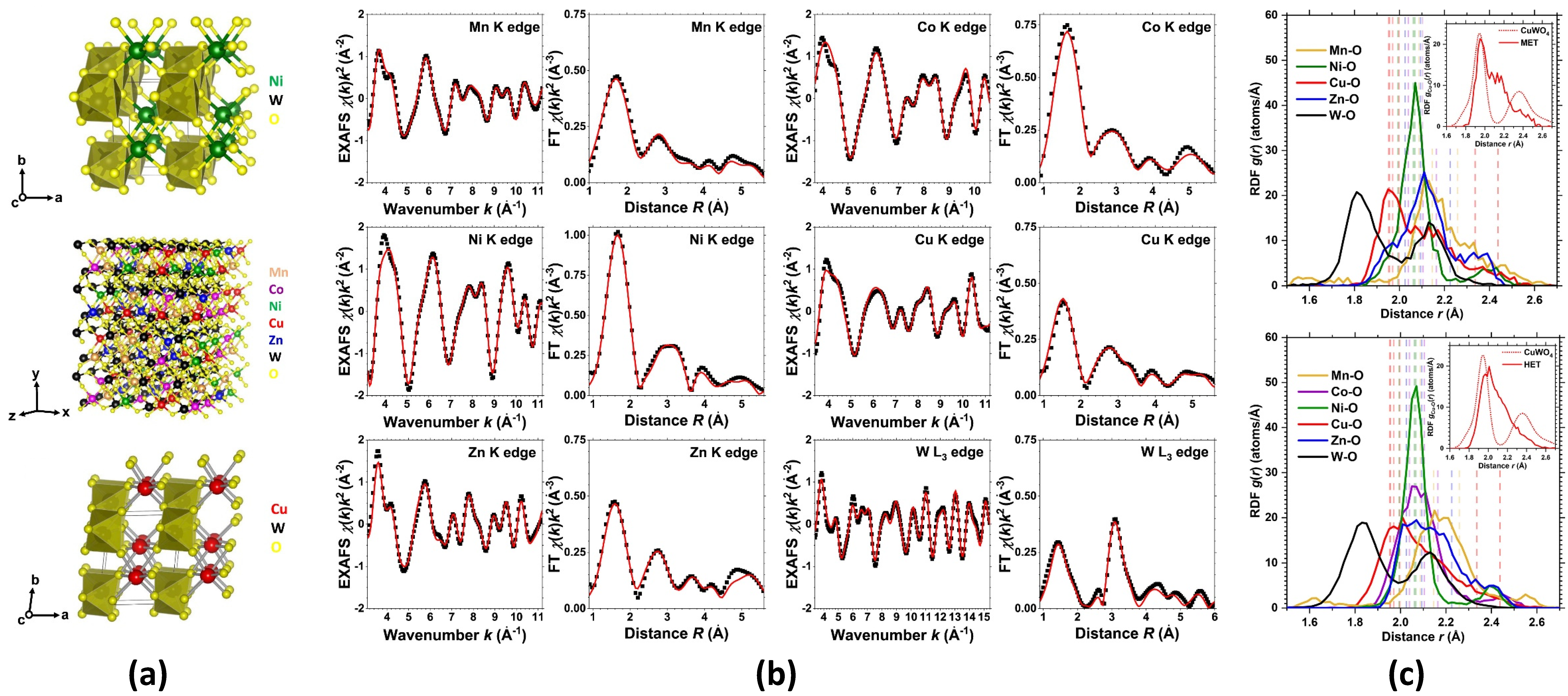}
	\caption{Results of the EXAFS study of MET (MnNiCuZn)WO$_4$ and HET (MnCoNiCuZn)WO$_4$ at 10~K using the RMC method:  		
		(a)  Crystallographic structures of monoclinic NiWO$_4$, monoclinic (MnCoNiCuZn)WO$_4$, and triclinic CuWO$_4$. 
		(b) Experimental (\textit{black dots}) and calculated (\textit{red lines}) EXAFS spectra $\chi(k)k^2$\ and their Fourier transforms (FTs) for HET (MnCoNiCuZn)WO$_4$ at K-edges of Mn, Co, Ni, Cu, and Zn, and at the L$_3$-edge of W at 10~K.  
		(c) Radial distribution functions (RDFs) $g(r)$ in (MnNiCuZn)WO$_4$ MET and (MnCoNiCuZn)WO$_4$ HET at 10~K. Vertical dashed lines indicate the respective bond lengths in pure MnWO$_4$, CoWO$_4$, NiWO$_4$, CuWO$_4$, and ZnWO$_4$ as determined by X-ray diffraction at 300~K. The inset shows partial RDFs for Cu--O atomic pairs in CuWO$_4$ and (MnNiCuZn)WO$_4$ or (MnCoNiCuZn)WO$_4$. 
		\textit{EXAFS}, Extended X-ray absorption fine structure; \textit{FTs}, Fourier transforms; \textit{HET}, high-entropy tungstate; \textit{MET}, medium-entropy tungstate; \textit{RDFs}, radial distribution functions; \textit{RMC}, reverse Monte Carlo.
		From (\protect\cite{Bakradze2023}). Copyright (2023), with permission from Elsevier.}	
	\label{fig10}	
\end{figure}

A medium-entropy (MnNiCuZn)WO$_4$ and high-entropy (MnCoNiCuZn)WO$_4$ tungstates with monoclinic crystal structure and a random distribution of 3d metal cations were recently  synthesized  (\cite{Bakradze2023}) and demonstrated different degree of local structure distortions. 
The XANES spectra at the Mn, Co, Ni, and Zn K-edges of both the medium-entropy tungstate (MET) and the high-entropy tungstate (HET) exhibit shapes similar to those found in pure tungstates, indicating a resemblance in the local atomic environment  around each of these metal ions and the proximity of their oxidation states  (\cite{Bakradze2023}).
However, the Cu K-edge XANES spectra of the MET and the HET differ from that of 
CuWO$_4$, suggesting differences in the distortion of the copper environment  (\cite{Bakradze2023}). 
EXAFS spectroscopy, combined with RMC simulations, allowed for a detailed  examination of the atomic environments around metal cations in these compounds by determining the partial RDFs for metal-oxygen distances   (Fig.\ \ref{fig10}) (\cite{Bakradze2023}). 
The EXAFS analysis revealed  that Ni$^{2+}$ ions have the strongest tendency to organize their local environment and form regular [NiO$_6$] octahedra, whereas Mn$^{2+}$, Co$^{2+}$, Zn$^{2+}$, and W$^{6+}$ ions exhibit distorted octahedral coordination (\cite{Bakradze2023}).  Perhaps the most intriguing result is that the shape of [CuO$_6$] octahedra in both tungstates differs from that in pure CuWO$_4$, where a strong Jahn--Teller distortion is present  (\cite{Bakradze2023}). The authors proposed that the increase in configurational entropy indirectly affects the longest Cu--O bonds in [CuO$_6$] octahedra  (\cite{Bakradze2023}).

\subsection{Conclusions}

XAS provides a great opportunity to study the local environment in multicomponent compounds such as HEMs. By tuning the X-ray energy across the absorption edge of each element in HEM, information on their oxidation state, chemical bonding, and local structural distortions can be obtained. However, care should be taken when analysis is extended beyond the nearest coordination shell(s), since MS contributions originating from many-atom distributions may become important and, thus, should be taken into account in the analysis. The similarity of atomic scattering amplitudes can also be a limiting factor when HEM include elements neighboring in the periodic table, such as in  Cantor alloys.

The use of advanced methods of EXAFS analysis based on atomistic simulations (\cite{Ferrari2020}) opens up many additional opportunities. The EXAFS spectra can be used to validate MD simulations, allowing the selection and optimization of potential models.  The RMC method allows one to perform multi-edge EXAFS analysis and to determine a structural model of HEM being in agreement with all available experimental EXAFS spectra. This provides the possibility of obtaining a set of partial distribution functions that can be used further to determine such structural parameters as average interatomic distances, MSRDs for pairs of atoms, and mean-squared displacements for atoms of each type.

Further development of methods for analyzing XANES/EXAFS spectra, such as ML techniques (\cite{Timoshenko2017,Liu2023}), is expected to offer new opportunities for obtaining information on the local structure and its distortion in HEMs.

\subsection*{Acknowledgment}

This work was supported by the Latvian Council of Science project No. lzp-2023/1-0476.

\newpage 


\bibliography{xafs,xasha,ha,md}{}

\begin{thebibliography}{149}
\expandafter\ifx\csname natexlab\endcsname\relax\def\natexlab#1{#1}\fi
\providecommand{\url}[1]{\texttt{#1}}
\providecommand{\href}[2]{#2}
\providecommand{\path}[1]{#1}
\providecommand{\DOIprefix}{doi:}
\providecommand{\ArXivprefix}{arXiv:}
\providecommand{\URLprefix}{URL: }
\providecommand{\Pubmedprefix}{pmid:}
\providecommand{\doi}[1]{\href{http://dx.doi.org/#1}{\path{#1}}}
\providecommand{\Pubmed}[1]{\href{pmid:#1}{\path{#1}}}
\providecommand{\bibinfo}[2]{#2}
\ifx\xfnm\relax \def\xfnm[#1]{\unskip,\space#1}\fi
\bibitem[{Abraham(1986)}]{Abraham1986}
\bibinfo{author}{Abraham, F.~F.} (\bibinfo{year}{1986}).
\newblock \bibinfo{title}{{Computational statistical mechanics methodology,
  applications and supercomputing}}.
\newblock {\it \bibinfo{journal}{Adv. Phys.}\/},  {\it \bibinfo{volume}{35}\/},
  \bibinfo{pages}{1--111}. \DOIprefix\doi{10.1080/00018738600101851}.
\bibitem[{Albedwawi et~al.(2021)Albedwawi, AlJaberi, Haidemenopoulos \&
  Polychronopoulou}]{Albedwawi2021}
\bibinfo{author}{Albedwawi, S.~H.}, \bibinfo{author}{AlJaberi, A.},
  \bibinfo{author}{Haidemenopoulos, G.~N.}, \&
  \bibinfo{author}{Polychronopoulou, K.} (\bibinfo{year}{2021}).
\newblock \bibinfo{title}{{High entropy oxides-exploring a paradigm of
  promising catalysts: A review}}.
\newblock {\it \bibinfo{journal}{Mater. Des.}\/},  {\it
  \bibinfo{volume}{202}\/}, \bibinfo{pages}{109534}.
  \DOIprefix\doi{10.1016/j.matdes.2021.109534}.
\bibitem[{Ankudinov et~al.(1998)Ankudinov, Ravel, Rehr \& Conradson}]{FEFF8}
\bibinfo{author}{Ankudinov, A.~L.}, \bibinfo{author}{Ravel, B.},
  \bibinfo{author}{Rehr, J.~J.}, \& \bibinfo{author}{Conradson, S.~D.}
  (\bibinfo{year}{1998}).
\newblock \bibinfo{title}{{Real-space multiple-scattering calculation and
  interpretation of X-ray-absorption near-edge structure}}.
\newblock {\it \bibinfo{journal}{Phys. Rev. B}\/},  {\it
  \bibinfo{volume}{58}\/}, \bibinfo{pages}{7565--7576}.
  \DOIprefix\doi{10.1103/PhysRevB.58.7565}.
\bibitem[{Bakradze et~al.(2023)Bakradze, Welter \& Kuzmin}]{Bakradze2023}
\bibinfo{author}{Bakradze, G.}, \bibinfo{author}{Welter, E.}, \&
  \bibinfo{author}{Kuzmin, A.} (\bibinfo{year}{2023}).
\newblock \bibinfo{title}{{Peculiarities of the local structure in new medium-
  and high-entropy, low-symmetry tungstates}}.
\newblock {\it \bibinfo{journal}{J. Phys. Chem. Solids}\/},  {\it
  \bibinfo{volume}{172}\/}, \bibinfo{pages}{111052}.
  \DOIprefix\doi{10.1016/j.jpcs.2022.111052}.
\bibitem[{Behler(2016)}]{Behler2016}
\bibinfo{author}{Behler, J.} (\bibinfo{year}{2016}).
\newblock \bibinfo{title}{{Perspective: Machine learning potentials for
  atomistic simulations}}.
\newblock {\it \bibinfo{journal}{J. Chem. Phys.}\/},  {\it
  \bibinfo{volume}{145}\/}. \DOIprefix\doi{10.1063/1.4966192}.
\bibitem[{Berne \& Thirumalai(1986)}]{Berne1986}
\bibinfo{author}{Berne, B.~J.}, \& \bibinfo{author}{Thirumalai, D.}
  (\bibinfo{year}{1986}).
\newblock \bibinfo{title}{{On the simulation of quantum systems: path integral
  methods}}.
\newblock {\it \bibinfo{journal}{Ann. Rev. Phys. Chem.}\/},  {\it
  \bibinfo{volume}{37}\/}, \bibinfo{pages}{401--424}.
  \DOIprefix\doi{10.1146/annurev.pc.37.100186.002153}.
\bibitem[{Bocharov et~al.(2017)Bocharov, Chollet, Krack, Bertsch, Grolimund,
  Martin, Kuzmin, Purans \& Kotomin}]{Bocharov2017}
\bibinfo{author}{Bocharov, D.}, \bibinfo{author}{Chollet, M.},
  \bibinfo{author}{Krack, M.}, \bibinfo{author}{Bertsch, J.},
  \bibinfo{author}{Grolimund, D.}, \bibinfo{author}{Martin, M.},
  \bibinfo{author}{Kuzmin, A.}, \bibinfo{author}{Purans, J.}, \&
  \bibinfo{author}{Kotomin, E.} (\bibinfo{year}{2017}).
\newblock \bibinfo{title}{{Analysis of the U L$_3$-edge X-ray absorption
  spectra in UO$_2$ using molecular dynamics simulations}}.
\newblock {\it \bibinfo{journal}{Prog. Nucl. Energy}\/},  {\it
  \bibinfo{volume}{94}\/}, \bibinfo{pages}{187--193}.
  \DOIprefix\doi{10.1016/j.pnucene.2016.07.017}.
\bibitem[{Bocharov et~al.(2020)Bocharov, Krack, Rafalskij, Kuzmin \&
  Purans}]{Bocharov2020}
\bibinfo{author}{Bocharov, D.}, \bibinfo{author}{Krack, M.},
  \bibinfo{author}{Rafalskij, Y.}, \bibinfo{author}{Kuzmin, A.}, \&
  \bibinfo{author}{Purans, J.} (\bibinfo{year}{2020}).
\newblock \bibinfo{title}{{Ab initio molecular dynamics simulations of negative
  thermal expansion in ScF$_3$: The effect of the supercell size}}.
\newblock {\it \bibinfo{journal}{Comput. Mater. Sci.}\/},  {\it
  \bibinfo{volume}{171}\/}, \bibinfo{pages}{109198}.
  \DOIprefix\doi{10.1016/j.commatsci.2019.109198}.
\bibitem[{Booth et~al.(1995)Booth, Bridges, Bauer, Li, Boyce, Claeson, Chu \&
  Xiong}]{Booth1995}
\bibinfo{author}{Booth, C.~H.}, \bibinfo{author}{Bridges, F.},
  \bibinfo{author}{Bauer, E.~D.}, \bibinfo{author}{Li, G.~G.},
  \bibinfo{author}{Boyce, J.~B.}, \bibinfo{author}{Claeson, T.},
  \bibinfo{author}{Chu, C.~W.}, \& \bibinfo{author}{Xiong, Q.}
  (\bibinfo{year}{1995}).
\newblock \bibinfo{title}{{XAFS measurements of negatively correlated atomic
  displacements in HgBa$_2$CuO$_4+\delta$}}.
\newblock {\it \bibinfo{journal}{Phys. Rev. B}\/},  {\it
  \bibinfo{volume}{52}\/}, \bibinfo{pages}{R15745--R15748}.
  \DOIprefix\doi{10.1103/PhysRevB.52.R15745}.
\bibitem[{Bordiga et~al.(2013)Bordiga, Groppo, Agostini, van Bokhoven \&
  Lamberti}]{Bordiga2013}
\bibinfo{author}{Bordiga, S.}, \bibinfo{author}{Groppo, E.},
  \bibinfo{author}{Agostini, G.}, \bibinfo{author}{van Bokhoven, J.~A.}, \&
  \bibinfo{author}{Lamberti, C.} (\bibinfo{year}{2013}).
\newblock \bibinfo{title}{{Reactivity of surface species in heterogeneous
  catalysts probed by in situ X-ray absorption techniques}}.
\newblock {\it \bibinfo{journal}{Chem. Rev.}\/},  {\it
  \bibinfo{volume}{113}\/}, \bibinfo{pages}{1736--1850}.
  \DOIprefix\doi{10.1021/cr2000898}.
\bibitem[{Bowron(2008)}]{Bowron2008epsr}
\bibinfo{author}{Bowron, D.~T.} (\bibinfo{year}{2008}).
\newblock \bibinfo{title}{{Experimentally consistent atomistic modeling of bulk
  and local structure in liquids and disordered materials by empirical
  potential structure refinement}}.
\newblock {\it \bibinfo{journal}{Pure Appl. Chem.}\/},  {\it
  \bibinfo{volume}{80}\/}, \bibinfo{pages}{1211--1227}.
  \DOIprefix\doi{10.1351/pac200880061211}.
\bibitem[{Braun et~al.(2018)Braun, Rost, Lim, Giri, Olson, Kotsonis, Stan,
  Brenner, Maria \& Hopkins}]{Braun2018}
\bibinfo{author}{Braun, J.~L.}, \bibinfo{author}{Rost, C.~M.},
  \bibinfo{author}{Lim, M.}, \bibinfo{author}{Giri, A.},
  \bibinfo{author}{Olson, D.~H.}, \bibinfo{author}{Kotsonis, G.~N.},
  \bibinfo{author}{Stan, G.}, \bibinfo{author}{Brenner, D.~W.},
  \bibinfo{author}{Maria, J.-P.}, \& \bibinfo{author}{Hopkins, P.~E.}
  (\bibinfo{year}{2018}).
\newblock \bibinfo{title}{{Charge-induced disorder controls the thermal
  conductivity of entropy-stabilized oxides}}.
\newblock {\it \bibinfo{journal}{Adv. Mater.}\/},  {\it
  \bibinfo{volume}{30}\/}, \bibinfo{pages}{1805004}.
  \DOIprefix\doi{10.1002/adma.201805004}.
\bibitem[{Brechtl \& Liaw(2021)}]{Brechtl2021}
\bibinfo{editor}{Brechtl, J.}, \& \bibinfo{editor}{Liaw, P.~K.} (Eds.)
  (\bibinfo{year}{2021}).
\newblock {\it \bibinfo{title}{{High-Entropy Materials: Theory, Experiments,
  and Applications}}\/}.
\newblock \bibinfo{publisher}{Springer Cham}.
\newblock \DOIprefix\doi{10.1007/978-3-030-77641-1}.
\bibitem[{Buckingham et~al.(2022)Buckingham, Ward-O'Brien, Xiao, Li, Qu \&
  Lewis}]{Buckingham2022}
\bibinfo{author}{Buckingham, M.~A.}, \bibinfo{author}{Ward-O'Brien, B.},
  \bibinfo{author}{Xiao, W.}, \bibinfo{author}{Li, Y.}, \bibinfo{author}{Qu,
  J.}, \& \bibinfo{author}{Lewis, D.~J.} (\bibinfo{year}{2022}).
\newblock \bibinfo{title}{{High entropy metal chalcogenides: synthesis,
  properties, applications and future directions}}.
\newblock {\it \bibinfo{journal}{Chem. Commun.}\/},  {\it
  \bibinfo{volume}{58}\/}, \bibinfo{pages}{8025--8037}.
  \DOIprefix\doi{10.1039/D2CC01796B}.
\bibitem[{Bunker(1983)}]{Bunker1983}
\bibinfo{author}{Bunker, G.} (\bibinfo{year}{1983}).
\newblock \bibinfo{title}{{Application of the ratio method of EXAFS analysis to
  disordered systems}}.
\newblock {\it \bibinfo{journal}{Nucl. Instrum. Methods}\/},  {\it
  \bibinfo{volume}{207}\/}, \bibinfo{pages}{437--444}.
  \DOIprefix\doi{10.1016/0167-5087(83)90655-5}.
\bibitem[{Cabaret et~al.(2001)Cabaret, Grand, Ramos, Flank, Rossano, Galoisy,
  Calas \& Ghaleb}]{Cabaret2001}
\bibinfo{author}{Cabaret, D.}, \bibinfo{author}{Grand, M.~L.},
  \bibinfo{author}{Ramos, A.}, \bibinfo{author}{Flank, A.-M.},
  \bibinfo{author}{Rossano, S.}, \bibinfo{author}{Galoisy, L.},
  \bibinfo{author}{Calas, G.}, \& \bibinfo{author}{Ghaleb, D.}
  (\bibinfo{year}{2001}).
\newblock \bibinfo{title}{{Medium range structure of borosilicate glasses from
  Si K-edge XANES: a combined approach based on multiple scattering and
  molecular dynamics calculations}}.
\newblock {\it \bibinfo{journal}{J. Non-Cryst. Solids}\/},  {\it
  \bibinfo{volume}{289}\/}, \bibinfo{pages}{1--8}.
  \DOIprefix\doi{10.1016/S0022-3093(01)00733-5}.
\bibitem[{Cantor et~al.(2004)Cantor, Chang, Knight \& Vincent}]{Cantor2004}
\bibinfo{author}{Cantor, B.}, \bibinfo{author}{Chang, I.},
  \bibinfo{author}{Knight, P.}, \& \bibinfo{author}{Vincent, A.}
  (\bibinfo{year}{2004}).
\newblock \bibinfo{title}{{Microstructural development in equiatomic
  multicomponent alloys}}.
\newblock {\it \bibinfo{journal}{Mater. Sci. Eng. A}\/},  {\it
  \bibinfo{volume}{375-377}\/}, \bibinfo{pages}{213--218}.
  \DOIprefix\doi{10.1016/j.msea.2003.10.257}.
\bibitem[{Cavin et~al.(2021)Cavin, Ahmadiparidari, Majidi, Thind, Misal,
  Prajapati, Hemmat, Rastegar, Beukelman, Singh, Unocic, Salehi-Khojin \&
  Mishra}]{Cavin2021}
\bibinfo{author}{Cavin, J.}, \bibinfo{author}{Ahmadiparidari, A.},
  \bibinfo{author}{Majidi, L.}, \bibinfo{author}{Thind, A.~S.},
  \bibinfo{author}{Misal, S.~N.}, \bibinfo{author}{Prajapati, A.},
  \bibinfo{author}{Hemmat, Z.}, \bibinfo{author}{Rastegar, S.},
  \bibinfo{author}{Beukelman, A.}, \bibinfo{author}{Singh, M.~R.},
  \bibinfo{author}{Unocic, K.~A.}, \bibinfo{author}{Salehi-Khojin, A.}, \&
  \bibinfo{author}{Mishra, R.} (\bibinfo{year}{2021}).
\newblock \bibinfo{title}{{2D high-entropy transition metal dichalcogenides for
  carbon dioxide electrocatalysis}}.
\newblock {\it \bibinfo{journal}{Adv. Mater.}\/},  {\it
  \bibinfo{volume}{33}\/}, \bibinfo{pages}{2100347}.
  \DOIprefix\doi{10.1002/adma.202100347}.
\bibitem[{Chen et~al.(2022)Chen, Li, Huang, Ma, Liu, Tang, Fang \&
  Dai}]{Chen2022}
\bibinfo{author}{Chen, H.}, \bibinfo{author}{Li, S.}, \bibinfo{author}{Huang,
  S.}, \bibinfo{author}{Ma, L.}, \bibinfo{author}{Liu, S.},
  \bibinfo{author}{Tang, F.}, \bibinfo{author}{Fang, Y.}, \&
  \bibinfo{author}{Dai, P.} (\bibinfo{year}{2022}).
\newblock \bibinfo{title}{{High-entropy structure design in layered transition
  metal dichalcogenides}}.
\newblock {\it \bibinfo{journal}{Acta Mater.}\/},  {\it
  \bibinfo{volume}{222}\/}, \bibinfo{pages}{117438}.
  \DOIprefix\doi{10.1016/j.actamat.2021.117438}.
\bibitem[{Chen et~al.(2009)Chen, Hu, Tsai, Hsieh, Kao, Yeh, Chin \&
  Chen}]{Chen2009}
\bibinfo{author}{Chen, Y.-L.}, \bibinfo{author}{Hu, Y.-H.},
  \bibinfo{author}{Tsai, C.-W.}, \bibinfo{author}{Hsieh, C.-A.},
  \bibinfo{author}{Kao, S.-W.}, \bibinfo{author}{Yeh, J.-W.},
  \bibinfo{author}{Chin, T.-S.}, \& \bibinfo{author}{Chen, S.-K.}
  (\bibinfo{year}{2009}).
\newblock \bibinfo{title}{{Alloying behavior of binary to octonary alloys based
  on Cu–Ni–Al–Co–Cr–Fe–Ti–Mo during mechanical alloying}}.
\newblock {\it \bibinfo{journal}{J. Alloys Compd.}\/},  {\it
  \bibinfo{volume}{477}\/}, \bibinfo{pages}{696--705}.
  \DOIprefix\doi{10.1016/j.jallcom.2008.10.111}.
\bibitem[{Clausen \& N{\o}rskov(2000)}]{Clausen2000}
\bibinfo{author}{Clausen, B.~S.}, \& \bibinfo{author}{N{\o}rskov, J.~K.}
  (\bibinfo{year}{2000}).
\newblock \bibinfo{title}{{Asymmetric pair distribution functions in
  catalysts}}.
\newblock {\it \bibinfo{journal}{Top. Catal.}\/},  {\it
  \bibinfo{volume}{10}\/}, \bibinfo{pages}{221--230}.
  \DOIprefix\doi{10.1023/A:1019196908404}.
\bibitem[{Dalba et~al.(1995)Dalba, Fornasini, Grazioli \& Rocca}]{Dalba1995age}
\bibinfo{author}{Dalba, G.}, \bibinfo{author}{Fornasini, P.},
  \bibinfo{author}{Grazioli, M.}, \& \bibinfo{author}{Rocca, F.}
  (\bibinfo{year}{1995}).
\newblock \bibinfo{title}{{Local disorder in crystalline and amorphous
  germanium}}.
\newblock {\it \bibinfo{journal}{Phys. Rev. B}\/},  {\it
  \bibinfo{volume}{52}\/}, \bibinfo{pages}{11034--11043}.
  \DOIprefix\doi{10.1103/PhysRevB.52.11034}.
\bibitem[{Dalba et~al.(1993)Dalba, Fornasini \& Rocca}]{Dalba1993}
\bibinfo{author}{Dalba, G.}, \bibinfo{author}{Fornasini, P.}, \&
  \bibinfo{author}{Rocca, F.} (\bibinfo{year}{1993}).
\newblock \bibinfo{title}{{Cumulant analysis of the extended X-ray-absorption
  fine structure of $\beta$-AgI}}.
\newblock {\it \bibinfo{journal}{Phys. Rev. B}\/},  {\it
  \bibinfo{volume}{47}\/}, \bibinfo{pages}{8502--8514}.
  \DOIprefix\doi{10.1103/PhysRevB.47.8502}.
\bibitem[{D'Angelo et~al.(2002)D'Angelo, Barone, Chillemi, Sanna, Meyer-Klaucke
  \& Pavel}]{Angelo2002}
\bibinfo{author}{D'Angelo, P.}, \bibinfo{author}{Barone, V.},
  \bibinfo{author}{Chillemi, G.}, \bibinfo{author}{Sanna, N.},
  \bibinfo{author}{Meyer-Klaucke, W.}, \& \bibinfo{author}{Pavel, N.~V.}
  (\bibinfo{year}{2002}).
\newblock \bibinfo{title}{{Hydrogen and higher shell contributions in
  Zn$^{2+}$, Ni$^{2+}$, and Co$^{2+}$ aqueous solutions: an X-ray absorption
  fine structure and molecular dynamics study}}.
\newblock {\it \bibinfo{journal}{J. Am. Chem. Soc.}\/},  {\it
  \bibinfo{volume}{124}\/}, \bibinfo{pages}{1958--1967}.
  \DOIprefix\doi{10.1021/ja015685x}.
\bibitem[{{D'Angelo} et~al.(1994){D'Angelo}, {Di Nola}, Filipponi, Pavel \&
  Roccatano}]{Angelo1994}
\bibinfo{author}{{D'Angelo}, P.}, \bibinfo{author}{{Di Nola}, A.},
  \bibinfo{author}{Filipponi, A.}, \bibinfo{author}{Pavel, N.~V.}, \&
  \bibinfo{author}{Roccatano, D.} (\bibinfo{year}{1994}).
\newblock \bibinfo{title}{{An extended x-ray absorption fine structure study of
  aqueous solutions by employing molecular dynamics simulations}}.
\newblock {\it \bibinfo{journal}{J. Chem. Phys.}\/},  {\it
  \bibinfo{volume}{100}\/}, \bibinfo{pages}{985--994}.
  \DOIprefix\doi{10.1063/1.466581}.
\bibitem[{Delaye(2001)}]{Delaye2001}
\bibinfo{author}{Delaye, J.-M.} (\bibinfo{year}{2001}).
\newblock \bibinfo{title}{{Modeling of multicomponent glasses: a review}}.
\newblock {\it \bibinfo{journal}{Curr. Opin. Solid State Mater. Sci.}\/},  {\it
  \bibinfo{volume}{5}\/}, \bibinfo{pages}{451--454}.
  \DOIprefix\doi{10.1016/S1359-0286(01)00028-6}.
\bibitem[{{Di Cicco}(1995)}]{DiCicco1995}
\bibinfo{author}{{Di Cicco}, A.} (\bibinfo{year}{1995}).
\newblock \bibinfo{title}{{EXAFS multiple-scattering data-analysis: GNXAS
  methodology and applications}}.
\newblock {\it \bibinfo{journal}{Physica B}\/},  {\it
  \bibinfo{volume}{208-209}\/}, \bibinfo{pages}{125--128}.
  \DOIprefix\doi{10.1016/0921-4526(94)00647-E}.
\bibitem[{{Di Cicco} \& Iesari(2022)}]{DiCicco2022}
\bibinfo{author}{{Di Cicco}, A.}, \& \bibinfo{author}{Iesari, F.}
  (\bibinfo{year}{2022}).
\newblock \bibinfo{title}{{Advances in modelling X-ray absorption spectroscopy
  data using reverse Monte Carlo}}.
\newblock {\it \bibinfo{journal}{Phys. Chem. Chem. Phys.}\/},  {\it
  \bibinfo{volume}{24}\/}, \bibinfo{pages}{6988--7000}.
  \DOIprefix\doi{10.1039/D1CP05525A}.
\bibitem[{{Di Cicco} \& Trapananti(2005)}]{DiCicco2005}
\bibinfo{author}{{Di Cicco}, A.}, \& \bibinfo{author}{Trapananti, A.}
  (\bibinfo{year}{2005}).
\newblock \bibinfo{title}{{Reverse Monte Carlo refinement of molecular and
  condensed systems by X-ray absorption spectroscopy}}.
\newblock {\it \bibinfo{journal}{J. Phys.: Condens. Matter}\/},  {\it
  \bibinfo{volume}{17}\/}, \bibinfo{pages}{S135--S144}.
  \DOIprefix\doi{10.1088/0953-8984/17/5/014}.
\bibitem[{Dupuy et~al.(2021)Dupuy, Chiu, Shafer, Arenholz, Takamura \&
  Schoenung}]{DUPUY2021}
\bibinfo{author}{Dupuy, A.~D.}, \bibinfo{author}{Chiu, I.-T.},
  \bibinfo{author}{Shafer, P.}, \bibinfo{author}{Arenholz, E.},
  \bibinfo{author}{Takamura, Y.}, \& \bibinfo{author}{Schoenung, J.~M.}
  (\bibinfo{year}{2021}).
\newblock \bibinfo{title}{{Hidden transformations in entropy-stabilized
  oxides}}.
\newblock {\it \bibinfo{journal}{J. Eur. Ceram. Soc.}\/},  {\it
  \bibinfo{volume}{41}\/}, \bibinfo{pages}{6660--6669}.
  \DOIprefix\doi{10.1016/j.jeurceramsoc.2021.06.014}.
\bibitem[{Fan et~al.(2016)Fan, Wang, Wu, Liu \& Lu}]{Fan2016}
\bibinfo{author}{Fan, Z.}, \bibinfo{author}{Wang, H.}, \bibinfo{author}{Wu,
  Y.}, \bibinfo{author}{Liu, X.~J.}, \& \bibinfo{author}{Lu, Z.~P.}
  (\bibinfo{year}{2016}).
\newblock \bibinfo{title}{{Thermoelectric high-entropy alloys with low lattice
  thermal conductivity}}.
\newblock {\it \bibinfo{journal}{RSC Adv.}\/},  {\it \bibinfo{volume}{6}\/},
  \bibinfo{pages}{52164--52170}. \DOIprefix\doi{10.1039/C5RA28088E}.
\bibitem[{Fantin et~al.(2022)Fantin, Cakir, Kasatikov, Schumacher \&
  Manzoni}]{Fantin2022}
\bibinfo{author}{Fantin, A.}, \bibinfo{author}{Cakir, C.},
  \bibinfo{author}{Kasatikov, S.}, \bibinfo{author}{Schumacher, G.}, \&
  \bibinfo{author}{Manzoni, A.} (\bibinfo{year}{2022}).
\newblock \bibinfo{title}{{Effects of heat treatment on microstructure,
  hardness and local structure in a compositionally complex alloy}}.
\newblock {\it \bibinfo{journal}{Mater. Chem. Phys.}\/},  {\it
  \bibinfo{volume}{276}\/}, \bibinfo{pages}{125432}.
  \DOIprefix\doi{10.1016/j.matchemphys.2021.125432}.
\bibitem[{Fantin et~al.(2020)Fantin, Lepore, Manzoni, Kasatikov, Scherb,
  Huthwelker, d'Acapito \& Schumacher}]{Fantin2020}
\bibinfo{author}{Fantin, A.}, \bibinfo{author}{Lepore, G.~O.},
  \bibinfo{author}{Manzoni, A.~M.}, \bibinfo{author}{Kasatikov, S.},
  \bibinfo{author}{Scherb, T.}, \bibinfo{author}{Huthwelker, T.},
  \bibinfo{author}{d'Acapito, F.}, \& \bibinfo{author}{Schumacher, G.}
  (\bibinfo{year}{2020}).
\newblock \bibinfo{title}{{Short-range chemical order and local lattice
  distortion in a compositionally complex alloy}}.
\newblock {\it \bibinfo{journal}{Acta Mater.}\/},  {\it
  \bibinfo{volume}{193}\/}, \bibinfo{pages}{329--337}.
  \DOIprefix\doi{10.1016/j.actamat.2020.04.034}.
\bibitem[{Farges et~al.(2004)Farges, Lefrere, Rossano, Berthereau, Calas \&
  Jr.}]{Farges2004}
\bibinfo{author}{Farges, F.}, \bibinfo{author}{Lefrere, Y.},
  \bibinfo{author}{Rossano, S.}, \bibinfo{author}{Berthereau, A.},
  \bibinfo{author}{Calas, G.}, \& \bibinfo{author}{Jr., G. E.~B.}
  (\bibinfo{year}{2004}).
\newblock \bibinfo{title}{{The effect of redox state on the local structural
  environment of iron in silicate glasses: a combined XAFS spectroscopy,
  molecular dynamics, and bond valence study}}.
\newblock {\it \bibinfo{journal}{J. Non-Cryst. Solids}\/},  {\it
  \bibinfo{volume}{344}\/}, \bibinfo{pages}{176--188}.
  \DOIprefix\doi{10.1016/j.jnoncrysol.2004.07.050}.
\bibitem[{Ferlat et~al.(2005)Ferlat, Soetens, Miguel \& Bopp}]{Ferlat2005}
\bibinfo{author}{Ferlat, G.}, \bibinfo{author}{Soetens, J.-C.},
  \bibinfo{author}{Miguel, A.~S.}, \& \bibinfo{author}{Bopp, P.~A.}
  (\bibinfo{year}{2005}).
\newblock \bibinfo{title}{{Combining extended X-ray absorption fine structure
  with numerical simulations for disordered systems}}.
\newblock {\it \bibinfo{journal}{J. Phys.: Condens. Matter}\/},  {\it
  \bibinfo{volume}{17}\/}, \bibinfo{pages}{S145}.
  \DOIprefix\doi{10.1088/0953-8984/17/5/015}.
\bibitem[{Ferrari et~al.(2020)Ferrari, Dutta, Gubaev, Ikeda, Srinivasan,
  Grabowski \& Körmann}]{Ferrari2020}
\bibinfo{author}{Ferrari, A.}, \bibinfo{author}{Dutta, B.},
  \bibinfo{author}{Gubaev, K.}, \bibinfo{author}{Ikeda, Y.},
  \bibinfo{author}{Srinivasan, P.}, \bibinfo{author}{Grabowski, B.}, \&
  \bibinfo{author}{Körmann, F.} (\bibinfo{year}{2020}).
\newblock \bibinfo{title}{{Frontiers in atomistic simulations of high entropy
  alloys}}.
\newblock {\it \bibinfo{journal}{J. Appl. Phys.}\/},  {\it
  \bibinfo{volume}{128}\/}, \bibinfo{pages}{150901}.
  \DOIprefix\doi{10.1063/5.0025310}.
\bibitem[{Filipponi \& {Di Cicco}(1995)}]{Filipponi1995b}
\bibinfo{author}{Filipponi, A.}, \& \bibinfo{author}{{Di Cicco}, A.}
  (\bibinfo{year}{1995}).
\newblock \bibinfo{title}{{X-ray-absorption spectroscopy and n-body
  distribution functions in condensed matter. II. Data analysis and
  applications}}.
\newblock {\it \bibinfo{journal}{Phys. Rev. B}\/},  {\it
  \bibinfo{volume}{52}\/}, \bibinfo{pages}{15135--15149}.
  \DOIprefix\doi{10.1103/PhysRevB.52.15135}.
\bibitem[{Filipponi \& {Di Cicco}(2000)}]{GNXAS}
\bibinfo{author}{Filipponi, A.}, \& \bibinfo{author}{{Di Cicco}, A.}
  (\bibinfo{year}{2000}).
\newblock \bibinfo{title}{{GNXAS: a software package for advanced EXAFS
  multiple-scattering calculations and data-analysis}}.
\newblock {\it \bibinfo{journal}{TASK Q.}\/},  {\it \bibinfo{volume}{4}\/},
  \bibinfo{pages}{575--669}.
\bibitem[{Filipponi et~al.(1995)Filipponi, {Di Cicco} \&
  Natoli}]{Filipponi1995a}
\bibinfo{author}{Filipponi, A.}, \bibinfo{author}{{Di Cicco}, A.}, \&
  \bibinfo{author}{Natoli, C.~R.} (\bibinfo{year}{1995}).
\newblock \bibinfo{title}{{X-ray-absorption spectroscopy and n-body
  distribution functions in condensed matter. I. Theory}}.
\newblock {\it \bibinfo{journal}{Phys. Rev. B}\/},  {\it
  \bibinfo{volume}{52}\/}, \bibinfo{pages}{15122--15134}.
  \DOIprefix\doi{10.1103/PhysRevB.52.15122}.
\bibitem[{Fornasini et~al.(2017)Fornasini, Grisenti, Dapiaggi, Agostini \&
  Miyanaga}]{Fornasini2017}
\bibinfo{author}{Fornasini, P.}, \bibinfo{author}{Grisenti, R.},
  \bibinfo{author}{Dapiaggi, M.}, \bibinfo{author}{Agostini, G.}, \&
  \bibinfo{author}{Miyanaga, T.} (\bibinfo{year}{2017}).
\newblock \bibinfo{title}{{Nearest-neighbour distribution of distances in
  crystals from extended X-ray absorption fine structure}}.
\newblock {\it \bibinfo{journal}{J. Chem. Phys.}\/},  {\it
  \bibinfo{volume}{147}\/}, \bibinfo{pages}{044503}.
  \DOIprefix\doi{10.1063/1.4995435}.
\bibitem[{Fujikawa et~al.(2014)Fujikawa, Ariga, Takakusagi, Uehara, Ohba \&
  Asakura}]{Fujikawa2014}
\bibinfo{author}{Fujikawa, K.}, \bibinfo{author}{Ariga, H.},
  \bibinfo{author}{Takakusagi, S.}, \bibinfo{author}{Uehara, H.},
  \bibinfo{author}{Ohba, T.}, \& \bibinfo{author}{Asakura, K.}
  (\bibinfo{year}{2014}).
\newblock \bibinfo{title}{{Micro Reverse Monte Carlo approach to EXAFS
  analysis}}.
\newblock {\it \bibinfo{journal}{e-J. Surf. Sci. Nanotechnol.}\/},  {\it
  \bibinfo{volume}{12}\/}, \bibinfo{pages}{322--329}.
  \DOIprefix\doi{10.1380/ejssnt.2014.322}.
\bibitem[{Gaboardi et~al.(2022)Gaboardi, Monteverde, Saraga, Aquilanti, Feng,
  Fahrenholtz \& Hilmas}]{GABOARDI2022}
\bibinfo{author}{Gaboardi, M.}, \bibinfo{author}{Monteverde, F.},
  \bibinfo{author}{Saraga, F.}, \bibinfo{author}{Aquilanti, G.},
  \bibinfo{author}{Feng, L.}, \bibinfo{author}{Fahrenholtz, W.}, \&
  \bibinfo{author}{Hilmas, G.} (\bibinfo{year}{2022}).
\newblock \bibinfo{title}{{Local structure in high-entropy transition metal
  diborides}}.
\newblock {\it \bibinfo{journal}{Acta Mater.}\/},  {\it
  \bibinfo{volume}{239}\/}, \bibinfo{pages}{118294}.
  \DOIprefix\doi{10.1016/j.actamat.2022.118294}.
\bibitem[{Gale \& Rohl(2003)}]{GULP2003}
\bibinfo{author}{Gale, J.~D.}, \& \bibinfo{author}{Rohl, A.~L.}
  (\bibinfo{year}{2003}).
\newblock \bibinfo{title}{{The General Utility Lattice Program (GULP)}}.
\newblock {\it \bibinfo{journal}{Mol. Simul.}\/},  {\it
  \bibinfo{volume}{29}\/}, \bibinfo{pages}{291--341}.
  \DOIprefix\doi{10.1080/0892702031000104887}.
\bibitem[{Geiss et~al.(2022)Geiss, Schulte \& Winterer}]{Winterer2022}
\bibinfo{author}{Geiss, J.}, \bibinfo{author}{Schulte, J.}, \&
  \bibinfo{author}{Winterer, M.} (\bibinfo{year}{2022}).
\newblock \bibinfo{title}{{Flash evaporation of low volatility solid precursors
  by a scanning infrared laser}}.
\newblock {\it \bibinfo{journal}{J. Nanopart. Res.}\/},  {\it
  \bibinfo{volume}{24}\/}, \bibinfo{pages}{248}.
  \DOIprefix\doi{10.1007/s11051-022-05611-3}.
\bibitem[{George et~al.(2019)George, Raabe \& Ritchie}]{George2019}
\bibinfo{author}{George, E.}, \bibinfo{author}{Raabe, D.}, \&
  \bibinfo{author}{Ritchie, R.} (\bibinfo{year}{2019}).
\newblock \bibinfo{title}{{High-entropy alloys}}.
\newblock {\it \bibinfo{journal}{Nat. Rev. Mater.}\/},  {\it
  \bibinfo{volume}{4}\/}, \bibinfo{pages}{515–534}.
  \DOIprefix\doi{10.1038/s41578-019-0121-4}.
\bibitem[{Gereben et~al.(2007)Gereben, Jovari, Temleitner \&
  Pusztai}]{Gereben2007}
\bibinfo{author}{Gereben, O.}, \bibinfo{author}{Jovari, P.},
  \bibinfo{author}{Temleitner, L.}, \& \bibinfo{author}{Pusztai, L.}
  (\bibinfo{year}{2007}).
\newblock \bibinfo{title}{{A new version of the RMC++ Reverse Monte Carlo
  programme, aimed at investigating the structure of covalent glasses}}.
\newblock {\it \bibinfo{journal}{J. Optoelectron. Adv. M.}\/},  {\it
  \bibinfo{volume}{9}\/}, \bibinfo{pages}{3021--3027}.
\bibitem[{Gereben \& Pusztai(2012)}]{Gereben2012rmc_pot}
\bibinfo{author}{Gereben, O.}, \& \bibinfo{author}{Pusztai, L.}
  (\bibinfo{year}{2012}).
\newblock \bibinfo{title}{{RMC\_POT: a computer code for reverse Monte Carlo
  modeling the structure of disordered systems containing molecules of
  arbitrary complexity}}.
\newblock {\it \bibinfo{journal}{J. Comput. Chem.}\/},  {\it
  \bibinfo{volume}{33}\/}, \bibinfo{pages}{2285--2291}.
  \DOIprefix\doi{10.1002/jcc.23058}.
\bibitem[{Ghigna et~al.(2020)Ghigna, Airoldi, Fracchia, Callegari,
  Anselmi-Tamburini, D’Angelo, Pianta, Ruffo, Cibin, de~Souza \&
  Quartarone}]{Ghigna2020}
\bibinfo{author}{Ghigna, P.}, \bibinfo{author}{Airoldi, L.},
  \bibinfo{author}{Fracchia, M.}, \bibinfo{author}{Callegari, D.},
  \bibinfo{author}{Anselmi-Tamburini, U.}, \bibinfo{author}{D’Angelo, P.},
  \bibinfo{author}{Pianta, N.}, \bibinfo{author}{Ruffo, R.},
  \bibinfo{author}{Cibin, G.}, \bibinfo{author}{de~Souza, D.~O.}, \&
  \bibinfo{author}{Quartarone, E.} (\bibinfo{year}{2020}).
\newblock \bibinfo{title}{{Lithiation mechanism in high-entropy oxides as anode
  materials for Li-ion batteries: An operando XAS study}}.
\newblock {\it \bibinfo{journal}{ACS Appl. Mater. Interfaces}\/},  {\it
  \bibinfo{volume}{12}\/}, \bibinfo{pages}{50344--50354}.
  \DOIprefix\doi{10.1021/acsami.0c13161}.
\bibitem[{Gornakova et~al.(2023)Gornakova, Straumal, Kuzmin, Tyurin,
  Chernyaeva, Druzhinin, Afonikova \& Davdian}]{Gornakova2023}
\bibinfo{author}{Gornakova, A.}, \bibinfo{author}{Straumal, B.},
  \bibinfo{author}{Kuzmin, A.}, \bibinfo{author}{Tyurin, A.},
  \bibinfo{author}{Chernyaeva, E.}, \bibinfo{author}{Druzhinin, A.},
  \bibinfo{author}{Afonikova, N.}, \& \bibinfo{author}{Davdian, G.}
  (\bibinfo{year}{2023}).
\newblock \bibinfo{title}{{Influence of heat treatment and high-pressure
  torsion on phase transformations in TiZrHfMoCr high-entropy alloy}}.
\newblock {\it \bibinfo{journal}{Metals}\/},  {\it \bibinfo{volume}{13}\/},
  \bibinfo{pages}{1030}. \DOIprefix\doi{10.3390/met13061030}.
\bibitem[{Gowthaman(2023)}]{Gowthaman2023}
\bibinfo{author}{Gowthaman, S.} (\bibinfo{year}{2023}).
\newblock \bibinfo{title}{{A review on mechanical and material characterisation
  through molecular dynamics using large-scale atomic/molecular massively
  parallel simulator (LAMMPS)}}.
\newblock {\it \bibinfo{journal}{Funct. Compos. Struct.}\/},  {\it
  \bibinfo{volume}{5}\/}, \bibinfo{pages}{012005}.
  \DOIprefix\doi{10.1088/2631-6331/acc3d5}.
\bibitem[{Gurman \& McGreevy(1990)}]{Gurman1990rmc}
\bibinfo{author}{Gurman, S.~J.}, \& \bibinfo{author}{McGreevy, R.~L.}
  (\bibinfo{year}{1990}).
\newblock \bibinfo{title}{{Reverse Monte Carlo simulation for the analysis of
  EXAFS data}}.
\newblock {\it \bibinfo{journal}{J. Phys.: Condens. Matter}\/},  {\it
  \bibinfo{volume}{2}\/}, \bibinfo{pages}{9463--9473}.
  \DOIprefix\doi{10.1088/0953-8984/2/48/001}.
\bibitem[{Han et~al.(2022)Han, Wang, Zhong, Han, Wang, Seifitokaldani, Yu, Liu,
  Sun, Vomiero \& Liang}]{Han2022}
\bibinfo{author}{Han, M.}, \bibinfo{author}{Wang, C.}, \bibinfo{author}{Zhong,
  J.}, \bibinfo{author}{Han, J.}, \bibinfo{author}{Wang, N.},
  \bibinfo{author}{Seifitokaldani, A.}, \bibinfo{author}{Yu, Y.},
  \bibinfo{author}{Liu, Y.}, \bibinfo{author}{Sun, X.},
  \bibinfo{author}{Vomiero, A.}, \& \bibinfo{author}{Liang, H.}
  (\bibinfo{year}{2022}).
\newblock \bibinfo{title}{{Promoted self-construction of $\beta$-NiOOH in
  amorphous high entropy electrocatalysts for the oxygen evolution reaction}}.
\newblock {\it \bibinfo{journal}{Appl. Catal. B}\/},  {\it
  \bibinfo{volume}{301}\/}, \bibinfo{pages}{120764}.
  \DOIprefix\doi{10.1016/j.apcatb.2021.120764}.
\bibitem[{Hao et~al.(2022)Hao, Zhuang, Cao, Gao, Wang, Lai, Lu, Ma, Dong, Liu,
  Du \& Zhu}]{Hao2022}
\bibinfo{author}{Hao, J.}, \bibinfo{author}{Zhuang, Z.}, \bibinfo{author}{Cao,
  K.}, \bibinfo{author}{Gao, G.}, \bibinfo{author}{Wang, C.},
  \bibinfo{author}{Lai, F.}, \bibinfo{author}{Lu, S.}, \bibinfo{author}{Ma,
  P.}, \bibinfo{author}{Dong, W.}, \bibinfo{author}{Liu, T.},
  \bibinfo{author}{Du, M.}, \& \bibinfo{author}{Zhu, H.}
  (\bibinfo{year}{2022}).
\newblock \bibinfo{title}{{Unraveling the electronegativity-dominated
  intermediate adsorption on high-entropy alloy electrocatalysts}}.
\newblock {\it \bibinfo{journal}{Nat. Commun.}\/},  {\it
  \bibinfo{volume}{13}\/}, \bibinfo{pages}{2662}.
  \DOIprefix\doi{10.1038/s41467-022-30379-4}.
\bibitem[{Harrington et~al.(2019)Harrington, Gild, Sarker, Toher, Rost, Dippo,
  McElfresh, Kaufmann, Marin, Borowski, Hopkins, Luo, Curtarolo, Brenner \&
  Vecchio}]{Harrington2019}
\bibinfo{author}{Harrington, T.~J.}, \bibinfo{author}{Gild, J.},
  \bibinfo{author}{Sarker, P.}, \bibinfo{author}{Toher, C.},
  \bibinfo{author}{Rost, C.~M.}, \bibinfo{author}{Dippo, O.~F.},
  \bibinfo{author}{McElfresh, C.}, \bibinfo{author}{Kaufmann, K.},
  \bibinfo{author}{Marin, E.}, \bibinfo{author}{Borowski, L.},
  \bibinfo{author}{Hopkins, P.~E.}, \bibinfo{author}{Luo, J.},
  \bibinfo{author}{Curtarolo, S.}, \bibinfo{author}{Brenner, D.~W.}, \&
  \bibinfo{author}{Vecchio, K.~S.} (\bibinfo{year}{2019}).
\newblock \bibinfo{title}{{Phase stability and mechanical properties of novel
  high entropy transition metal carbides}}.
\newblock {\it \bibinfo{journal}{Acta Mater.}\/},  {\it
  \bibinfo{volume}{166}\/}, \bibinfo{pages}{271--280}.
  \DOIprefix\doi{10.1016/j.actamat.2018.12.054}.
\bibitem[{He et~al.(2021)He, Wang, Lan, Ruff, Sun, Naeem, Liu \& Wang}]{He2021}
\bibinfo{author}{He, H.}, \bibinfo{author}{Wang, B.}, \bibinfo{author}{Lan,
  S.}, \bibinfo{author}{Ruff, J. P.~C.}, \bibinfo{author}{Sun, C.},
  \bibinfo{author}{Naeem, M.}, \bibinfo{author}{Liu, C.-T.}, \&
  \bibinfo{author}{Wang, X.-L.} (\bibinfo{year}{2021}).
\newblock \bibinfo{title}{{Anomalous X-ray scattering and extended X-ray
  absorption fine structure study of the local structure of CrFeCoNiMo$_x$ (x =
  0.11, 0.18, and 0.23) high-entropy alloys}}.
\newblock {\it \bibinfo{journal}{JOM}\/},  {\it \bibinfo{volume}{73}\/},
  \bibinfo{pages}{3285--3290}. \DOIprefix\doi{10.1007/s11837-021-04871-z}.
\bibitem[{Hu et~al.(2021)Hu, Cao, Wang, Liu, Zhang, Cao, Lu \& Cheng}]{Hu2021}
\bibinfo{author}{Hu, J.}, \bibinfo{author}{Cao, L.}, \bibinfo{author}{Wang,
  Z.}, \bibinfo{author}{Liu, J.}, \bibinfo{author}{Zhang, J.},
  \bibinfo{author}{Cao, Y.}, \bibinfo{author}{Lu, Z.}, \&
  \bibinfo{author}{Cheng, H.} (\bibinfo{year}{2021}).
\newblock \bibinfo{title}{{Hollow high-entropy metal organic framework derived
  nanocomposite as efficient electrocatalyst for oxygen reduction reaction}}.
\newblock {\it \bibinfo{journal}{Compos. Commun.}\/},  {\it
  \bibinfo{volume}{27}\/}, \bibinfo{pages}{100866}.
  \DOIprefix\doi{10.1016/j.coco.2021.100866}.
\bibitem[{Huang et~al.(2021)Huang, Yeh \& Yang}]{Huang2021}
\bibinfo{author}{Huang, Y.}, \bibinfo{author}{Yeh, J.-W.}, \&
  \bibinfo{author}{Yang, A. C.-M.} (\bibinfo{year}{2021}).
\newblock \bibinfo{title}{{``High-entropy polymers'': A new route of polymer
  mixing with suppressed phase separation}}.
\newblock {\it \bibinfo{journal}{Materialia}\/},  {\it \bibinfo{volume}{15}\/},
  \bibinfo{pages}{100978}. \DOIprefix\doi{10.1016/j.mtla.2020.100978}.
\bibitem[{Jacobson et~al.(2023)Jacobson, Huang, Titus, Smaha, Papac, Lee,
  Zakutayev \& Brennecka}]{Jacobson2023}
\bibinfo{author}{Jacobson, V.}, \bibinfo{author}{Huang, J.},
  \bibinfo{author}{Titus, C.}, \bibinfo{author}{Smaha, R.},
  \bibinfo{author}{Papac, M.}, \bibinfo{author}{Lee, S.},
  \bibinfo{author}{Zakutayev, A.}, \& \bibinfo{author}{Brennecka, G.}
  (\bibinfo{year}{2023}).
\newblock \bibinfo{title}{{The role of Co valence in charge transport in the
  entropy-stabilized oxide
  (Mg$_{0.2}$Co$_{0.2}$Ni$_{0.2}$Cu$_{0.2}$Zn$_{0.2}$)O}}.
\newblock {\it \bibinfo{journal}{J. Am. Ceram. Soc.}\/},  {\it
  \bibinfo{volume}{106}\/}, \bibinfo{pages}{1531--1539}.
  \DOIprefix\doi{10.1111/jace.18820}.
\bibitem[{Jeong et~al.(2003)Jeong, Heffner, Graf \& Billinge}]{Jeong2003}
\bibinfo{author}{Jeong, I.-K.}, \bibinfo{author}{Heffner, R.~H.},
  \bibinfo{author}{Graf, M.~J.}, \& \bibinfo{author}{Billinge, S. J.~L.}
  (\bibinfo{year}{2003}).
\newblock \bibinfo{title}{{Lattice dynamics and correlated atomic motion from
  the atomic pair distribution function}}.
\newblock {\it \bibinfo{journal}{Phys. Rev. B}\/},  {\it
  \bibinfo{volume}{67}\/}, \bibinfo{pages}{104301}.
  \DOIprefix\doi{10.1103/PhysRevB.67.104301}.
\bibitem[{Jonane et~al.(2018)Jonane, Anspoks \& Kuzmin}]{Jonane2018}
\bibinfo{author}{Jonane, I.}, \bibinfo{author}{Anspoks, A.}, \&
  \bibinfo{author}{Kuzmin, A.} (\bibinfo{year}{2018}).
\newblock \bibinfo{title}{{Advanced approach to the local structure
  reconstruction and theory validation on the example of the W L$_3$-edge
  extended X-ray absorption fine structure of tungsten}}.
\newblock {\it \bibinfo{journal}{Model. Simul. Mater. Sci. Eng.}\/},  {\it
  \bibinfo{volume}{26}\/}, \bibinfo{pages}{025004}.
  \DOIprefix\doi{10.1088/1361-651X/aa9bab}.
\bibitem[{Keen \& McGreevy(1990)}]{Keen1990}
\bibinfo{author}{Keen, D.}, \& \bibinfo{author}{McGreevy, R.}
  (\bibinfo{year}{1990}).
\newblock \bibinfo{title}{{Structural modelling of glasses using reverse Monte
  Carlo simulation}}.
\newblock {\it \bibinfo{journal}{Nature}\/},  {\it \bibinfo{volume}{344}\/},
  \bibinfo{pages}{423–425}. \DOIprefix\doi{10.1038/344423a0}.
\bibitem[{Keski-Rahkonen \& Krause(1974)}]{KESKI1974}
\bibinfo{author}{Keski-Rahkonen, O.}, \& \bibinfo{author}{Krause, M.~O.}
  (\bibinfo{year}{1974}).
\newblock \bibinfo{title}{Total and partial atomic-level widths}.
\newblock {\it \bibinfo{journal}{Atomic Data and Nuclear Data Tables}\/},  {\it
  \bibinfo{volume}{14}\/}, \bibinfo{pages}{139--146}.
  \DOIprefix\doi{10.1016/S0092-640X(74)80020-3}.
\bibitem[{Kim et~al.(2022)Kim, Oh, Choi, Jang, Song, Kim, Yoo \& Cho}]{Kim2022}
\bibinfo{author}{Kim, M.}, \bibinfo{author}{Oh, I.}, \bibinfo{author}{Choi,
  H.}, \bibinfo{author}{Jang, W.}, \bibinfo{author}{Song, J.},
  \bibinfo{author}{Kim, C.~S.}, \bibinfo{author}{Yoo, J.-W.}, \&
  \bibinfo{author}{Cho, S.} (\bibinfo{year}{2022}).
\newblock \bibinfo{title}{{A solution-based route to compositionally complex
  metal oxide structures using high-entropy layered double hydroxides}}.
\newblock {\it \bibinfo{journal}{Cell Rep. Phys. Sci.}\/},  {\it
  \bibinfo{volume}{3}\/}, \bibinfo{pages}{100702}.
  \DOIprefix\doi{10.1016/j.xcrp.2021.100702}.
\bibitem[{Krayzman \& Levin(2010)}]{Krayzman2010}
\bibinfo{author}{Krayzman, V.}, \& \bibinfo{author}{Levin, I.}
  (\bibinfo{year}{2010}).
\newblock \bibinfo{title}{{Reverse Monte Carlo refinements of local displacive
  order in perovskites: AgNbO$_3$ case study}}.
\newblock {\it \bibinfo{journal}{J. Phys.: Condens. Matter}\/},  {\it
  \bibinfo{volume}{22}\/}, \bibinfo{pages}{404201}.
  \DOIprefix\doi{10.1088/0953-8984/22/40/404201}.
\bibitem[{Krayzman et~al.(2009)Krayzman, Levin, Woicik, Proffen, Vanderah \&
  Tucker}]{Krayzman2009}
\bibinfo{author}{Krayzman, V.}, \bibinfo{author}{Levin, I.},
  \bibinfo{author}{Woicik, J.~C.}, \bibinfo{author}{Proffen, T.},
  \bibinfo{author}{Vanderah, T.~A.}, \& \bibinfo{author}{Tucker, M.~G.}
  (\bibinfo{year}{2009}).
\newblock \bibinfo{title}{{A combined fit of total scattering and extended
  X-ray absorption fine structure data for local-structure determination in
  crystalline materials}}.
\newblock {\it \bibinfo{journal}{J. Appl. Crystallogr.}\/},  {\it
  \bibinfo{volume}{42}\/}, \bibinfo{pages}{867--877}.
  \DOIprefix\doi{10.1107/S0021889809023541}.
\bibitem[{Kuzmin et~al.(2016)Kuzmin, Anspoks, Kalinko \&
  Timoshenko}]{Kuzmin2016}
\bibinfo{author}{Kuzmin, A.}, \bibinfo{author}{Anspoks, A.},
  \bibinfo{author}{Kalinko, A.}, \& \bibinfo{author}{Timoshenko, J.}
  (\bibinfo{year}{2016}).
\newblock \bibinfo{title}{{The use of X-ray absorption spectra for validation
  of classical force-field models}}.
\newblock {\it \bibinfo{journal}{Z. Phys. Chem.}\/},  {\it
  \bibinfo{volume}{230}\/}, \bibinfo{pages}{537--549}.
  \DOIprefix\doi{10.1515/zpch-2015-0664}.
\bibitem[{Kuzmin \& Chaboy(2014)}]{Kuzmin2014exafs}
\bibinfo{author}{Kuzmin, A.}, \& \bibinfo{author}{Chaboy, J.}
  (\bibinfo{year}{2014}).
\newblock \bibinfo{title}{{EXAFS and XANES analysis of oxides at the
  nanoscale}}.
\newblock {\it \bibinfo{journal}{IUCrJ}\/},  {\it \bibinfo{volume}{1}\/},
  \bibinfo{pages}{571--589}. \DOIprefix\doi{10.1107/S205225251402110}.
\bibitem[{Kuzmin \& Evarestov(2009)}]{Kuzmin2009}
\bibinfo{author}{Kuzmin, A.}, \& \bibinfo{author}{Evarestov, R.~A.}
  (\bibinfo{year}{2009}).
\newblock \bibinfo{title}{{Quantum mechanics-molecular dynamics approach to the
  interpretation of X-ray absorption spectra}}.
\newblock {\it \bibinfo{journal}{J. Phys.: Condens. Matter}\/},  {\it
  \bibinfo{volume}{21}\/}, \bibinfo{pages}{055401}.
  \DOIprefix\doi{10.1088/0953-8984/21/5/055401}.
\bibitem[{Kuzmin \& Purans(1993)}]{Kuzmin1993}
\bibinfo{author}{Kuzmin, A.}, \& \bibinfo{author}{Purans, J.}
  (\bibinfo{year}{1993}).
\newblock \bibinfo{title}{{The influence of the focusing effect on the X-ray
  absorption fine structure above all the tungsten L edges in
  non-stoichiometric tungsten oxides}}.
\newblock {\it \bibinfo{journal}{J. Phys.: Condens. Matter}\/},  {\it
  \bibinfo{volume}{5}\/}, \bibinfo{pages}{9423--9430}.
  \DOIprefix\doi{10.1088/0953-8984/5/50/023}.
\bibitem[{Kuzmin et~al.(2020)Kuzmin, Timoshenko, Kalinko, Jonane \&
  Anspoks}]{KUZMIN2020rev}
\bibinfo{author}{Kuzmin, A.}, \bibinfo{author}{Timoshenko, J.},
  \bibinfo{author}{Kalinko, A.}, \bibinfo{author}{Jonane, I.}, \&
  \bibinfo{author}{Anspoks, A.} (\bibinfo{year}{2020}).
\newblock \bibinfo{title}{{Treatment of disorder effects in X-ray absorption
  spectra beyond the conventional approach}}.
\newblock {\it \bibinfo{journal}{Radiation Physics and Chemistry}\/},  {\it
  \bibinfo{volume}{175}\/}, \bibinfo{pages}{108112}.
  \DOIprefix\doi{10.1016/j.radphyschem.2018.12.032}.
\bibitem[{Lee et~al.(1981)Lee, Citrin, Eisenberger \& Kincaid}]{Lee1981}
\bibinfo{author}{Lee, P.~A.}, \bibinfo{author}{Citrin, P.~H.},
  \bibinfo{author}{Eisenberger, P.}, \& \bibinfo{author}{Kincaid, B.~M.}
  (\bibinfo{year}{1981}).
\newblock \bibinfo{title}{{Extended X-ray absorption fine structure - its
  strengths and limitations as a structural tool}}.
\newblock {\it \bibinfo{journal}{Rev. Mod. Phys.}\/},  {\it
  \bibinfo{volume}{53}\/}, \bibinfo{pages}{769--806}.
  \DOIprefix\doi{10.1103/RevModPhys.53.769}.
\bibitem[{Lee \& Pendry(1975)}]{Lee1975}
\bibinfo{author}{Lee, P.~A.}, \& \bibinfo{author}{Pendry, J.~B.}
  (\bibinfo{year}{1975}).
\newblock \bibinfo{title}{{Theory of the extended X-ray absorption fine
  structure}}.
\newblock {\it \bibinfo{journal}{Phys. Rev. B}\/},  {\it
  \bibinfo{volume}{11}\/}, \bibinfo{pages}{2795--2811}.
  \DOIprefix\doi{10.1103/PhysRevB.11.2795}.
\bibitem[{Leetmaa et~al.(2010)Leetmaa, Wikfeldt \&
  Pettersson}]{Leetmaa2010specswap}
\bibinfo{author}{Leetmaa, M.}, \bibinfo{author}{Wikfeldt, K.~T.}, \&
  \bibinfo{author}{Pettersson, L. G.~M.} (\bibinfo{year}{2010}).
\newblock \bibinfo{title}{{SpecSwap-RMC: a novel reverse Monte Carlo approach
  using a discrete set of local configurations and pre-computed properties}}.
\newblock {\it \bibinfo{journal}{J. Phys.: Condens. Matter}\/},  {\it
  \bibinfo{volume}{22}\/}, \bibinfo{pages}{135001}.
  \DOIprefix\doi{10.1088/0953-8984/22/13/135001}.
\bibitem[{Lei et~al.(2019)Lei, Liu, Wang, Wu, Jiang \& Lu}]{Lei2019}
\bibinfo{author}{Lei, Z.}, \bibinfo{author}{Liu, X.}, \bibinfo{author}{Wang,
  H.}, \bibinfo{author}{Wu, Y.}, \bibinfo{author}{Jiang, S.}, \&
  \bibinfo{author}{Lu, Z.} (\bibinfo{year}{2019}).
\newblock \bibinfo{title}{{Development of advanced materials via entropy
  engineering}}.
\newblock {\it \bibinfo{journal}{Scr. Mater.}\/},  {\it
  \bibinfo{volume}{165}\/}, \bibinfo{pages}{164--169}.
  \DOIprefix\doi{10.1016/j.scriptamat.2019.02.015}.
\bibitem[{Levin et~al.(2014)Levin, Krayzman \& Woicik}]{Levin2014}
\bibinfo{author}{Levin, I.}, \bibinfo{author}{Krayzman, V.}, \&
  \bibinfo{author}{Woicik, J.~C.} (\bibinfo{year}{2014}).
\newblock \bibinfo{title}{{Local structure in perovskite BaSrTiO$_3$: Reverse
  Monte Carlo refinements from multiple measurement techniques}}.
\newblock {\it \bibinfo{journal}{Phys. Rev. B}\/},  {\it
  \bibinfo{volume}{89}\/}, \bibinfo{pages}{024106}.
  \DOIprefix\doi{10.1103/PhysRevB.89.024106}.
\bibitem[{Liu et~al.(2023)Liu, Zhang \& Pei}]{Liu2023}
\bibinfo{author}{Liu, X.}, \bibinfo{author}{Zhang, J.}, \&
  \bibinfo{author}{Pei, Z.} (\bibinfo{year}{2023}).
\newblock \bibinfo{title}{{Machine learning for high-entropy alloys: Progress,
  challenges and opportunities}}.
\newblock {\it \bibinfo{journal}{Prog. Mater. Sci.}\/},  {\it
  \bibinfo{volume}{131}\/}, \bibinfo{pages}{101018}.
  \DOIprefix\doi{10.1016/j.pmatsci.2022.101018}.
\bibitem[{Luo et~al.(2022)Luo, Patra, Chuang, Nguyen, Ting, Li, Pao \&
  Chang}]{Luo2022}
\bibinfo{author}{Luo, X.-F.}, \bibinfo{author}{Patra, J.},
  \bibinfo{author}{Chuang, W.-T.}, \bibinfo{author}{Nguyen, T.~X.},
  \bibinfo{author}{Ting, J.-M.}, \bibinfo{author}{Li, J.},
  \bibinfo{author}{Pao, C.-W.}, \& \bibinfo{author}{Chang, J.-K.}
  (\bibinfo{year}{2022}).
\newblock \bibinfo{title}{{Charge–discharge mechanism of high-entropy Co-free
  spinel oxide toward Li$^+$ storage examined using operando quick-scanning
  X-ray absorption spectroscopy}}.
\newblock {\it \bibinfo{journal}{Adv. Sci.}\/},  {\it \bibinfo{volume}{9}\/},
  \bibinfo{pages}{2201219}. \DOIprefix\doi{10.1002/advs.202201219}.
\bibitem[{Ma et~al.(2021)Ma, Ma, Dreyer, Wang, Wang, Goonetilleke, Omar,
  Mikhailova, Hahn, Breitung \& Brezesinski}]{Ma2021_2101342}
\bibinfo{author}{Ma, Y.}, \bibinfo{author}{Ma, Y.}, \bibinfo{author}{Dreyer,
  S.~L.}, \bibinfo{author}{Wang, Q.}, \bibinfo{author}{Wang, K.},
  \bibinfo{author}{Goonetilleke, D.}, \bibinfo{author}{Omar, A.},
  \bibinfo{author}{Mikhailova, D.}, \bibinfo{author}{Hahn, H.},
  \bibinfo{author}{Breitung, B.}, \& \bibinfo{author}{Brezesinski, T.}
  (\bibinfo{year}{2021}).
\newblock \bibinfo{title}{{High-entropy metal-organic frameworks for highly
  reversible sodium storage}}.
\newblock {\it \bibinfo{journal}{Adv. Mater.}\/},  {\it
  \bibinfo{volume}{33}\/}, \bibinfo{pages}{2101342}.
  \DOIprefix\doi{10.1002/adma.202101342}.
\bibitem[{Markland \& Ceriotti(2018)}]{Markland2018}
\bibinfo{author}{Markland, T.}, \& \bibinfo{author}{Ceriotti, M.}
  (\bibinfo{year}{2018}).
\newblock \bibinfo{title}{{Nuclear quantum effects enter the mainstream}}.
\newblock {\it \bibinfo{journal}{Nat. Rev. Chem.}\/},  {\it
  \bibinfo{volume}{2}\/}, \bibinfo{pages}{0109}.
  \DOIprefix\doi{10.1038/s41570-017-0109}.
\bibitem[{Marx \& Parrinello(1996)}]{Marx1996}
\bibinfo{author}{Marx, D.}, \& \bibinfo{author}{Parrinello, M.}
  (\bibinfo{year}{1996}).
\newblock \bibinfo{title}{{Ab initio path integral molecular dynamics: basic
  ideas}}.
\newblock {\it \bibinfo{journal}{J. Chem. Phys.}\/},  {\it
  \bibinfo{volume}{104}\/}, \bibinfo{pages}{4077}.
  \DOIprefix\doi{10.1063/1.471221}.
\bibitem[{Maulik et~al.(2017)Maulik, Patra, Bhattacharyya, Jha \&
  Kumar}]{Maulik2017}
\bibinfo{author}{Maulik, O.}, \bibinfo{author}{Patra, N.},
  \bibinfo{author}{Bhattacharyya, D.}, \bibinfo{author}{Jha, S.}, \&
  \bibinfo{author}{Kumar, V.} (\bibinfo{year}{2017}).
\newblock \bibinfo{title}{{Local atomic structure investigation of
  AlFeCuCrMg$_x$ (0.5, 1, 1.7) high entropy alloys: X-ray absorption
  spectroscopy study}}.
\newblock {\it \bibinfo{journal}{Solid State Commun.}\/},  {\it
  \bibinfo{volume}{252}\/}, \bibinfo{pages}{73--77}.
  \DOIprefix\doi{10.1016/j.ssc.2017.01.018}.
\bibitem[{McGreevy(2001)}]{McGreevy2001}
\bibinfo{author}{McGreevy, R.~L.} (\bibinfo{year}{2001}).
\newblock \bibinfo{title}{{Reverse Monte Carlo modelling}}.
\newblock {\it \bibinfo{journal}{J. Phys.: Condens. Matter}\/},  {\it
  \bibinfo{volume}{13}\/}, \bibinfo{pages}{R877--R913}.
  \DOIprefix\doi{10.1088/0953-8984/13/46/201}.
\bibitem[{McGreevy \& Pusztai(1988)}]{McGreevy1988}
\bibinfo{author}{McGreevy, R.~L.}, \& \bibinfo{author}{Pusztai, L.}
  (\bibinfo{year}{1988}).
\newblock \bibinfo{title}{{Reverse Monte Carlo simulation: A new technique for
  the determination of disordered structures}}.
\newblock {\it \bibinfo{journal}{Mol. Simul.}\/},  {\it \bibinfo{volume}{1}\/},
  \bibinfo{pages}{359--367}. \DOIprefix\doi{10.1080/08927028808080958}.
\bibitem[{Merkling et~al.(2001)Merkling, Mu\~noz P\'aez, Pappalardo \&
  S\'anchez~Marcos}]{Merkling2001}
\bibinfo{author}{Merkling, P.~J.}, \bibinfo{author}{Mu\~noz P\'aez, A.},
  \bibinfo{author}{Pappalardo, R.~R.}, \& \bibinfo{author}{S\'anchez~Marcos,
  E.} (\bibinfo{year}{2001}).
\newblock \bibinfo{title}{{Combination of XANES spectroscopy and molecular
  dynamics to probe the local structure in disordered systems}}.
\newblock {\it \bibinfo{journal}{Phys. Rev. B}\/},  {\it
  \bibinfo{volume}{64}\/}, \bibinfo{pages}{092201}.
  \DOIprefix\doi{10.1103/PhysRevB.64.092201}.
\bibitem[{Metropolis et~al.(1953)Metropolis, Rosenbluth, Rosenbluth, Teller \&
  Teller}]{Metropolis1953}
\bibinfo{author}{Metropolis, N.}, \bibinfo{author}{Rosenbluth, A.~W.},
  \bibinfo{author}{Rosenbluth, M.~N.}, \bibinfo{author}{Teller, A.~H.}, \&
  \bibinfo{author}{Teller, E.} (\bibinfo{year}{1953}).
\newblock \bibinfo{title}{{Equation of state calculations by fast computing
  machines}}.
\newblock {\it \bibinfo{journal}{J. Chem. Phys.}\/},  {\it
  \bibinfo{volume}{21}\/}, \bibinfo{pages}{1087--1092}.
  \DOIprefix\doi{10.1063/1.1699114}.
\bibitem[{Miracle \& Senkov(2017)}]{Miracle2017}
\bibinfo{author}{Miracle, D.}, \& \bibinfo{author}{Senkov, O.}
  (\bibinfo{year}{2017}).
\newblock \bibinfo{title}{A critical review of high entropy alloys and related
  concepts}.
\newblock {\it \bibinfo{journal}{Acta Mater.}\/},  {\it
  \bibinfo{volume}{122}\/}, \bibinfo{pages}{448--511}.
  \DOIprefix\doi{10.1016/j.actamat.2016.08.081}.
\bibitem[{Molenda et~al.(2023)Molenda, Milewska, Zaj{\k{a}}c, Walczak, Wolczko,
  Komenda \& Tobola}]{Molenda2023}
\bibinfo{author}{Molenda, J.}, \bibinfo{author}{Milewska, A.},
  \bibinfo{author}{Zaj{\k{a}}c, W.}, \bibinfo{author}{Walczak, K.},
  \bibinfo{author}{Wolczko, M.}, \bibinfo{author}{Komenda, A.}, \&
  \bibinfo{author}{Tobola, J.} (\bibinfo{year}{2023}).
\newblock \bibinfo{title}{{Impact of O3/P3 phase transition on the performance
  of the Na$_x$Ti$_{1/6}$Mn$_{1/6}$Fe1/6Co$_{1/6}$Ni$_{1/6}$Cu$_{1/6}$O$_2$
  cathode material for Na-ion batteries}}.
\newblock {\it \bibinfo{journal}{J. Mater. Chem. A}\/},  {\it
  \bibinfo{volume}{11}\/}, \bibinfo{pages}{4248--4260}.
  \DOIprefix\doi{10.1039/D2TA08431G}.
\bibitem[{Morris et~al.(2021)Morris, Yao, Finfrock, Huang, Shahbazian-Yassar,
  Hu \& Zhang}]{MORRIS2021}
\bibinfo{author}{Morris, D.}, \bibinfo{author}{Yao, Y.},
  \bibinfo{author}{Finfrock, Y.~Z.}, \bibinfo{author}{Huang, Z.},
  \bibinfo{author}{Shahbazian-Yassar, R.}, \bibinfo{author}{Hu, L.}, \&
  \bibinfo{author}{Zhang, P.} (\bibinfo{year}{2021}).
\newblock \bibinfo{title}{{Composition-dependent structure and properties of 5-
  and 15-element high-entropy alloy nanoparticles}}.
\newblock {\it \bibinfo{journal}{Cell Rep. Phys. Sci.}\/},  {\it
  \bibinfo{volume}{2}\/}, \bibinfo{pages}{100641}.
  \DOIprefix\doi{10.1016/j.xcrp.2021.100641}.
\bibitem[{M\"{u}ller et~al.(1982)M\"{u}ller, Jepsen \& Wilkins}]{Muller1982}
\bibinfo{author}{M\"{u}ller, J.}, \bibinfo{author}{Jepsen, O.}, \&
  \bibinfo{author}{Wilkins, J.} (\bibinfo{year}{1982}).
\newblock \bibinfo{title}{{X-ray absorption spectra: K-edges of 3d transition
  metals, L-edges of 3d and 4d metals, and M-edges of palladium}}.
\newblock {\it \bibinfo{journal}{Solid State Communications}\/},  {\it
  \bibinfo{volume}{42}\/}, \bibinfo{pages}{365--368}.
  \DOIprefix\doi{10.1016/0038-1098(82)90154-5}.
\bibitem[{Muller et~al.(1978)Muller, Jepsen, Andersen \& Wilkins}]{Muller1978}
\bibinfo{author}{Muller, J.~E.}, \bibinfo{author}{Jepsen, O.},
  \bibinfo{author}{Andersen, O.~K.}, \& \bibinfo{author}{Wilkins, J.~W.}
  (\bibinfo{year}{1978}).
\newblock \bibinfo{title}{{Systematic Structure in the K-Edge Photoabsorption
  Spectra of the 4d Transition Metals: Theory}}.
\newblock {\it \bibinfo{journal}{Phys. Rev. Lett.}\/},  {\it
  \bibinfo{volume}{40}\/}, \bibinfo{pages}{720--722}.
  \DOIprefix\doi{10.1103/PhysRevLett.40.720}.
\bibitem[{Natoli et~al.(1990)Natoli, Benfatto, Brouder, L\'opez \&
  Foulis}]{Natoli1990}
\bibinfo{author}{Natoli, C.~R.}, \bibinfo{author}{Benfatto, M.},
  \bibinfo{author}{Brouder, C.}, \bibinfo{author}{L\'opez, M. F.~R.}, \&
  \bibinfo{author}{Foulis, D.~L.} (\bibinfo{year}{1990}).
\newblock \bibinfo{title}{{Multichannel multiple-scattering theory with general
  potentials}}.
\newblock {\it \bibinfo{journal}{Phys. Rev. B}\/},  {\it
  \bibinfo{volume}{42}\/}, \bibinfo{pages}{1944--1968}.
  \DOIprefix\doi{10.1103/PhysRevB.42.1944}.
\bibitem[{Nemani et~al.(2023)Nemani, Torkamanzadeh, Wyatt, Presser \&
  Anasori}]{Nemani2023}
\bibinfo{author}{Nemani, S.}, \bibinfo{author}{Torkamanzadeh, M.},
  \bibinfo{author}{Wyatt, B.}, \bibinfo{author}{Presser, V.}, \&
  \bibinfo{author}{Anasori, B.} (\bibinfo{year}{2023}).
\newblock \bibinfo{title}{{Functional two-dimensional high-entropy materials}}.
\newblock {\it \bibinfo{journal}{Commun. Mater.}\/},  {\it
  \bibinfo{volume}{4}\/}, \bibinfo{pages}{16}.
  \DOIprefix\doi{10.1038/s43246-023-00341-y}.
\bibitem[{Okamoto(2004)}]{Okamoto2004}
\bibinfo{author}{Okamoto, Y.} (\bibinfo{year}{2004}).
\newblock \bibinfo{title}{{XAFS simulation of highly disordered materials}}.
\newblock {\it \bibinfo{journal}{Nucl. Instrum. Methods Phys. Res. A}\/},  {\it
  \bibinfo{volume}{526}\/}, \bibinfo{pages}{572--583}.
  \DOIprefix\doi{10.1016/j.nima.2004.02.025}.
\bibitem[{Okamoto et~al.(2002)Okamoto, Akabori, Motohashi, Itoh \&
  Ogawa}]{Okamoto2002}
\bibinfo{author}{Okamoto, Y.}, \bibinfo{author}{Akabori, M.},
  \bibinfo{author}{Motohashi, H.}, \bibinfo{author}{Itoh, A.}, \&
  \bibinfo{author}{Ogawa, T.} (\bibinfo{year}{2002}).
\newblock \bibinfo{title}{{High-temperature XAFS measurement of molten salt
  systems}}.
\newblock {\it \bibinfo{journal}{Nucl. Instrum. Meth. Phys. Res. A}\/},  {\it
  \bibinfo{volume}{487}\/}, \bibinfo{pages}{605--611}.
  \DOIprefix\doi{10.1016/S0168-9002(01)02202-1}.
\bibitem[{Pierce et~al.(2012)Pierce, Salomon-Ferrer, Augusto F.~de Oliveira,
  McCammon \& Walker}]{Pierce2012}
\bibinfo{author}{Pierce, L.~C.}, \bibinfo{author}{Salomon-Ferrer, R.},
  \bibinfo{author}{Augusto F.~de Oliveira, C.}, \bibinfo{author}{McCammon,
  J.~A.}, \& \bibinfo{author}{Walker, R.~C.} (\bibinfo{year}{2012}).
\newblock \bibinfo{title}{{Routine access to millisecond time scale events with
  accelerated molecular dynamics}}.
\newblock {\it \bibinfo{journal}{J. Chem. Theory Comput.}\/},  {\it
  \bibinfo{volume}{8}\/}, \bibinfo{pages}{2997--3002}.
  \DOIprefix\doi{10.1021/ct300284c}.
\bibitem[{Price et~al.(2012)Price, Zonias, Skylaris, Hyde, Ravel \&
  Russell}]{Price2012}
\bibinfo{author}{Price, S. W.~T.}, \bibinfo{author}{Zonias, N.},
  \bibinfo{author}{Skylaris, C.-K.}, \bibinfo{author}{Hyde, T.~I.},
  \bibinfo{author}{Ravel, B.}, \& \bibinfo{author}{Russell, A.~E.}
  (\bibinfo{year}{2012}).
\newblock \bibinfo{title}{{Fitting EXAFS data using molecular dynamics outputs
  and a histogram approach}}.
\newblock {\it \bibinfo{journal}{Phys. Rev. B}\/},  {\it
  \bibinfo{volume}{85}\/}, \bibinfo{pages}{075439}.
  \DOIprefix\doi{10.1103/PhysRevB.85.075439}.
\bibitem[{Pugliese et~al.(2023)Pugliese, Tortora, Tomassucci, Kasem, Mizokawa,
  Mizuguchi \& Saini}]{Pugliese2023}
\bibinfo{author}{Pugliese, G.}, \bibinfo{author}{Tortora, L.},
  \bibinfo{author}{Tomassucci, G.}, \bibinfo{author}{Kasem, R.},
  \bibinfo{author}{Mizokawa, T.}, \bibinfo{author}{Mizuguchi, Y.}, \&
  \bibinfo{author}{Saini, N.} (\bibinfo{year}{2023}).
\newblock \bibinfo{title}{{Possible local order in the high entropy TrZr$_2$
  superconductors}}.
\newblock {\it \bibinfo{journal}{J. Phys. Chem. Solids}\/},  {\it
  \bibinfo{volume}{174}\/}, \bibinfo{pages}{111154}.
  \DOIprefix\doi{10.1016/j.jpcs.2022.111154}.
\bibitem[{Qi et~al.(1987)Qi, Perez, Ansari, Lu \& Croft}]{Qi1987}
\bibinfo{author}{Qi, B.}, \bibinfo{author}{Perez, I.}, \bibinfo{author}{Ansari,
  P.~H.}, \bibinfo{author}{Lu, F.}, \& \bibinfo{author}{Croft, M.}
  (\bibinfo{year}{1987}).
\newblock \bibinfo{title}{{L$_2$ and L$_3$ measurements of transition-metal 5d
  orbital occupancy, spin-orbit effects, and chemical bonding}}.
\newblock {\it \bibinfo{journal}{Phys. Rev. B}\/},  {\it
  \bibinfo{volume}{36}\/}, \bibinfo{pages}{2972--2975}.
  \DOIprefix\doi{10.1103/PhysRevB.36.2972}.
\bibitem[{Qian et~al.(2021)Qian, Han, Zheng, Chen, Tyagi, Li, Du, Zheng, Huang,
  Zhang, Shi, Huang, Shi, Chen, Qin, Bernholc, Chen, Chen, Hong \&
  Zhang}]{Qian2021}
\bibinfo{author}{Qian, X.}, \bibinfo{author}{Han, D.}, \bibinfo{author}{Zheng,
  L.}, \bibinfo{author}{Chen, J.}, \bibinfo{author}{Tyagi, M.},
  \bibinfo{author}{Li, Q.}, \bibinfo{author}{Du, F.}, \bibinfo{author}{Zheng,
  S.}, \bibinfo{author}{Huang, X.}, \bibinfo{author}{Zhang, S.},
  \bibinfo{author}{Shi, J.}, \bibinfo{author}{Huang, H.}, \bibinfo{author}{Shi,
  X.}, \bibinfo{author}{Chen, J.}, \bibinfo{author}{Qin, H.},
  \bibinfo{author}{Bernholc, J.}, \bibinfo{author}{Chen, X.},
  \bibinfo{author}{Chen, L.-Q.}, \bibinfo{author}{Hong, L.}, \&
  \bibinfo{author}{Zhang, Q.~M.} (\bibinfo{year}{2021}).
\newblock \bibinfo{title}{{High-entropy polymer produces a giant electrocaloric
  effect at low fields}}.
\newblock {\it \bibinfo{journal}{Nature}\/},  {\it \bibinfo{volume}{600}\/},
  \bibinfo{pages}{664--669}. \DOIprefix\doi{10.1038/s41586-021-04189-5}.
\bibitem[{Rehr \& Albers(2000)}]{Rehr2000}
\bibinfo{author}{Rehr, J.~J.}, \& \bibinfo{author}{Albers, R.~C.}
  (\bibinfo{year}{2000}).
\newblock \bibinfo{title}{{Theoretical approaches to X-ray absorption fine
  structure}}.
\newblock {\it \bibinfo{journal}{Rev. Mod. Phys.}\/},  {\it
  \bibinfo{volume}{72}\/}, \bibinfo{pages}{621--654}.
  \DOIprefix\doi{10.1103/RevModPhys.72.621}.
\bibitem[{Rehr et~al.(2009)Rehr, Kas, Prange, Sorini, Takimoto \&
  Vila}]{Rehr2009}
\bibinfo{author}{Rehr, J.~J.}, \bibinfo{author}{Kas, J.~J.},
  \bibinfo{author}{Prange, M.~P.}, \bibinfo{author}{Sorini, A.~P.},
  \bibinfo{author}{Takimoto, Y.}, \& \bibinfo{author}{Vila, F.}
  (\bibinfo{year}{2009}).
\newblock \bibinfo{title}{{Ab initio theory and calculations of X-ray
  spectra}}.
\newblock {\it \bibinfo{journal}{C. R. Phys.}\/},  {\it
  \bibinfo{volume}{10}\/}, \bibinfo{pages}{548--559}.
  \DOIprefix\doi{10.1016/j.crhy.2008.08.004}.
\bibitem[{Rehr et~al.(2010)Rehr, Kas, Vila, Prange \& Jorissen}]{FEFF9}
\bibinfo{author}{Rehr, J.~J.}, \bibinfo{author}{Kas, J.~J.},
  \bibinfo{author}{Vila, F.~D.}, \bibinfo{author}{Prange, M.~P.}, \&
  \bibinfo{author}{Jorissen, K.} (\bibinfo{year}{2010}).
\newblock \bibinfo{title}{{Parameter-free calculations of X-ray spectra with
  FEFF9}}.
\newblock {\it \bibinfo{journal}{Phys. Chem. Chem. Phys.}\/},  {\it
  \bibinfo{volume}{12}\/}, \bibinfo{pages}{5503--5513}.
  \DOIprefix\doi{10.1039/B926434E}.
\bibitem[{Ritter et~al.(2022)Ritter, Phakatkar, Rasul, Saray, Sorokina,
  Shokuhfar, Gon\c{c}alves \& Shahbazian-Yassar}]{Ritter2022}
\bibinfo{author}{Ritter, T.~G.}, \bibinfo{author}{Phakatkar, A.~H.},
  \bibinfo{author}{Rasul, M.~G.}, \bibinfo{author}{Saray, M.~T.},
  \bibinfo{author}{Sorokina, L.~V.}, \bibinfo{author}{Shokuhfar, T.},
  \bibinfo{author}{Gon\c{c}alves, J.~M.}, \&
  \bibinfo{author}{Shahbazian-Yassar, R.} (\bibinfo{year}{2022}).
\newblock \bibinfo{title}{{Electrochemical synthesis of high entropy hydroxides
  and oxides boosted by hydrogen evolution reaction}}.
\newblock {\it \bibinfo{journal}{Cell Rep. Phys. Sci.}\/},  {\it
  \bibinfo{volume}{3}\/}, \bibinfo{pages}{100847}.
  \DOIprefix\doi{10.1016/j.xcrp.2022.100847}.
\bibitem[{Rost et~al.(2015)Rost, Sachet, Borman, Moballegh, Dickey, Hou, Jones,
  Curtarolo \& Maria}]{Rost2015}
\bibinfo{author}{Rost, C.~M.}, \bibinfo{author}{Sachet, E.},
  \bibinfo{author}{Borman, T.}, \bibinfo{author}{Moballegh, A.},
  \bibinfo{author}{Dickey, E.~C.}, \bibinfo{author}{Hou, D.},
  \bibinfo{author}{Jones, J.~L.}, \bibinfo{author}{Curtarolo, S.}, \&
  \bibinfo{author}{Maria, J.-P.} (\bibinfo{year}{2015}).
\newblock \bibinfo{title}{{Entropy-stabilized oxides}}.
\newblock {\it \bibinfo{journal}{Nat. Commun.}\/},  {\it
  \bibinfo{volume}{6}\/}, \bibinfo{pages}{8485}.
  \DOIprefix\doi{10.1038/ncomms9485}.
\bibitem[{Ruiz-Lopez et~al.(1988)Ruiz-Lopez, Loos, Goulon, Benfatto \&
  Natoli}]{RuizLopez1988}
\bibinfo{author}{Ruiz-Lopez, M.}, \bibinfo{author}{Loos, M.},
  \bibinfo{author}{Goulon, J.}, \bibinfo{author}{Benfatto, M.}, \&
  \bibinfo{author}{Natoli, C.} (\bibinfo{year}{1988}).
\newblock \bibinfo{title}{{Reinvestigation of the EXAFS and XANES spectra of
  ferrocene and nickelocene in the framework of the multiple scattering
  theory}}.
\newblock {\it \bibinfo{journal}{Chem. Phys.}\/},  {\it
  \bibinfo{volume}{121}\/}, \bibinfo{pages}{419--437}.
  \DOIprefix\doi{10.1016/0301-0104(88)87246-X}.
\bibitem[{Sapelkin \& Bayliss(2002)}]{Sapelkin2002}
\bibinfo{author}{Sapelkin, A.~V.}, \& \bibinfo{author}{Bayliss, S.~C.}
  (\bibinfo{year}{2002}).
\newblock \bibinfo{title}{{Distance dependence of mean-square relative
  displacements in EXAFS}}.
\newblock {\it \bibinfo{journal}{Phys. Rev. B}\/},  {\it
  \bibinfo{volume}{65}\/}, \bibinfo{pages}{172104}.
  \DOIprefix\doi{10.1103/PhysRevB.65.172104}.
\bibitem[{Sarkar et~al.(2018)Sarkar, Velasco, Wang, Wang, Talasila, de~Biasi,
  K\"{u}bel, Brezesinski, Bhattacharya, Hahn \& Breitung}]{Sarkar2018}
\bibinfo{author}{Sarkar, A.}, \bibinfo{author}{Velasco, L.},
  \bibinfo{author}{Wang, D.}, \bibinfo{author}{Wang, Q.},
  \bibinfo{author}{Talasila, G.}, \bibinfo{author}{de~Biasi, L.},
  \bibinfo{author}{K\"{u}bel, C.}, \bibinfo{author}{Brezesinski, T.},
  \bibinfo{author}{Bhattacharya, S.~S.}, \bibinfo{author}{Hahn, H.}, \&
  \bibinfo{author}{Breitung, B.} (\bibinfo{year}{2018}).
\newblock \bibinfo{title}{{High entropy oxides for reversible energy storage}}.
\newblock {\it \bibinfo{journal}{Nat. Commun.}\/},  {\it
  \bibinfo{volume}{9}\/}, \bibinfo{pages}{3400}.
  \DOIprefix\doi{10.1038/s41467-018-05774-5}.
\bibitem[{Sayers et~al.(1971)Sayers, Stern \& Lytle}]{Sayers1971}
\bibinfo{author}{Sayers, D.~E.}, \bibinfo{author}{Stern, E.~A.}, \&
  \bibinfo{author}{Lytle, F.~W.} (\bibinfo{year}{1971}).
\newblock \bibinfo{title}{{New technique for investigating noncrystalline
  structures: Fourier analysis of the extended X-ray-absorption fine
  structure}}.
\newblock {\it \bibinfo{journal}{Phys. Rev. Lett.}\/},  {\it
  \bibinfo{volume}{27}\/}, \bibinfo{pages}{1204--1207}.
  \DOIprefix\doi{10.1103/PhysRevLett.27.1204}.
\bibitem[{Shapeev et~al.(2022)Shapeev, Bocharov \& Kuzmin}]{Shapeev2022}
\bibinfo{author}{Shapeev, A.~V.}, \bibinfo{author}{Bocharov, D.}, \&
  \bibinfo{author}{Kuzmin, A.} (\bibinfo{year}{2022}).
\newblock \bibinfo{title}{{Validation of moment tensor potentials for fcc and
  bcc metals using EXAFS spectra}}.
\newblock {\it \bibinfo{journal}{Comput. Mater. Sci.}\/},  {\it
  \bibinfo{volume}{210}\/}, \bibinfo{pages}{111028}.
  \DOIprefix\doi{10.1016/j.commatsci.2021.111028}.
\bibitem[{Smekhova et~al.(2022{\natexlab{a}})Smekhova, Kuzmin, Siemensmeyer,
  Abrudan, Reinholz, Buzanich, Schneider, Laplanche \& Yusenko}]{Smekhova2022a}
\bibinfo{author}{Smekhova, A.}, \bibinfo{author}{Kuzmin, A.},
  \bibinfo{author}{Siemensmeyer, K.}, \bibinfo{author}{Abrudan, R.},
  \bibinfo{author}{Reinholz, U.}, \bibinfo{author}{Buzanich, A.~G.},
  \bibinfo{author}{Schneider, M.}, \bibinfo{author}{Laplanche, G.}, \&
  \bibinfo{author}{Yusenko, K.~V.} (\bibinfo{year}{2022}{\natexlab{a}}).
\newblock \bibinfo{title}{{Inner relaxations in equiatomic single-phase
  high-entropy cantor alloy}}.
\newblock {\it \bibinfo{journal}{J. Alloys Compd.}\/},  {\it
  \bibinfo{volume}{920}\/}, \bibinfo{pages}{165999}.
  \DOIprefix\doi{10.1016/j.jallcom.2022.165999}.
\bibitem[{Smekhova et~al.(2022{\natexlab{b}})Smekhova, Kuzmin, Siemensmeyer,
  Luo, Chen, Radu, Weschke, Reinholz, Buzanich \& Yusenko}]{Smekhova2022b}
\bibinfo{author}{Smekhova, A.}, \bibinfo{author}{Kuzmin, A.},
  \bibinfo{author}{Siemensmeyer, K.}, \bibinfo{author}{Luo, C.},
  \bibinfo{author}{Chen, K.}, \bibinfo{author}{Radu, F.},
  \bibinfo{author}{Weschke, E.}, \bibinfo{author}{Reinholz, U.},
  \bibinfo{author}{Buzanich, A.~G.}, \& \bibinfo{author}{Yusenko, K.~V.}
  (\bibinfo{year}{2022}{\natexlab{b}}).
\newblock \bibinfo{title}{{Al-driven peculiarities of local coordination and
  magnetic properties in single-phase Al$_x$-CrFeCoNi high-entropy alloys}}.
\newblock {\it \bibinfo{journal}{Nano Res.}\/},  {\it \bibinfo{volume}{15}\/},
  \bibinfo{pages}{4845--4858}. \DOIprefix\doi{10.1007/s12274-021-3704-5}.
\bibitem[{Smekhova et~al.(2023)Smekhova, Kuzmin, Siemensmeyer, Luo, Taylor,
  Thakur, Radu, Weschke, Buzanich, Xiao, Savan, Yusenko \&
  Ludwig}]{Smekhova2023}
\bibinfo{author}{Smekhova, A.}, \bibinfo{author}{Kuzmin, A.},
  \bibinfo{author}{Siemensmeyer, K.}, \bibinfo{author}{Luo, C.},
  \bibinfo{author}{Taylor, J.}, \bibinfo{author}{Thakur, S.},
  \bibinfo{author}{Radu, F.}, \bibinfo{author}{Weschke, E.},
  \bibinfo{author}{Buzanich, A.~G.}, \bibinfo{author}{Xiao, B.},
  \bibinfo{author}{Savan, A.}, \bibinfo{author}{Yusenko, K.~V.}, \&
  \bibinfo{author}{Ludwig, A.} (\bibinfo{year}{2023}).
\newblock \bibinfo{title}{{Local structure and magnetic properties of a
  nanocrystalline Mn-rich Cantor alloy thin film down to the atomic scale}}.
\newblock {\it \bibinfo{journal}{Nano Res.}\/},  {\it \bibinfo{volume}{16}\/},
  \bibinfo{pages}{5626--5639}. \DOIprefix\doi{10.1007/s12274-022-5135-3}.
\bibitem[{Sun et~al.(2017)Sun, Sun, Sun, Chen, Du, Wang, Jiang \&
  Huang}]{Sun2017}
\bibinfo{author}{Sun, X.}, \bibinfo{author}{Sun, F.}, \bibinfo{author}{Sun,
  Z.}, \bibinfo{author}{Chen, J.}, \bibinfo{author}{Du, X.},
  \bibinfo{author}{Wang, J.}, \bibinfo{author}{Jiang, Z.}, \&
  \bibinfo{author}{Huang, Y.} (\bibinfo{year}{2017}).
\newblock \bibinfo{title}{{Disorder effects on EXAFS modeling for catalysts
  working at elevated temperatures}}.
\newblock {\it \bibinfo{journal}{Rad. Phys. Chem.}\/},  {\it
  \bibinfo{volume}{137}\/}, \bibinfo{pages}{93--98}.
  \DOIprefix\doi{10.1016/j.radphyschem.2016.01.039}.
\bibitem[{Sushil et~al.(2021)Sushil, Kumar, Gautam \& Ahmad}]{Sushil2021}
\bibinfo{author}{Sushil, J.}, \bibinfo{author}{Kumar, A.},
  \bibinfo{author}{Gautam, A.}, \& \bibinfo{author}{Ahmad, M.~I.}
  (\bibinfo{year}{2021}).
\newblock \bibinfo{title}{{High entropy phase evolution and fine structure of
  five component oxide (Mg, Co, Ni, Cu, Zn)O by citrate gel method}}.
\newblock {\it \bibinfo{journal}{Mater. Chem. Phys.}\/},  {\it
  \bibinfo{volume}{259}\/}, \bibinfo{pages}{124014}.
  \DOIprefix\doi{10.1016/j.matchemphys.2020.124014}.
\bibitem[{Tamm et~al.(2015)Tamm, Aabloo, Klintenberg, Stocks \&
  Caro}]{Tamm2015}
\bibinfo{author}{Tamm, A.}, \bibinfo{author}{Aabloo, A.},
  \bibinfo{author}{Klintenberg, M.}, \bibinfo{author}{Stocks, M.}, \&
  \bibinfo{author}{Caro, A.} (\bibinfo{year}{2015}).
\newblock \bibinfo{title}{{Atomic-scale properties of Ni-based FCC ternary, and
  quaternary alloys}}.
\newblock {\it \bibinfo{journal}{Acta Mater.}\/},  {\it
  \bibinfo{volume}{99}\/}, \bibinfo{pages}{307--312}.
  \DOIprefix\doi{10.1016/j.actamat.2015.08.015}.
\bibitem[{Tan et~al.(2023)Tan, Li, Chen, Chen, Su, Zhang, Gong, Wu, Wang \&
  Dai}]{Tan2023}
\bibinfo{author}{Tan, Y.-Y.}, \bibinfo{author}{Li, T.}, \bibinfo{author}{Chen,
  Y.}, \bibinfo{author}{Chen, Z.-J.}, \bibinfo{author}{Su, M.-Y.},
  \bibinfo{author}{Zhang, J.}, \bibinfo{author}{Gong, Y.}, \bibinfo{author}{Wu,
  T.}, \bibinfo{author}{Wang, H.-Y.}, \& \bibinfo{author}{Dai, L.-H.}
  (\bibinfo{year}{2023}).
\newblock \bibinfo{title}{{Uncovering heterogeneity of local lattice distortion
  in TiZrHfNbTa refractory high entropy alloy by SR-XRD and EXAFS}}.
\newblock {\it \bibinfo{journal}{Scr. Mater.}\/},  {\it
  \bibinfo{volume}{223}\/}, \bibinfo{pages}{115079}.
  \DOIprefix\doi{10.1016/j.scriptamat.2022.115079}.
\bibitem[{Tan et~al.(2021)Tan, Su, Xie, Chen, Gong, Zheng, Shi, Mo, Li, Li,
  Wang \& Dai}]{Tan2021}
\bibinfo{author}{Tan, Y.-Y.}, \bibinfo{author}{Su, M.-Y.},
  \bibinfo{author}{Xie, Z.-C.}, \bibinfo{author}{Chen, Z.-J.},
  \bibinfo{author}{Gong, Y.}, \bibinfo{author}{Zheng, L.-R.},
  \bibinfo{author}{Shi, Z.}, \bibinfo{author}{Mo, G.}, \bibinfo{author}{Li,
  Y.}, \bibinfo{author}{Li, L.-W.}, \bibinfo{author}{Wang, H.-Y.}, \&
  \bibinfo{author}{Dai, L.-H.} (\bibinfo{year}{2021}).
\newblock \bibinfo{title}{{Chemical composition dependent local lattice
  distortions and magnetism in high entropy alloys}}.
\newblock {\it \bibinfo{journal}{Intermetallics}\/},  {\it
  \bibinfo{volume}{129}\/}, \bibinfo{pages}{107050}.
  \DOIprefix\doi{10.1016/j.intermet.2020.107050}.
\bibitem[{Tavani et~al.(2020)Tavani, Fracchia, Pianta, Ghigna, Quartarone \&
  D’Angelo}]{Tavani2020}
\bibinfo{author}{Tavani, F.}, \bibinfo{author}{Fracchia, M.},
  \bibinfo{author}{Pianta, N.}, \bibinfo{author}{Ghigna, P.},
  \bibinfo{author}{Quartarone, E.}, \& \bibinfo{author}{D’Angelo, P.}
  (\bibinfo{year}{2020}).
\newblock \bibinfo{title}{{Multivariate curve resolution analysis of operando
  XAS data for the investigation of the lithiation mechanisms in high entropy
  oxides}}.
\newblock {\it \bibinfo{journal}{Chem. Phys. Lett.}\/},  {\it
  \bibinfo{volume}{760}\/}, \bibinfo{pages}{137968}.
  \DOIprefix\doi{10.1016/j.cplett.2020.137968}.
\bibitem[{Tavani et~al.(2021)Tavani, Fracchia, Tofoni, Braglia, Jouve, Morandi,
  Manzoli, Torelli, Ghigna \& D{'}Angelo}]{Tavani2021}
\bibinfo{author}{Tavani, F.}, \bibinfo{author}{Fracchia, M.},
  \bibinfo{author}{Tofoni, A.}, \bibinfo{author}{Braglia, L.},
  \bibinfo{author}{Jouve, A.}, \bibinfo{author}{Morandi, S.},
  \bibinfo{author}{Manzoli, M.}, \bibinfo{author}{Torelli, P.},
  \bibinfo{author}{Ghigna, P.}, \& \bibinfo{author}{D{'}Angelo, P.}
  (\bibinfo{year}{2021}).
\newblock \bibinfo{title}{{Structural and mechanistic insights into
  low-temperature CO oxidation over a prototypical high entropy oxide by Cu
  L-edge operando soft X-ray absorption spectroscopy}}.
\newblock {\it \bibinfo{journal}{Phys. Chem. Chem. Phys.}\/},  {\it
  \bibinfo{volume}{23}\/}, \bibinfo{pages}{26575--26584}.
  \DOIprefix\doi{10.1039/D1CP03946F}.
\bibitem[{Teo(1986)}]{Teo1986}
\bibinfo{author}{Teo, B.~K.} (\bibinfo{year}{1986}).
\newblock {\it \bibinfo{title}{{EXAFS: Basic Principles and Data Analysis}}\/}.
\newblock \bibinfo{address}{Berlin}: \bibinfo{publisher}{Springer}.
\bibitem[{Teplonogova et~al.(2022)Teplonogova, Yapryntsev, Baranchikov \&
  Ivanov}]{Teplonogova2022}
\bibinfo{author}{Teplonogova, M.}, \bibinfo{author}{Yapryntsev, A.},
  \bibinfo{author}{Baranchikov, A.}, \& \bibinfo{author}{Ivanov, V.}
  (\bibinfo{year}{2022}).
\newblock \bibinfo{title}{{High-entropy layered rare earth hydroxides}}.
\newblock {\it \bibinfo{journal}{Inorg. Chem.}\/},  {\it
  \bibinfo{volume}{61}\/}, \bibinfo{pages}{19817--19827}.
  \DOIprefix\doi{10.1021/acs.inorgchem.2c02950}.
\bibitem[{Timoshenko et~al.(2014{\natexlab{a}})Timoshenko, Anspoks, Kalinko \&
  Kuzmin}]{Timoshenko2014cuwo4}
\bibinfo{author}{Timoshenko, J.}, \bibinfo{author}{Anspoks, A.},
  \bibinfo{author}{Kalinko, A.}, \& \bibinfo{author}{Kuzmin, A.}
  (\bibinfo{year}{2014}{\natexlab{a}}).
\newblock \bibinfo{title}{{Analysis of extended X-ray absorption fine structure
  data from copper tungstate by the reverse Monte Carlo method}}.
\newblock {\it \bibinfo{journal}{Phys. Scr.}\/},  {\it \bibinfo{volume}{89}\/},
  \bibinfo{pages}{044006}. \DOIprefix\doi{10.1088/0031-8949/89/04/044006}.
\bibitem[{Timoshenko et~al.(2014{\natexlab{b}})Timoshenko, Anspoks, Kalinko \&
  Kuzmin}]{Timoshenko2014zno}
\bibinfo{author}{Timoshenko, J.}, \bibinfo{author}{Anspoks, A.},
  \bibinfo{author}{Kalinko, A.}, \& \bibinfo{author}{Kuzmin, A.}
  (\bibinfo{year}{2014}{\natexlab{b}}).
\newblock \bibinfo{title}{{Temperature dependence of the local structure and
  lattice dynamics of wurtzite-type ZnO}}.
\newblock {\it \bibinfo{journal}{Acta Mater.}\/},  {\it
  \bibinfo{volume}{79}\/}, \bibinfo{pages}{194--202}.
  \DOIprefix\doi{10.1016/j.actamat.2014.07.029}.
\bibitem[{Timoshenko \& Kuzmin(2009)}]{Timoshenko2009wt}
\bibinfo{author}{Timoshenko, J.}, \& \bibinfo{author}{Kuzmin, A.}
  (\bibinfo{year}{2009}).
\newblock \bibinfo{title}{{Wavelet data analysis of EXAFS spectra}}.
\newblock {\it \bibinfo{journal}{Comput. Phys. Commun.}\/},  {\it
  \bibinfo{volume}{180}\/}, \bibinfo{pages}{920--925}.
  \DOIprefix\doi{10.1016/j.cpc.2008.12.020}.
\bibitem[{Timoshenko et~al.(2012)Timoshenko, Kuzmin \&
  Purans}]{Timoshenko2012rmc}
\bibinfo{author}{Timoshenko, J.}, \bibinfo{author}{Kuzmin, A.}, \&
  \bibinfo{author}{Purans, J.} (\bibinfo{year}{2012}).
\newblock \bibinfo{title}{{Reverse Monte Carlo modeling of thermal disorder in
  crystalline materials from EXAFS spectra}}.
\newblock {\it \bibinfo{journal}{Comput. Phys. Commun.}\/},  {\it
  \bibinfo{volume}{183}\/}, \bibinfo{pages}{1237--1245}.
  \DOIprefix\doi{10.1016/j.cpc.2012.02.002}.
\bibitem[{Timoshenko et~al.(2014{\natexlab{c}})Timoshenko, Kuzmin \&
  Purans}]{Timoshenko2014rmc}
\bibinfo{author}{Timoshenko, J.}, \bibinfo{author}{Kuzmin, A.}, \&
  \bibinfo{author}{Purans, J.} (\bibinfo{year}{2014}{\natexlab{c}}).
\newblock \bibinfo{title}{{EXAFS study of hydrogen intercalation into ReO$_3$
  using the evolutionary algorithm}}.
\newblock {\it \bibinfo{journal}{J. Phys.: Condens. Matter}\/},  {\it
  \bibinfo{volume}{26}\/}, \bibinfo{pages}{055401}.
  \DOIprefix\doi{10.1088/0953-8984/26/5/055401}.
\bibitem[{Timoshenko et~al.(2017)Timoshenko, Lu, Lin \&
  Frenkel}]{Timoshenko2017}
\bibinfo{author}{Timoshenko, J.}, \bibinfo{author}{Lu, D.},
  \bibinfo{author}{Lin, Y.}, \& \bibinfo{author}{Frenkel, A.~I.}
  (\bibinfo{year}{2017}).
\newblock \bibinfo{title}{{Supervised machine-learning-based determination of
  three-dimensional structure of metallic nanoparticles}}.
\newblock {\it \bibinfo{journal}{J. Phys. Chem. Lett.}\/},  {\it
  \bibinfo{volume}{8}\/}, \bibinfo{pages}{5091--5098}.
  \DOIprefix\doi{10.1021/acs.jpclett.7b02364}.
\bibitem[{Tranquada \& Ingalls(1983)}]{Tranquada1983}
\bibinfo{author}{Tranquada, J.~M.}, \& \bibinfo{author}{Ingalls, R.}
  (\bibinfo{year}{1983}).
\newblock \bibinfo{title}{{Extended x-ray-absorption fine-structure study of
  anharmonicity in CuBr}}.
\newblock {\it \bibinfo{journal}{Phys. Rev. B}\/},  {\it
  \bibinfo{volume}{28}\/}, \bibinfo{pages}{3520--3528}.
  \DOIprefix\doi{10.1103/PhysRevB.28.3520}.
\bibitem[{Tucker et~al.(2007)Tucker, Keen, Dove, Goodwin \&
  Hui}]{Tucker2007rmcprofile}
\bibinfo{author}{Tucker, M.~G.}, \bibinfo{author}{Keen, D.~A.},
  \bibinfo{author}{Dove, M.~T.}, \bibinfo{author}{Goodwin, A.~L.}, \&
  \bibinfo{author}{Hui, Q.} (\bibinfo{year}{2007}).
\newblock \bibinfo{title}{{RMCProfile: reverse Monte Carlo for polycrystalline
  materials}}.
\newblock {\it \bibinfo{journal}{J. Phys.: Condens. Matter}\/},  {\it
  \bibinfo{volume}{19}\/}, \bibinfo{pages}{335218}.
  \DOIprefix\doi{10.1088/0953-8984/19/33/335218}.
\bibitem[{Tuckerman et~al.(1996)Tuckerman, Ungar, von Rosenvinge \&
  Klein}]{Tuckerman1996}
\bibinfo{author}{Tuckerman, M.~E.}, \bibinfo{author}{Ungar, P.~J.},
  \bibinfo{author}{von Rosenvinge, T.}, \& \bibinfo{author}{Klein, M.~L.}
  (\bibinfo{year}{1996}).
\newblock \bibinfo{title}{{Ab initio molecular dynamics simulations}}.
\newblock {\it \bibinfo{journal}{J. Phys. Chem.}\/},  {\it
  \bibinfo{volume}{100}\/}, \bibinfo{pages}{12878--12887}.
  \DOIprefix\doi{10.1021/jp960480+}.
\bibitem[{Walczak et~al.(2022)Walczak, Plewa, Ghica, Zając, Trenczek-Zając,
  Zając, Toboła \& Molenda}]{Walczak2022}
\bibinfo{author}{Walczak, K.}, \bibinfo{author}{Plewa, A.},
  \bibinfo{author}{Ghica, C.}, \bibinfo{author}{Zając, W.},
  \bibinfo{author}{Trenczek-Zając, A.}, \bibinfo{author}{Zając, M.},
  \bibinfo{author}{Toboła, J.}, \& \bibinfo{author}{Molenda, J.}
  (\bibinfo{year}{2022}).
\newblock
  \bibinfo{title}{{NaMn$_{0.2}$Fe$_{0.2}$Co$_{0.2}$Ni$_{0.2}$Ti$_{0.2}$O$_2$
  high-entropy layered oxide – experimental and theoretical evidence of high
  electrochemical performance in sodium batteries}}.
\newblock {\it \bibinfo{journal}{Energy Storage Mater.}\/},  {\it
  \bibinfo{volume}{47}\/}, \bibinfo{pages}{500--514}.
  \DOIprefix\doi{10.1016/j.ensm.2022.02.038}.
\bibitem[{Wang et~al.(2022)Wang, Dreyer, Wang, Ding, Diemant, Karkera, Ma,
  Sarkar, Zhou, Gorbunov, Omar, Mikhailova, Presser, Fichtner, Hahn,
  Brezesinski, Breitung \& Wang}]{Wang2022}
\bibinfo{author}{Wang, J.}, \bibinfo{author}{Dreyer, S.~L.},
  \bibinfo{author}{Wang, K.}, \bibinfo{author}{Ding, Z.},
  \bibinfo{author}{Diemant, T.}, \bibinfo{author}{Karkera, G.},
  \bibinfo{author}{Ma, Y.}, \bibinfo{author}{Sarkar, A.},
  \bibinfo{author}{Zhou, B.}, \bibinfo{author}{Gorbunov, M.~V.},
  \bibinfo{author}{Omar, A.}, \bibinfo{author}{Mikhailova, D.},
  \bibinfo{author}{Presser, V.}, \bibinfo{author}{Fichtner, M.},
  \bibinfo{author}{Hahn, H.}, \bibinfo{author}{Brezesinski, T.},
  \bibinfo{author}{Breitung, B.}, \& \bibinfo{author}{Wang, Q.}
  (\bibinfo{year}{2022}).
\newblock \bibinfo{title}{{P2-type layered high-entropy oxides as sodium-ion
  cathode materials}}.
\newblock {\it \bibinfo{journal}{Mater. Futures}\/},  {\it
  \bibinfo{volume}{1}\/}, \bibinfo{pages}{035104}.
  \DOIprefix\doi{10.1088/2752-5724/ac8ab9}.
\bibitem[{Wang et~al.(2023)Wang, Hua, Huang, Stenzel, Wang, Ding, Cui, Wang,
  Ehrenberg, Breitung, K\"{u}bel \& Mu}]{Wang2023}
\bibinfo{author}{Wang, K.}, \bibinfo{author}{Hua, W.}, \bibinfo{author}{Huang,
  X.}, \bibinfo{author}{Stenzel, D.}, \bibinfo{author}{Wang, J.},
  \bibinfo{author}{Ding, Z.}, \bibinfo{author}{Cui, Y.}, \bibinfo{author}{Wang,
  Q.}, \bibinfo{author}{Ehrenberg, H.}, \bibinfo{author}{Breitung, B.},
  \bibinfo{author}{K\"{u}bel, C.}, \& \bibinfo{author}{Mu, X.}
  (\bibinfo{year}{2023}).
\newblock \bibinfo{title}{{Synergy of cations in high entropy oxide lithium ion
  battery anode}}.
\newblock {\it \bibinfo{journal}{Nat. Commun.}\/},  {\it
  \bibinfo{volume}{14}\/}, \bibinfo{pages}{1487}.
  \DOIprefix\doi{10.1038/s41467-023-37034-6}.
\bibitem[{Wei et~al.(2020)Wei, Liao, Wu, Yang, He, Biesold-McGee, Liang, Yen,
  Tang, Yeh, Lin \& He}]{Wei2020}
\bibinfo{author}{Wei, P.-C.}, \bibinfo{author}{Liao, C.-N.},
  \bibinfo{author}{Wu, H.-J.}, \bibinfo{author}{Yang, D.}, \bibinfo{author}{He,
  J.}, \bibinfo{author}{Biesold-McGee, G.~V.}, \bibinfo{author}{Liang, S.},
  \bibinfo{author}{Yen, W.-T.}, \bibinfo{author}{Tang, X.},
  \bibinfo{author}{Yeh, J.-W.}, \bibinfo{author}{Lin, Z.}, \&
  \bibinfo{author}{He, J.-H.} (\bibinfo{year}{2020}).
\newblock \bibinfo{title}{Thermodynamic routes to ultralow thermal conductivity
  and high thermoelectric performance}.
\newblock {\it \bibinfo{journal}{Adv. Mater.}\/},  {\it
  \bibinfo{volume}{32}\/}, \bibinfo{pages}{1906457}.
  \DOIprefix\doi{10.1002/adma.201906457}.
\bibitem[{Winterer(2000)}]{Winterer2000}
\bibinfo{author}{Winterer, M.} (\bibinfo{year}{2000}).
\newblock \bibinfo{title}{{Reverse Monte Carlo analysis of extended X-ray
  absorption fine structure spectra of monoclinic and amorphous zirconia}}.
\newblock {\it \bibinfo{journal}{J. Appl. Phys.}\/},  {\it
  \bibinfo{volume}{88}\/}, \bibinfo{pages}{5635--5644}.
  \DOIprefix\doi{10.1063/1.1319167}.
\bibitem[{Woicik et~al.(2023)Woicik, Cockayne, Shirley, Levin, Weiland, Ravel
  \& Abeykoon}]{Woicik2023}
\bibinfo{author}{Woicik, J.~C.}, \bibinfo{author}{Cockayne, E.},
  \bibinfo{author}{Shirley, E.~L.}, \bibinfo{author}{Levin, I.},
  \bibinfo{author}{Weiland, C.}, \bibinfo{author}{Ravel, B.}, \&
  \bibinfo{author}{Abeykoon, A. M.~M.} (\bibinfo{year}{2023}).
\newblock \bibinfo{title}{{Lattice vibrations and energy landscape of the
  isoelectronic semiconductor series CuBr, ZnSe, GaAs, and Ge: The special case
  of CuBr and its $d$-level chemistry}}.
\newblock {\it \bibinfo{journal}{Phys. Rev. B}\/},  {\it
  \bibinfo{volume}{108}\/}, \bibinfo{pages}{195202}.
  \DOIprefix\doi{10.1103/PhysRevB.108.195202}.
\bibitem[{Wu et~al.(2021)Wu, Chen, Shi, Wu, Gui, Tan, Li \& Wu}]{Wu2021}
\bibinfo{author}{Wu, T.}, \bibinfo{author}{Chen, Y.}, \bibinfo{author}{Shi,
  S.}, \bibinfo{author}{Wu, M.}, \bibinfo{author}{Gui, W.},
  \bibinfo{author}{Tan, Y.}, \bibinfo{author}{Li, J.}, \& \bibinfo{author}{Wu,
  Y.} (\bibinfo{year}{2021}).
\newblock \bibinfo{title}{{Effects of W alloying on the lattice distortion and
  wear behavior of laser cladding AlCoCrFeNiW$_x$ high-entropy alloy
  coatings}}.
\newblock {\it \bibinfo{journal}{Materials}\/},  {\it \bibinfo{volume}{14}\/},
  \bibinfo{pages}{5450}. \DOIprefix\doi{10.3390/ma14185450}.
\bibitem[{Xiang et~al.(2021)Xiang, Xing, Dai, Wang, Su, Miao, Zhang, Wang, Qi,
  Yao, Wang, Zhao, Li \& Zhou}]{Xiang2021}
\bibinfo{author}{Xiang, H.}, \bibinfo{author}{Xing, Y.}, \bibinfo{author}{Dai,
  F.}, \bibinfo{author}{Wang, H.}, \bibinfo{author}{Su, L.},
  \bibinfo{author}{Miao, L.}, \bibinfo{author}{Zhang, G.},
  \bibinfo{author}{Wang, Y.}, \bibinfo{author}{Qi, X.}, \bibinfo{author}{Yao,
  L.}, \bibinfo{author}{Wang, H.}, \bibinfo{author}{Zhao, B.},
  \bibinfo{author}{Li, J.}, \& \bibinfo{author}{Zhou, Y.}
  (\bibinfo{year}{2021}).
\newblock \bibinfo{title}{{High-entropy ceramics: Present status, challenges,
  and a look forward}}.
\newblock {\it \bibinfo{journal}{J. Adv. Ceram.}\/},  {\it
  \bibinfo{volume}{10}\/}, \bibinfo{pages}{385--441}.
  \DOIprefix\doi{10.1007/s40145-021-0477-y}.
\bibitem[{Xu et~al.(2020)Xu, Zhang, Liu, Do-Thanh, Chen, Xu, Lin, Jiao, Wang,
  Wang, Chen \& Dai}]{Xu2020}
\bibinfo{author}{Xu, H.}, \bibinfo{author}{Zhang, Z.}, \bibinfo{author}{Liu,
  J.}, \bibinfo{author}{Do-Thanh, C.-L.}, \bibinfo{author}{Chen, H.},
  \bibinfo{author}{Xu, S.}, \bibinfo{author}{Lin, Q.}, \bibinfo{author}{Jiao,
  Y.}, \bibinfo{author}{Wang, J.}, \bibinfo{author}{Wang, Y.},
  \bibinfo{author}{Chen, Y.}, \& \bibinfo{author}{Dai, S.}
  (\bibinfo{year}{2020}).
\newblock \bibinfo{title}{{Entropy-stabilized single-atom Pd catalysts via
  high-entropy fluorite oxide supports}}.
\newblock {\it \bibinfo{journal}{Nat. Commun.}\/},  {\it
  \bibinfo{volume}{11}\/}, \bibinfo{pages}{3908}.
  \DOIprefix\doi{10.1038/s41467-020-17738-9}.
\bibitem[{Yancey et~al.(2013)Yancey, Chill, Zhang, Frenkel, Henkelman \&
  Crooks}]{Yancey2013}
\bibinfo{author}{Yancey, D.~F.}, \bibinfo{author}{Chill, S.~T.},
  \bibinfo{author}{Zhang, L.}, \bibinfo{author}{Frenkel, A.~I.},
  \bibinfo{author}{Henkelman, G.}, \& \bibinfo{author}{Crooks, R.~M.}
  (\bibinfo{year}{2013}).
\newblock \bibinfo{title}{{A theoretical and experimental examination of
  systematic ligand-induced disorder in Au dendrimer-encapsulated
  nanoparticles}}.
\newblock {\it \bibinfo{journal}{Chem. Sci.}\/},  {\it \bibinfo{volume}{4}\/},
  \bibinfo{pages}{2912--2921}. \DOIprefix\doi{10.1039/C3SC50614B}.
\bibitem[{Yang \& Kawazoe(2012)}]{Yang2012}
\bibinfo{author}{Yang, Y.}, \& \bibinfo{author}{Kawazoe, Y.}
  (\bibinfo{year}{2012}).
\newblock \bibinfo{title}{{Characterization of zero-point vibration in
  one-component crystals}}.
\newblock {\it \bibinfo{journal}{EPL}\/},  {\it \bibinfo{volume}{98}\/},
  \bibinfo{pages}{66007}. \DOIprefix\doi{10.1209/0295-5075/98/66007}.
\bibitem[{Yeh et~al.(2004)Yeh, Chen, Lin, Gan, Chin, Shun, Tsau \&
  Chang}]{Yeh2004}
\bibinfo{author}{Yeh, J.-W.}, \bibinfo{author}{Chen, S.-K.},
  \bibinfo{author}{Lin, S.-J.}, \bibinfo{author}{Gan, J.-Y.},
  \bibinfo{author}{Chin, T.-S.}, \bibinfo{author}{Shun, T.-T.},
  \bibinfo{author}{Tsau, C.-H.}, \& \bibinfo{author}{Chang, S.-Y.}
  (\bibinfo{year}{2004}).
\newblock \bibinfo{title}{{Nanostructured high-entropy alloys with multiple
  principal elements: novel alloy design concepts and outcomes}}.
\newblock {\it \bibinfo{journal}{Adv. Eng. Mater.}\/},  {\it
  \bibinfo{volume}{6}\/}, \bibinfo{pages}{299--303}.
  \DOIprefix\doi{10.1002/adem.200300567}.
\bibitem[{Yuan et~al.(2022)Yuan, Qin, Li, Huang, Feng, Fang, Lollar, Pang,
  Zhang, Sun, Alsalme, Cagin \& Zhou}]{Yuan2022}
\bibinfo{author}{Yuan, S.}, \bibinfo{author}{Qin, J.-S.}, \bibinfo{author}{Li,
  J.}, \bibinfo{author}{Huang, L.}, \bibinfo{author}{Feng, L.},
  \bibinfo{author}{Fang, Y.}, \bibinfo{author}{Lollar, C.},
  \bibinfo{author}{Pang, J.}, \bibinfo{author}{Zhang, L.},
  \bibinfo{author}{Sun, D.}, \bibinfo{author}{Alsalme, A.},
  \bibinfo{author}{Cagin, T.}, \& \bibinfo{author}{Zhou, H.-C.}
  (\bibinfo{year}{2022}).
\newblock \bibinfo{title}{{Retrosynthesis of multi-component metal-organic
  frameworks}}.
\newblock {\it \bibinfo{journal}{Nat. Commun.}\/},  {\it
  \bibinfo{volume}{9}\/}, \bibinfo{pages}{808}.
  \DOIprefix\doi{10.1038/s41467-018-03102-5}.
\bibitem[{Zhang \& Song(2022)}]{Zhang2022a}
\bibinfo{author}{Zhang, F.}, \& \bibinfo{author}{Song, H.-Q.}
  (\bibinfo{year}{2022}).
\newblock \bibinfo{title}{{Effect of atomic size mismatch and chemical
  complexity on the local lattice distortion of BCC solid solution alloys}}.
\newblock {\it \bibinfo{journal}{Mater. Today Commun.}\/},  {\it
  \bibinfo{volume}{33}\/}, \bibinfo{pages}{104367}.
  \DOIprefix\doi{10.1016/j.mtcomm.2022.104367}.
\bibitem[{Zhang et~al.(2018)Zhang, Tong, Jin, Bei, Weber, Huq, Lanzirotti,
  Newville, Pagan, Ko \& Zhang}]{Zhang2018}
\bibinfo{author}{Zhang, F.}, \bibinfo{author}{Tong, Y.}, \bibinfo{author}{Jin,
  K.}, \bibinfo{author}{Bei, H.}, \bibinfo{author}{Weber, W.~J.},
  \bibinfo{author}{Huq, A.}, \bibinfo{author}{Lanzirotti, A.},
  \bibinfo{author}{Newville, M.}, \bibinfo{author}{Pagan, D.~C.},
  \bibinfo{author}{Ko, J. Y.~P.}, \& \bibinfo{author}{Zhang, Y.}
  (\bibinfo{year}{2018}).
\newblock \bibinfo{title}{{Chemical complexity induced local structural
  distortion in NiCoFeMnCr high-entropy alloy}}.
\newblock {\it \bibinfo{journal}{Mater. Res. Lett.}\/},  {\it
  \bibinfo{volume}{6}\/}, \bibinfo{pages}{450--455}.
  \DOIprefix\doi{10.1080/21663831.2018.1478332}.
\bibitem[{Zhang et~al.(2017)Zhang, Zhao, Jin, Xue, Velisa, Bei, Huang, Ko,
  Pagan, Neuefeind, Weber \& Zhang}]{Zhang2017}
\bibinfo{author}{Zhang, F.~X.}, \bibinfo{author}{Zhao, S.},
  \bibinfo{author}{Jin, K.}, \bibinfo{author}{Xue, H.},
  \bibinfo{author}{Velisa, G.}, \bibinfo{author}{Bei, H.},
  \bibinfo{author}{Huang, R.}, \bibinfo{author}{Ko, J. Y.~P.},
  \bibinfo{author}{Pagan, D.~C.}, \bibinfo{author}{Neuefeind, J.~C.},
  \bibinfo{author}{Weber, W.~J.}, \& \bibinfo{author}{Zhang, Y.}
  (\bibinfo{year}{2017}).
\newblock \bibinfo{title}{{Local structure and short-range order in a NiCoCr
  solid solution alloy}}.
\newblock {\it \bibinfo{journal}{Phys. Rev. Lett.}\/},  {\it
  \bibinfo{volume}{118}\/}, \bibinfo{pages}{205501}.
  \DOIprefix\doi{10.1103/PhysRevLett.118.205501}.
\bibitem[{Zhang et~al.(2022{\natexlab{a}})Zhang, Luan, Lou, Liang, Chen, Xu,
  Yin, Wang, Zeng, Ren, Zeng, Shao, Yao \& Zeng}]{Zhang2022b}
\bibinfo{author}{Zhang, X.}, \bibinfo{author}{Luan, H.}, \bibinfo{author}{Lou,
  H.}, \bibinfo{author}{Liang, T.}, \bibinfo{author}{Chen, S.},
  \bibinfo{author}{Xu, D.}, \bibinfo{author}{Yin, Z.}, \bibinfo{author}{Wang,
  L.}, \bibinfo{author}{Zeng, J.}, \bibinfo{author}{Ren, Y.},
  \bibinfo{author}{Zeng, Z.}, \bibinfo{author}{Shao, Y.}, \bibinfo{author}{Yao,
  K.-F.}, \& \bibinfo{author}{Zeng, Q.} (\bibinfo{year}{2022}{\natexlab{a}}).
\newblock \bibinfo{title}{{Highly variable chemical short-range order in a
  high-entropy metallic glass}}.
\newblock {\it \bibinfo{journal}{Mater. Today Phys.}\/},  {\it
  \bibinfo{volume}{27}\/}, \bibinfo{pages}{100799}.
  \DOIprefix\doi{10.1016/j.mtphys.2022.100799}.
\bibitem[{Zhang et~al.(2020)Zhang, Eremenko, Krayzman, Tucker \&
  Levin}]{Zhang2020}
\bibinfo{author}{Zhang, Y.}, \bibinfo{author}{Eremenko, M.},
  \bibinfo{author}{Krayzman, V.}, \bibinfo{author}{Tucker, M.~G.}, \&
  \bibinfo{author}{Levin, I.} (\bibinfo{year}{2020}).
\newblock \bibinfo{title}{{New capabilities for enhancement of RMCProfile:
  instrumental profiles with arbitrary peak shapes for structural refinements
  using the reverse Monte Carlo method}}.
\newblock {\it \bibinfo{journal}{Journal of Applied Crystallography}\/},  {\it
  \bibinfo{volume}{53}\/}, \bibinfo{pages}{1509--1518}.
  \DOIprefix\doi{10.1107/S1600576720013254}.
\bibitem[{Zhang et~al.(2022{\natexlab{b}})Zhang, Osetsky \& Weber}]{Zhang2022}
\bibinfo{author}{Zhang, Y.}, \bibinfo{author}{Osetsky, Y.~N.}, \&
  \bibinfo{author}{Weber, W.~J.} (\bibinfo{year}{2022}{\natexlab{b}}).
\newblock \bibinfo{title}{{Tunable chemical disorder in concentrated alloys:
  defect physics and radiation performance}}.
\newblock {\it \bibinfo{journal}{Chem. Rev.}\/},  {\it
  \bibinfo{volume}{122}\/}, \bibinfo{pages}{789--829}.
  \DOIprefix\doi{10.1021/acs.chemrev.1c00387}.

\end{thebibliography}

	
\end{document}